\documentclass[11pt,preprint,prd,aps,superscriptaddress,nofootinbib,showpacs,a4paper]{revtex4}
\usepackage[applemac]{inputenc}
\usepackage{color}
\usepackage{graphicx}
\usepackage{epsfig}
\usepackage{amsmath}
\usepackage{amssymb}
\usepackage[normalem]{ulem} 
\newcommand{\be}{\begin{equation}}
\newcommand{\ee}{\end{equation}}
\newcommand{\bqa}{\begin{eqnarray}}
\newcommand{\eqa}{\end{eqnarray}}

\date{\today}
\begin{document}

\date{\today}

\title{Dalitz plot studies of  \mbox{\boldmath $D^0 \to K^0_S K^+ K^-$} decays in a factorization approach}

\affiliation{Sorbonne Universit\'es, Universit\'e Pierre et Marie Curie,\\
Universit\'e de Paris et IN2P3-CNRS, UMR 7585, \\
 Laboratoire de Physique Nucl\'eaire et de Hautes \'Energies,  4 place Jussieu, 75252 Paris, France}

\affiliation{Division of Theoretical Physics, The Henryk Niewodnicza\'nski Institute of Nuclear Physics,\\
                  Polish Academy of Sciences, 31-342 Krak\'ow, Poland\\}

\thanks{Retired Professor}

\author{\bf {J.-P.~Dedonder$^{1*}$}}

\author{\bf {R.~Kami\'nski$^2$}}
                  
\author{\bf {L.~Le\'sniak$^2$}}

\author{\bf {B.~Loiseau$^1$}}

\begin{abstract}
\noindent The \textit{BABAR} Collaboration data of the $D^0 \to K^0_S K^+ K^-$ process are analyzed within a quasi-two-body factorization framework. Starting from the weak effective Hamiltonian, one has  to evaluate  matrix elements of $D^0$ transitions to  two kaons for the tree amplitudes and  the transitions between one kaon and two kaons for the annihilation ones ($W$-exchange). In earlier studies, assuming these transitions to proceed through the dominant intermediate resonances, we approximated them as being proportional to the kaon form factors. Here, to obtain a good fit, one has to multiply the scalar-kaon form factors, derived from unitary relativistic coupled-channel models  or in a dispersion relation approach, by phenomenological energy-dependent functions. The final state kaon-kaon interactions in the $S$-, $P$- and $D$- waves are taken into account. All $S$-wave channels are treated in a unitary way.  In other respects, it is shown in a model-independent manner that the $K^+K^-$ and $\bar K^0 K^+$ $S$-wave effective mass squared distributions, corrected for phase space, are significantly different. At variance with the \textit{BABAR} analysis, it means that the $f_0(980)$ resonance must be included in the phenomenological analysis of the $D^0 \to K^0_S K^+ K^-$ data. The best fit described in the main text has 19 free parameters and indicates i) the dominance of annihilation amplitudes, ii) a large dominance of the  $f_0(980)$ meson in the near threshold $K^+K^-$ invariant mass distribution, and iii) a sizable branching fraction to the  $[\rho(770)^+ + \rho(1450)^+ + \rho(1700)^+] K_S^0$ final states. A first appendix provides an update of the determination of the isoscalar-scalar meson-meson amplitudes based on an enlarged set of data embodying new precise low energy $\pi \pi$ data.  A second appendix proposes two alternative fits using the scalar-kaon form factors calculated from the Muskhelishvili-Omn\`es dispersion relation approach. These fits have $\chi^2$ quite close to  that of the best fit but they show important contributions from both the $f_0$ and $a_0^0$ mesons and  a weaker role of the $\rho^+$ mesons.
\end{abstract}

\maketitle

\section{Introduction} \label{Introduction}

Measurements of the $D^0$-$\overline{D}^0$ mixing parameters, through Dalitz-plot time dependent amplitude 
analyses of the weak process $D^0 \to K^0_S K^+ K^-$, have been performed by the  Belle~\cite{Zupanc2009} and 
\textit{BABAR} collaborations~\cite{B10}. Such studies could help in the understanding of the origin of mixing and may indicate the presence of new physics contribution. As predicted by the standard model in the charm sector, the violation of the \textit{CP} symmetry should be small for these $D^0$ decays. In Refs.~\cite{Zupanc2009,B10} the description of the $D^0 \to K^0_S K^+ K^-$ decay amplitude has been performed using the isobar model developed in~\cite{B5}, extended in~\cite{Aubert2008} and~\cite{B10,supmaPRL105}.  The isobar model has also been applied in the experimental analysis based on the data taken from the BESIII experiment~\cite{Weidenkaff, Ablikim2006.02800}.
 
The Cabibbo-Kobayashi-Maskawa (CKM) angle $\gamma$ (or $\phi_3$) has been evaluated from the analyses of the  $B^\pm \to D^0 K^\pm$, with $D^0 \to K^0_S \pi^+ \pi^-$ and $D^0 \to K^0_S K^+ K^-$decays
\cite{SanchezPRL105121801,Poluektov2010,J.P.Lees_PRD87_052015_BABAR,LHCbNP2014}.
This angle can be also measured using some knowledge on the strong-phase difference between $D^0$ and
 $\bar{D}^0 \to K^0_S K^+ K^-$ decay amplitudes obtained by the CLEO Collaboration~\cite{J.Libby_PRD82_CLEO}. This method has been used by the Belle~\cite{H.Aihara_Belle2012}, LHCb~\cite{R.Aaij_JHEP2018} and BESIII~\cite{BES2007}  collaborations.

A good knowledge of the final state meson interactions in the $D^0 \to K^0_S K^+ K^-$ decays is important to reduce the uncertainties in the determination of  the $D^0$-$\overline{D}^0$ mixing parameters and of the CKM angle $\gamma$.
The structures seen in the Dalitz plot spectra point out to  the complexity of these final state strong interactions. Their studies can provide a better understanding of the strange meson interactions and of the $D^0$ decay mechanism into $K^0_S K^+ K^-$.

The experimental analyses like that of Ref.~\cite{B10} rely mainly upon the use of the isobar model.
For a given reaction, this model has basically two fitted parameters for each part of the decay amplitude. 
In this approach one can take into account  many existing resonances coupled to the interacting pairs of mesons. In Refs.~\cite{B10,supmaPRL105}, the authors introduce explicitly eight resonances $a_0(980)^0$
, $a_0(980)^{+}$
, $a_0(980)^{-}$
, $\phi(1020)$
, $f_2(1270)$
, $f_0(1370)$
, $a_0(1450)^0$
, $a_0(1450)^{+}$
 . Their analysis rely on 17 free parameters. However, the decay amplitudes are not unitary and unitarity is not preserved in the three-body decay channels; it is also violated in the two-body subchannels. Furthermore, it is  difficult  to differentiate the $S$-wave amplitudes from the nonresonant background terms. Their interferences are  noteworthy and then some two-body branching fractions, extracted from the data, could be unreliable. One of the difficulty in the experimental analyses based on the isobar model is the choice of the resonances needed to reach a good agreement with the Dalitz plot data.  In Ref.~\cite{Aubert2008} the \textit{BABAR} collaboration authors have added the scalar $a_0(1450)$ to their model developed in 2005~\cite{B5}.
In the recent BESIII analysis~\cite{Ablikim2006.02800} the Dalitz plot is described with six resonances: $a_0(980)^0$
,  $a_0(980)^{+}$
, $\phi(1020)$
, $a_2(1320)^+$
,  $a_2(1320)^-$
, $a_0(1450)^-$.

Extending our previous work on the  $D^0 \to K^0_S \pi^+ \pi^-$ decays~\cite{JPD_PRD89}, we analyze, in the quasi-two-body factorization framework, the  $D^0 \to K^0_S K^+ K^-$ data provided by the \textit{BABAR} collaboration~\cite{B10}.
As in our earlier studies, we assume that two of the three final-state mesons form a single state which originates from a quark-antiquark, $q \bar q$, pair and then apply the factorization procedure to these quasi-two-body final states.
Starting from the weak effective Hamiltonian,  we derive tree and annihilation ($W$-meson exchange) amplitudes both being either Cabibbo favored (CF) with $c \to s\bar{d} u$ transition or doubly Cabibbo suppressed (DCS) with $c \to du \bar s$ transition.

In the factorization approach, the CF and DCS  amplitudes are expressed as superpositions of appropriate effective coefficients and two products of two transition matrix elements. 
The kaon form factors do not appear explicitly except
the isovector ones that enter in only one term of the CF tree amplitude\footnote{ See the $a_1(m_c)$ term of Eq.~(\ref{TCF}).}.
In all other terms of our amplitudes,
one has to evaluate either, for the tree ones, the matrix elements of $D^0$ transitions to two-kaon states or, for the annihilation ones, the transitions between one kaon and two kaon-states. 
Similarly to previous studies~\cite{Boito}, assuming these transitions to proceed through the dominant intermediate resonances, we have approximated them as being proportional to the isoscalar or isovector  kaon form factors.  
 
 Taking advantage of the coupling between the $\pi \pi$ and the $K\overline{K}$ channels and extending the derivation of the unitary isoscalar-scalar pion form factor~\cite{DedonderPol} to that of the kaon one, two of the present authors (L.~L. and R.~K.) together with two collaborators, have recently studied, in the quasi-two-body QCD factorization approach, the $B^\pm \to K^+ K^- K^\pm$ decays~\cite{PLB699_102,KKK}. These $S$-wave form factors are derived using a unitary relativistic three coupled-channel model including $\pi \pi$, $K \bar K$ and effective $(2\pi)(2\pi)$ interactions together with chiral symmetry constraints. They include the contributions of the $f_0(980)$ and $f_0(1400)$ resonances and require the knowledge of the  isoscalar-scalar meson-meson amplitudes from the two kaon threshold to energies above the $D^0$ mass.
The parameters of these amplitudes derived in the late nineties by three of us (R.~K., L.~L. and B.~L.)~\cite{KLL,EPJ} 
have been updated using new precise low energy  $\pi \pi$ data together with an enlarged set of data as is shown in Appendix~\ref{UpdatedTpipi}.

The calculation requires also the knowledge of a contribution proportional to the isovector-scalar form factor; it is represented by a function calculated from a unitary model with relativistic two-coupled channel $\pi \eta$ and $K \bar K$ equations based on the  two-channel model of the  $a_0(980)$ and $a_0(1450)$ resonances built in~\cite{AFLL, AFLL2}.  

The vector form factors have been calculated using vector dominance, quark model assumptions and isospin symmetry in Ref.~\cite{Bruch2005}. They receive contributions from the vector mesons: $\rho(770)$, $\rho(1450)$, $\rho(1700)$, $\omega(782)$, $\omega(1420)$, $\omega(1850)$, $\phi(1020)$ and $\phi(1680)$. The isoscalar-tensor amplitude, saturated by the $f_2(1270)$, is described by a relativistic Breit-Wigner term. The isovector-tensor resonance $a_2(1320)$ has a mass close to that of the $f_2(1270)$.
If the contribution of these tensor mesons are described by relativistic Breit-Wigner components, it is difficult to disentangle them because of the degeneracy in their masses, widths and partial decay widths  into $K \overline{K}$~\cite{PDG2020}.
Consequently, as in Ref.~\cite{B10}, we consider only the $f_2(1270)$ to represent the $D$ wave.
It is an ``effective" $f_2(1270)$ which takes into account both tensor mesons.

In the present approach, a best fit is achieved in which the data are reproduced with amplitudes that are optimized notably by adjusting the isoscalar- and isovector-scalar form factors. It is interesting then to see what could be the  outcome of a model   
 based, for instance, on the scalar form factors calculated from the Muskhelishvili-Omn\`es dispersion relation approach~\cite{MO, Moussallam_2000, Moussallam_2019, Bachir2015}. 
As in our best fit model ($\chi^2/ndf = 1.25$), we have to introduce energy dependent phenomenological functions multiplying the scalar form factors to obtain two fits with almost as good $\chi^2/ndf $. 
{Branching fractions of these two alternative models are compared to those of our best fit model.

The paper is organized as follows. A full derivation of the  $D^0 \to K^0_S K^+ K^-$ decay amplitude, in the framework of the quasi-two-body factorization approach is given in Sec.~\ref{amplitudes}. Based only on the experimental data of the \textit{BABAR} Collaboration~\cite{B10} and without any model assumptions, we show in Sec.~\ref{comp} that the $f_0(980)$ contribution, at variance with the \textit{BABAR} analysis, has to be included in the decay amplitude.  Section~\ref{results} presents the result of our  best fit to the Dalitz plot data sample of Ref.~\cite{B10}. Some discussion and concluding remarks can be found in Sec.~\ref{conclusions}. Appendix~\ref{UpdatedTpipi} presents the update of the description of the $\pi \pi$, $\bar K K$ and effective $(2\pi)(2\pi)$ S-wave isospin zero scattering amplitudes. Two alternative models for the $D^0 \to K^0_S K^+ K^-$ decay amplitude, using kaon scalar-form factors derived from the dispersion relation approach, are presented in Appendix~\ref{fitswMOBM}.

\section{The $D^0 \to K^0_S K^+ K^-$ decay amplitudes in a factorization framework} \label{amplitudes}

The decay amplitudes for  the $D^0 \to K^0_S K^+ K^-$ process 
 can be evaluated as matrix elements of the effective weak Hamiltonian \cite{Buchalla1996}
\be \label{Heff}
H_{eff}=\frac{G_F}{\sqrt{2}} V_{CKM} \Big[ C_1(\mu) O_1(\mu)+ C_2(\mu) O_2(\mu) \Big] + h.c., 
\ee
 where the coefficients $V_{CKM}$ are given in terms of Cabibbo-Kobayashi-Maskawa quark-mixing matrix 
elements  and $G_F$ denotes the Fermi coupling constant. The $C_i(\mu)$ are the Wilson coefficients of 
the  four-quark operators  $O_i(\mu)$ at a renormalization scale $\mu$, chosen to be equal to the 
$c$-quark mass $m_c$. The left-handed current-current operators $O_{1,2}(\mu)$ arise from $W$-boson exchange.

The transition matrix elements that occur in the present work require two specific values of  the $V_{CKM}$ coupling matrix elements~\cite{JPD_PRD89}:
 \be \label{lambda}
\Lambda_1\equiv V^*_{cs} V_{ud} = 1- \lambda^2  \hspace{1cm}{\rm and}\hspace{1cm}
 \Lambda_2\equiv V^*_{cd} V_{us}= - \lambda^2,
\ee
 where $\lambda
 $ is the sine of the Cabibbo angle and where $\Lambda_1$ is associated to Cabibbo favored (CF) transitions while $\Lambda_2$ is associated to doubly Cabibbo suppressed (DCS) amplitudes. 
The amplitudes are functions of the Mandelstam invariants 
 \be \label{3sa}
 s_{\pm}= m_{\pm}^2 =(p_0+p_\pm)^2,\hspace{2cm}  s_{0}=m_0^2=(p_+ + p_-)^2,
 \ee
 where $p_0$, $p_+$ and $p_-$ are the four-momenta of the $K_S^0$, $K^+$ and $K^-$ mesons, 
respectively. Energy-momentum conservation implies  
\be \label{pD0}
 p_{D^0}=p_0+p_++p_-\hspace{1cm}
{\rm and} \hspace{1cm}
s_0+s_++s_-=m_{D^0}^2+m_{K^0}^2 + 2 m_K^2,\ee
where $p_{D^0}$ is the $D^0$ four-momentum and  $m_{D^0}=1864.83$ MeV, $m_{K^0}=497.611$ MeV and $m_K=493.677$ MeV denote the  $D^0$, $K^0$  and charged kaon masses (Ref.~\cite{PDG2020}).

\subsection{Tree and annihilation CF and DCS amplitudes} \label{Tree}

The full amplitude is the superposition of two tree CF and DCS amplitudes, $T^{CF}(s_0,s_-,s_+)$ and $T^{DCS}(s_0,s_-,s_+)$ and of two annihilation (i.e., exchange of W meson between the $c$ and $\overline  u$ quarks of the $D^0$) CF and DCS amplitudes, $A^{CF}(s_0,s_-,s_+)$ and $ A^{DCS}(s_0,s_-,s_+)$. Thus,  one writes the full amplitude as

\bqa
\label{fulamp}
\mathcal{M}(s_0,s_-,s_+)&=&
\left \langle K^0_S(p_0)\ K^+(p_+){K^-}(p_-) \vert H_{eff}\vert D^0(p_{D^0}) \right \rangle \nonumber \\
&=& T^{CF}(s_0,s_-,s_+) + T^{DCS}(s_0,s_-,s_+)+ A^{CF}(s_0,s_-,s_+) + A^{DCS}(s_0,s_-,s_+).
\eqa
 Although the three variables $s_0,s_-,s_+$ appear as arguments of the amplitudes, all amplitudes depend only on two of them because of the relation~(\ref{pD0}).

\begin{figure}[h]  \begin{center}
\includegraphics[scale=0.7]{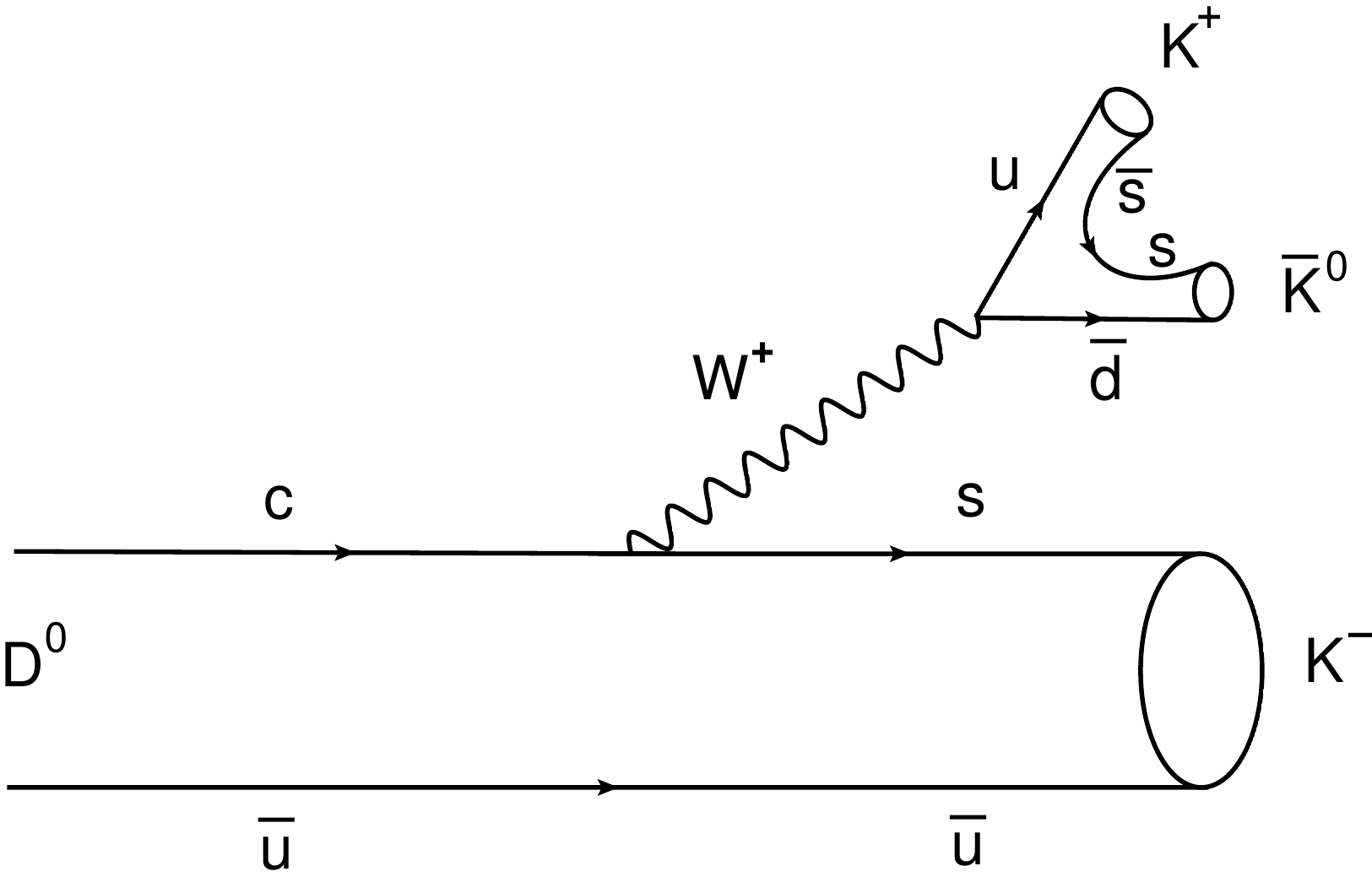}
\caption{ Tree diagrams for Cabibbo favored amplitudes  proportional to  $\Lambda_1 \ a_1$ with  $K^- \left [\overline{K}^0\  K^+\right ]$ final states}\label{F1}
\end{center} \end{figure}

Assuming that the factorization approach \cite{Beneke2003,BurasNPB434_606,Buchalla1996,AliPRD58} holds,
the diagram illustrated in Fig.~\ref{F1} is proportional to $\Lambda_1  a_1(m_c)$ with quasi-two-body $K^-[\overline{K}^0  K^+]_L^{I=1}$ final states and the diagram in Fig.~\ref{F2}, proportional to  $\Lambda_1 a_2(m_c)$ with quasi-two-body $\overline{K}^0[K^+  K^-]_{L}^I $ with angular momentum $L=S,P,D$ and isospin $I=0,1$ states, yield the tree CF amplitude $T^{CF}(s_0,s_-,s_+) $  which reads, with $\vert 0 \rangle$ denoting the vacuum state, 
\bqa \label{TCF}
T^{CF}(s_0,s_-,s_+) \!
&\simeq & \frac{G_F}{2} \Lambda_1 \! \sum_{L={S,P,D}}  \left \{ a_1(m_c) 
 \langle \left [ \overline{K}^0 (p_0) \ K^+(p_+) \right ]_L^{I=1} \vert (\overline{u} \ d)_{V-A}\vert 0 \rangle 
\langle K^-(p_-)\vert (\overline  s \ c)_{V-A}\vert D^0(p_{D^0})\right .\rangle \nonumber \\  
& +&  \left . a_2(m_c)   \sum_{I={0,1}} \  \langle \overline{K}^0(p_0)\vert (\overline{s} \ d)_{V-A} \vert 0 \rangle   
  \langle  [K^+(p_+){K^-}(p_-) ]_{L,u}^I \vert (\overline{u} \ c)_{V-A} \vert D^0(p_{D^0})\rangle \right \} \nonumber \\ 
&=&\sum_{L={S,P,D}} \! \! \left [T^{CF}_{[\overline{K}^0 K^+]_L^1 K^-}(s_0,s_-,s_+)
 +  T^{CF}_{\overline{K}^0 [K^+K^-]_{L,u}^{0} }(s_0,s_-,s_+) 
 +  T^{CF}_{\overline{K}^0 [K^+K^-]_{L,u}^{1} }(s_0,s_-,s_+)\right ]\nonumber \\
&=&T^{CF}_{[\overline{K}^0 K^+]^1 K^-}(s_0,s_-,s_+) + \sum_{I=0,1} T^{CF}_{\overline{K}^0 [K^+K^-]_u^{I} }(s_0,s_-,s_+).
\eqa
The short hand notation of the last line of Eq.~(\ref{TCF}) implies the L summation\footnote{ {\it e.g.}, 
$$T^{CF}_{[\overline{K}^0 K^+]^1 K^-}(s_0,s_-,s_+) = \sum_{L={S,P,D}} T^{CF}_{[\overline{K}^0 K^+]_L^1 K^-}(s_0,s_-,s_+), etc.$$}. It will be used wherever necessary. In the case of the creation of a $K^+K^-$ pair we indicate by a subscript $q$ the $q \overline{q}$ pair from which it originates (here a $u \overline{u}$ one, as seen in Fig.~\ref{F2}). We shall therefore use the notation $[K^+K^-]_{L,q}^I$ and/or $[K^+K^-]_q^I$ whenever necessary.  We have also introduced the short-hand notation 
\be \label{qbarqVA}
(\overline {q}\ q)_{V-A}=\overline {q} \gamma\ (1 -\gamma_5)\ q 
\ee
which will be used throughout the text ($\gamma$ and $\gamma_5$ are Dirac's matrices). In deriving Eq.~(\ref{TCF}) small $CP$ violation effects in $K^0_S$ decays are neglected and we use 
 \be \label{K0S}
\vert K^0_S \rangle \approx \frac{1}{\sqrt{2}}\  \left (\vert K^0 \rangle+ \vert \overline{K}^0 \rangle \right ).
\ee
 At leading order in the strong coupling constant $\alpha_S$, the effective QCD factorization 
coefficients $a_1(m_c)$ and $a_2(m_c)$ are expressed as 
\be  \label{a12}
 a_1(m_c)=C_1(m_c)+\frac{C_{2}(m_c)}{N_C},\hspace{1cm} a_2(m_c)=C_2(m_c)+\frac{C_{1}(m_c)}{N_C},
 \ee
 where  $N_C=3$ is the number of colors.  Higher order vertex and hard scattering corrections are not discussed in the present work and we introduce effective values for these coefficients. From now on, the simplified notation 
 $a_1 \equiv  a_1(m_c)$ and $a_2 \equiv  a_2(m_c)$ will be used. As in Ref.~\cite{JPD_PRD89}, we take $a_1=1.1$ and $a_2=-0.5$.
 
 \begin{figure}[h]  \begin{center}
\includegraphics[scale=0.7]{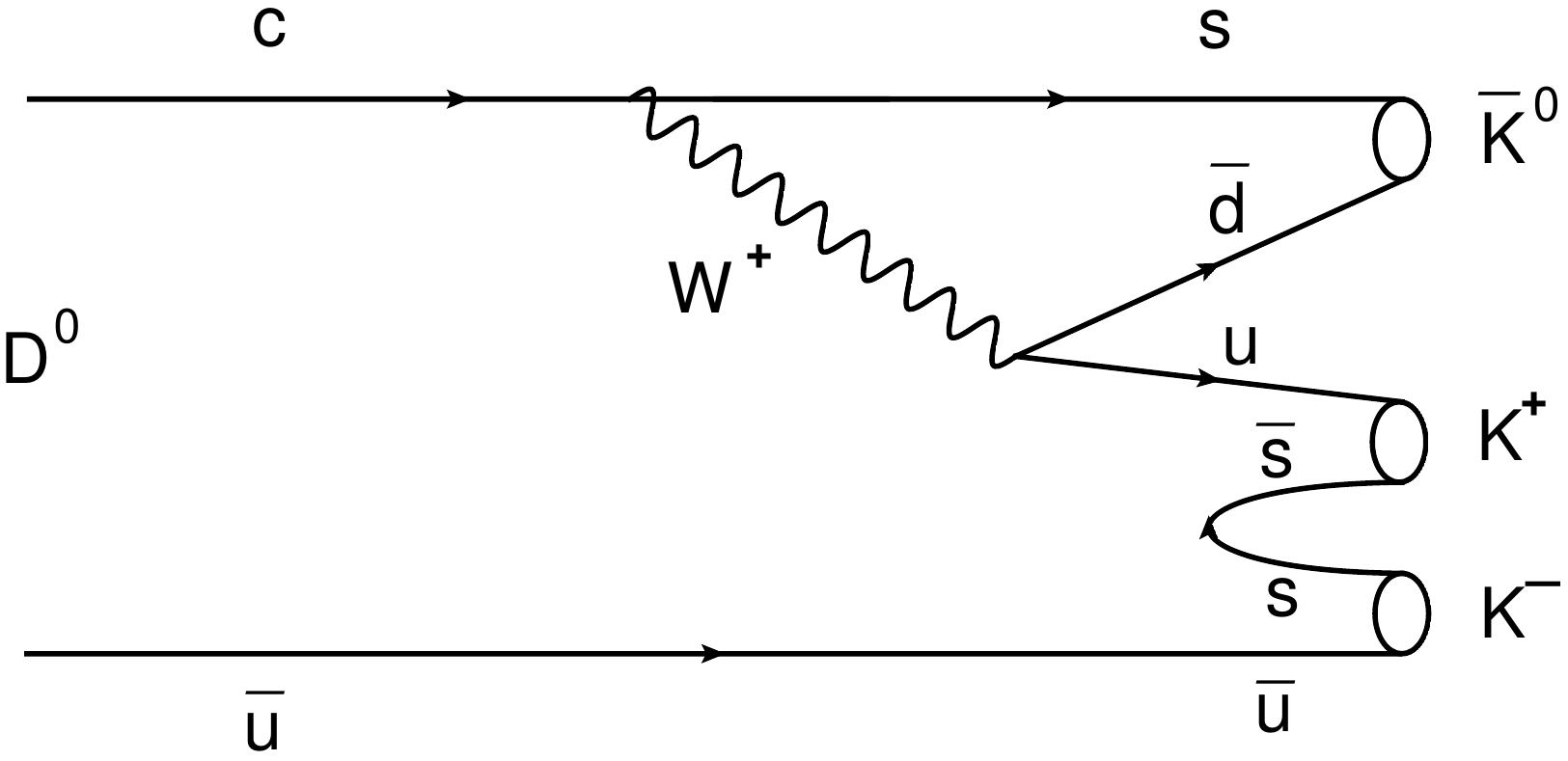}
\caption{ Tree diagram for Cabibbo favored amplitudes  proportional to  $\Lambda_1 a_2$ with  $\overline{K}^0\  \left [K^+K^-\right ]_u^{I=0,1}$ final states}\label{F2}
\end{center} \end{figure}

Similarly, the DCS tree amplitudes illustrated in Figs.~\ref{F3} and~\ref{F4}, yield the amplitude $T^{DCS}(s_0,s_-,s_+)$
\bqa \label{TDCS}
T^{DCS}(s_0,s_-,s_+) \!
&\simeq &\! \frac{G_F}{2}\ \Lambda_2 \sum_{L={S,P,D}} \!  \Big \{ a_1  \langle K^+(p_+) 
\vert (\overline  u \ s)_{V-A}\vert 0   \rangle\ \langle [K^0(p_0 )\ K^-(p_-)]_L^1 \vert (\overline  d \ c)_{V-A}\vert D^0(p_{D^0}) \rangle \nonumber \\
& +&   a_2  \sum_{I=0,1}  \langle K^0(p_0)\vert (\overline  d \ s)_{V-A} \vert 0  \rangle \ \langle
[K^+(p_+)\ K^-(p_-) ]_{L,u}^I\vert (\overline  u \ c)_{V-A} \vert D^0(p_{D^0}) \rangle \Big \} \nonumber \\ 
  &=&\! \sum_{L={S,P,D}} \! \left [ T^{DCS}_{K^+[{ K}^0 K^-]_L^1}(s_0,s_-,s_+)  
  + T^{DCS}_{{K}^0[K^+ K^-]_{L,u}^{0}}(s_0,s_-,s_+)
+ T^{DCS}_{{K}^0[K^+ K^-]_{L,u}^{1}}(s_0,s_-,s_+) \right ]\nonumber \\
&=&\!  T^{DCS}_{K^+[{ K}^0 K^-]^1}(s_0,s_-,s_+) \! + \sum_{I=0,1} T^{DCS}_{{K}^0[K^+ K^-]_u^{I}}(s_0,s_-,s_+).
\eqa

\begin{figure}[h]  \begin{center}
\includegraphics[scale=0.7]{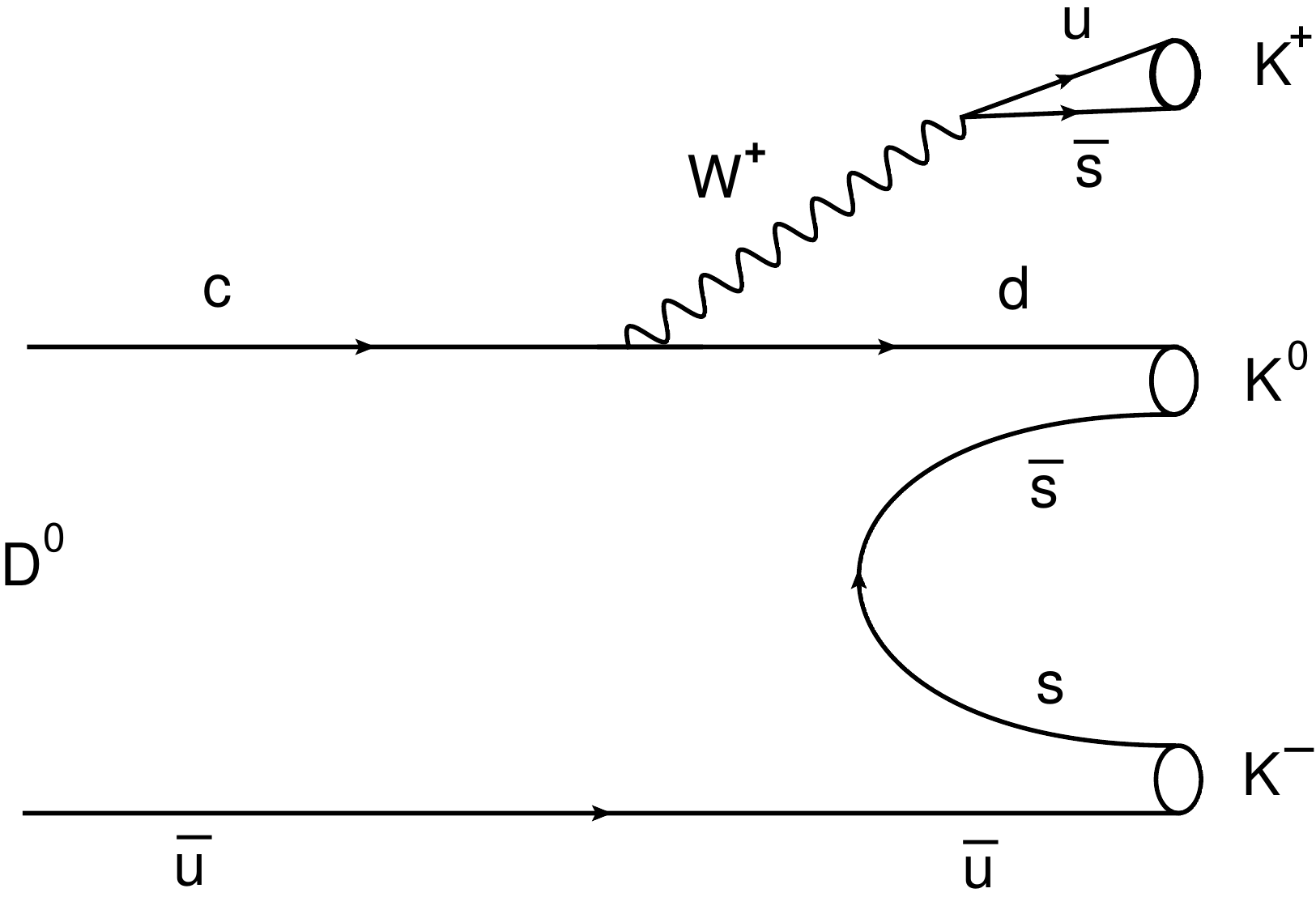}
\caption{Tree diagram for doubly Cabibbo suppressed amplitudes for $K^+ \left [{K}^0 K^-\right ] ^{I=1}$ final states}\label{F3}
\end{center} \end{figure}

\begin{figure}[]  \begin{center}
\includegraphics[scale=0.7]{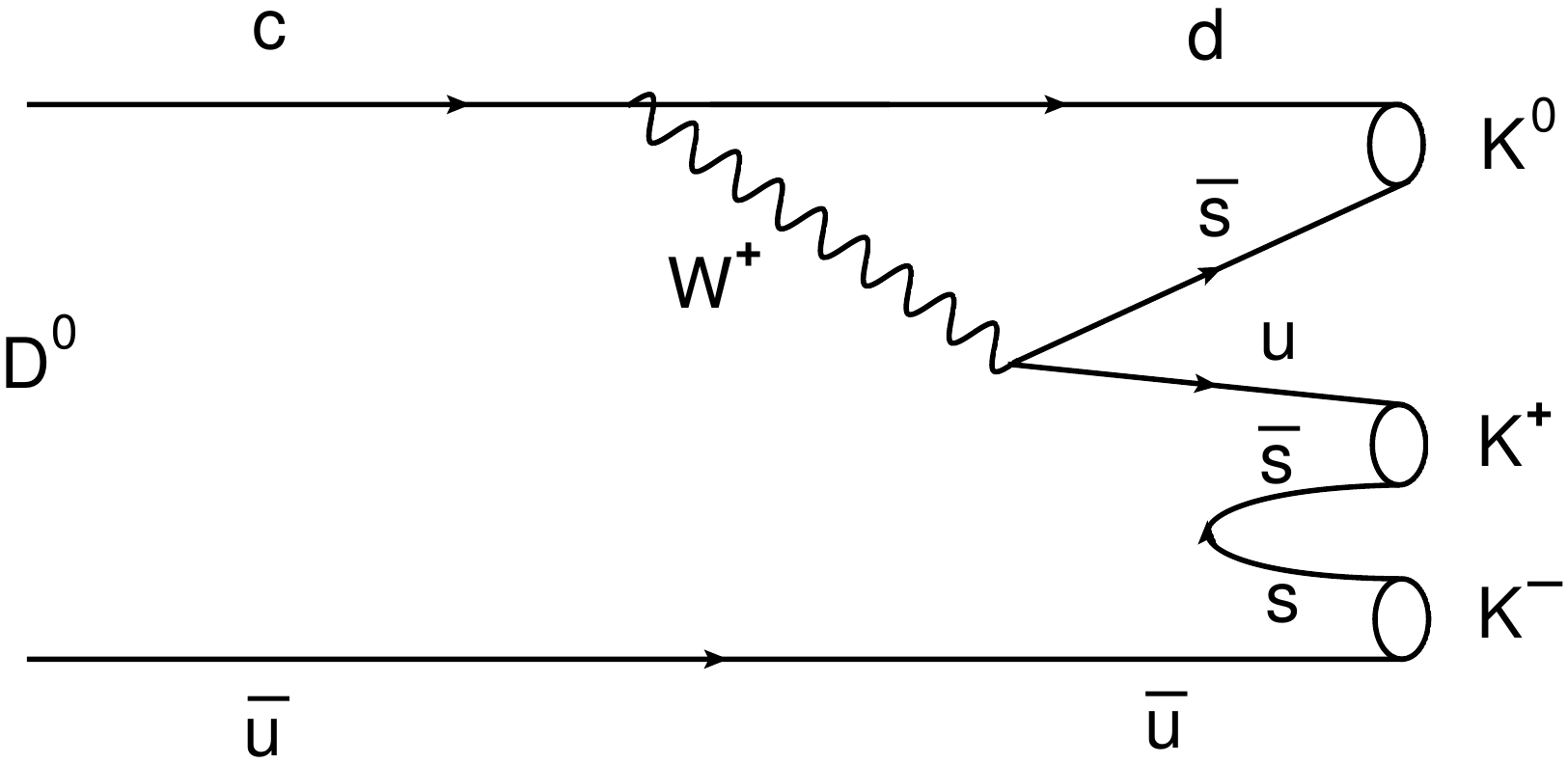}
\caption{Tree diagram for doubly Cabibbo suppressed amplitudes for ${K}^0 \left [ K^+\ K^- \right ] _u^{I=0,1}$ final states}\label{F4}
\end{center} \end{figure}

A similar derivation goes for the CF annihilation amplitudes illustrated by the diagram of Fig.~\ref{F5}, $A^{CF}(s_0,s_-,s_+)$, so that one has\footnote{In the amplitudes~(\ref{ANNCF}) and (\ref{ANNDCS}) we neglect the terms with the quasi two-boby $K^+(p_+)  [ \overline{K}^0 (p_0)  K^-(p_-) ]_{L}^{1}$ and $K^-(p_-) [K^+(p_+)K^0(p_0)]_{L}^{1} $ final states. One expects a small contribution of these terms because there exist no strangeness -2 ( $[ \overline{K}^0   K^- ]_{L}^{1}]$ state) and +2 ($ [K^+K^0]_{L}^{1}]$ state)  resonances.}
 
\bqa
\label{ANNCF}
A^{CF}(s_0,s_-,s_+)&\simeq&
 \frac{G_F}{2}\ \Lambda_1\ a_2   \sum_{L={S,P,D}}  \Big \{ \sum_{I={0,1}}
  \langle \overline{K}^0 (p_0) [K^-(p_-) K^+(p_+) ]_{L,s}^I
  \vert (\overline{s} \ d)_{V-A}\vert 0  \rangle \nonumber \\
&+&  \langle K^-(p_+)  [ \overline{K}^0 (p_0)  K^+(p_-) ]_{L}^{1}
  \vert (\overline{s} \ d)_{V-A}\vert 0  \rangle  \Big \} \ 
 \langle 0\vert (\overline{c} \ u)_{V-A} \vert D^0(p_{D^0}) \rangle \nonumber \\ 
&=& \sum_{L={S,P,D}} \Big [A^{CF}_{K^-  [\overline{K}^0 K^+]_{L}^{1}} (s_0,s_-,s_+) +
 A^{CF}_{\overline{K}^0 [K^+ K^-]_{L,s}^{0}}(s_0,s_-,s_+)\Big ]   \nonumber \\
 &=& A^{CF}_{K^- [\overline{K}^0 K^+]^1} (s_0,s_-,s_+) +  A^{CF}_{\overline{K}^0 [K^+ K^-]_{s}^{0}}(s_0,s_-,s_+).
\eqa

\begin{figure}[h]  \begin{center}
\includegraphics[scale=0.7]{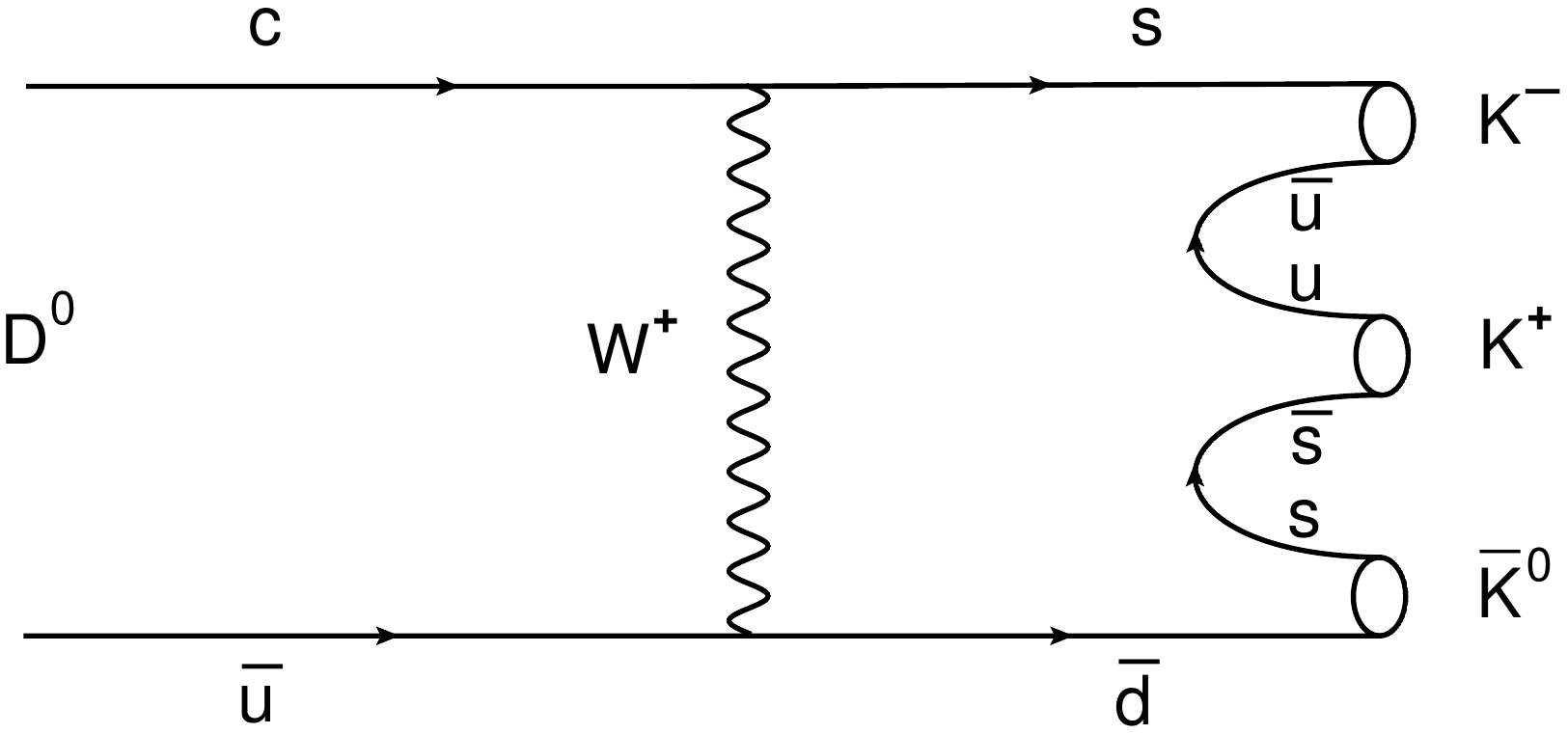}
\caption{Diagram for  CF  annihilation ($W$-exchange) amplitudes with $\overline{K}^0 \ \left [K^+\ K^- \right ]_s^{I=0} $ 
or $K^-  [K^+\overline{K}^0 ]^{I=1} $
final states. }\label{F5}
\end{center} \end{figure}

\begin{figure}[h]  \begin{center}
\includegraphics[scale=0.7]{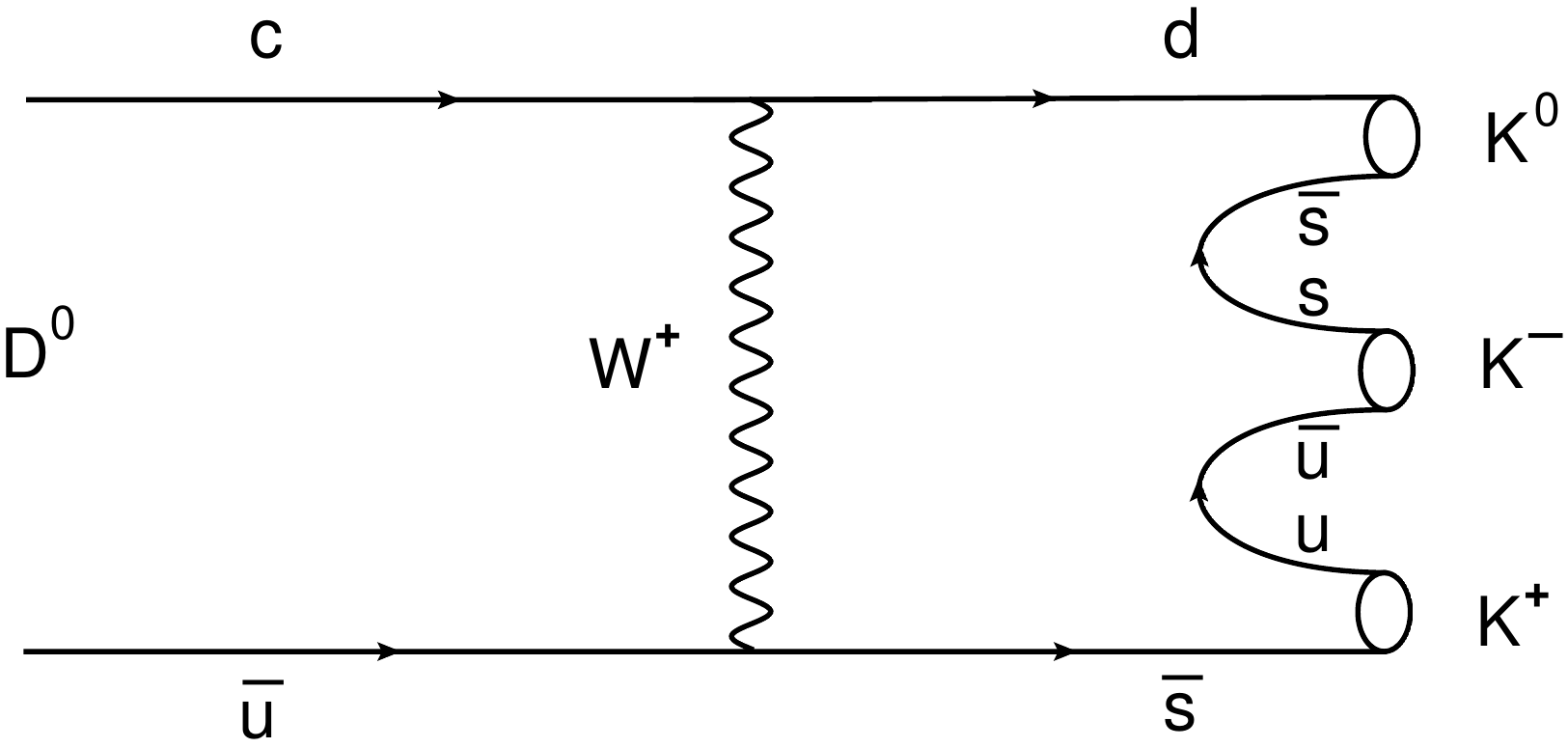}
\caption{Diagram for  DCS annihilation ($W$-exchange) amplitudes with ${K}^0 \ \left [K^+\ K^- \right ]_s^{I=0} $ 
or $K^+\left [K^0 K^-\right ]^{I=1} $ final states. }
\label{F6W}
\end{center} \end{figure}
The corresponding  DCS annihilation amplitudes, $A^{DCS}(s_0,s_-,s_+)$,  (see Fig.~\ref{F6W}), are easily obtained from the CF amplitudes in Eq.~(\ref{ANNCF})
\bqa
\label{ANNDCS}
A^{DCS}(s_0,s_-,s_+)&\approx&
 \frac{G_F}{2}\ \Lambda_2\ a_2 \ \sum_{L={S,P,D}} \sum_{I=0,1} 
 \Big [ \langle  [K^-(p_-) K^+(p_+) ]_{L,s}^I  {K}^0 (p_0)
  \vert (\overline{d} \ s)_{V-A}\vert 0  \rangle  \nonumber \\
&+&  \langle  K^+(p_+) [K^0(p_0) K^-(p_-)]_{L}^{1} 
 \vert (\overline {d} \ s)_{V-A} \vert 0  \rangle \Big ]
 \langle 0\vert (\overline{c} \ u)_{V-A} \vert D^0(p_{D^0}) \rangle \nonumber \\  
&=& \sum_{L={S,P,D}} \Big [A^{DCS}_{K^+ [K^0 K^-]_{L}^{1}}(s_0,s_-,s_+) + 
A^{DCS}_{K^0[K^+ K^-]_{L,s}^{0}}(s_0,s_-,s_+) \Big ]   \nonumber \\
&=&   A^{DCS}_{K^+ [ K^0 K^-]^{1}}(s_0,s_-,s_+) +  A^{DCS}_{ K^0[K^+ K^-]_s^{0}}(s_0,s_-,s_+). 
\eqa
Let us now review in detail all the amplitudes that will have to be evaluated. We will follow closely the construction detailed in Ref. \cite{JPD_PRD89}.\\

\subsection{Explicit tree amplitudes} \label{STA}

In the following, starting from the expressions given in Eqs.~(\ref{TCF}) for the CF amplitudes and in~(\ref{TDCS}) for the DCS ones, we will express  the different three-body matrix elements entering in the amplitudes in terms of vertex functions noted $G_{R_{S,P,D}[\overline{K}^0K^+]^{1}}(s)$ in the case of a $[\overline{K}^0K^+]^{1}$ final state,
$G_{R_{S,P,D}[{K}^0K^-]^{1}}(s)$ in the case of a $[{K}^0K^-]^{1}$ final state, and 
$G_{R_{S,P,D}[{K}^+K^-]_q^{0,1}}(s)$ in the case of a $[{K}^+K^-]_q^{0,1}$ final state. 
The vertex functions describe the decays into $K \overline{K}$ of the possibly present intermediate resonances $R_{S, P, D}$ which contribute to the process.\\

Further on, we will need to introduce transition form factors for which, as in Ref.~\cite{JPD_PRD89}, we will assume isospin and charge conjugation symmetry so that the following equations arise:
\bqa
F_0^{D^0a_0^-}(s) &=&  \sqrt{2} \ F_0^{ D^0a_0^0}(s), \nonumber \\
F_{0}^{[{\overline{K}^0K^+}]^1}(s) &=& F_{0}^{[{{K}^0K^-}]^1}(s), \label{FSKbarK} \nonumber \\
F_{1}^{\overline{K}^0K^+}(s) &=& -F_{1}^{{K}^0K^-}(s), \nonumber \\
F_0^{K^-a_0^+}(m_{D^0}^2) &=& \sqrt{2}\ F_0^{K^0a_0^0}(m_{D^0}^2),\nonumber\\
A_0^{D^0\rho^-}(s) &=&  \sqrt{2} \  A_0^{D^0\rho^0}(s), \nonumber \\
A_0^{\overline{K}^0 \phi}(m_{D^0}^2) &=&  A_0^{K^0 \phi}(m_{D^0}^2). 
\eqa
In the above equations $F_0$ and $F_1$ denote scalar and vector transition form factors of two pseudoscalar mesons while $A_0$'s are transition form factors of pseudoscalar and vector mesons.

\subsubsection {\it Scalar amplitudes}

Following a derivation similar to that developped in Ref.~\cite{JPD_PRD89}, the isoscalar-scalar CF amplitude associated to the $\overline{K}^0 \ [K^+ K^-]_{S,u}^{I}$ final states can be described by (see Fig.~\ref{F2})
\bqa
\label{ISSKPKM}
T^{CF}_{\overline{K}^0 \ [K^+ K^-]_{S,u}^{0}}(s_0,s_-,s_+) &=& - \frac{G_F}{2}\ \Lambda_1 \ a_2\  \frac{f_{K^0}}{\sqrt{2}}\ ( m_{D^0}^2 - s_0)\  \sum_{R_S} F_0^{D^0 R_S[K^+K^-]_u^{0}}(m_{K^0}^2) \nonumber \\
&\times&\ G_{R_S[K^+K^-]_u^{0}}(s_0) \ \langle R_S[K^+K^-]_u^{0} \vert u \overline{u}\rangle, 
\eqa
where $f_{K^0}$ is the $K^0$ decay constant and the sum over $R_S$ runs over the possibly contributing resonances in the isoscalar-scalar channel. It can be seen here that we have approximated  the three-body matrix element 
$\langle [K^+(p_+){K^-}(p_-) ]_{S,u}^0 \vert (\overline{u} \ c)_{V-A} \vert D^0(p_{D^0})\rangle$ entering Eq.~(\ref{TCF}) by the above sum over $R_S$. It thus includes the contributions of the $f_0(500)$, $f_0(980)$  and of the $f_0(1370)$ and $f_0(1500)$ resonances. The $D^0$ to $R_S$ transition form factor entering Eq.~(\ref{ISSKPKM}) could have a different value for each resonance $R_S$; here we can assume that its variation from one resonance to the other is small and we can choose for its value that of the transition to $f_0(980)$. Unless otherwise specified, by $f_0$ in $F_0^{D^0 f_0}(m_{K^0}^2)$ we mean $f_0(980)$. We may parametrize the sum over $R_S$ by introducing the isoscalar-scalar form factor, $ \Gamma_2^{n*}(s_0)$, where $n$ denotes a nonstrange quark pair and which can be built following  the method discussed in Ref.~\cite{DedonderPol}. We then apply the following approximation 
\be \label{SUMFFD0}
\sum_{R_S}  F_0^{D^0 R_S[K^+K^-]_u^{0}}(m_{K^0}^2) \ G_{R_S[K^+K^-]_u^{0}}(s_0) \ \langle R_S[K^+K^-]_u^{0} \vert u \overline{u}\rangle =  \chi^{n}\ \frac{\Gamma_2^{n*}(s_0)}{\sqrt{2}} \ F_0^{D^0 f_0}(m_{K^0}^2),
\ee
where  $\chi^{n}$ is a constant complex factor.  Hence 
\be
\label{ISSKPKM1}
T^{CF}_{\overline{K}^0 \ [K^+ K^-]_{S,u}^{0}}(s_0,s_-,s_+) = - \frac{G_F}{2}\ \Lambda_1 \ a_2\  
\ f_{K^0}\ ( m_{D^0}^2 - s_0) \  F_0^{D^0 f_{0}}(m_{K^0}^2)\ \frac{\chi^{n}}{2} \Gamma_2^{n*}(s_0).
\ee
The real transition form factor, $F_0^{D^0 f_{0}}(m_{K^0}^2)$, can be obtained from Ref.~\cite{El-Bennich_PRD79}. This amplitude has to be associated with the corresponding isoscalar-scalar  $K^0 [K^+K^-]_{S,u}^{0}$ DCS amplitude (see Fig.~\ref{F4}) approximated by 
\be \label{ISSDCSKPKM}
T_{K^0[K^+K^-]_{S,u}^0}^{DCS}(s_0,s_-,s_+) =  \frac{\Lambda_2}{\Lambda_1}
~T^{CF}_{\overline{K}^0 \ [K^+ K^-]_{S,u}^{0}}(s_0,s_-,s_+)
\ee
Recombining the two amplitudes~(\ref{ISSKPKM1}) and~(\ref{ISSDCSKPKM}), we have 
\bqa \label{T1S}
T_1 &=& T_{K^0[K^+K^-]_{S,u}^0}^{CF}(s_0,s_-,s_+) + T_{K^0[K^+K^-]_{S,u}^0}^{DCS}(s_0,s_-,s_+)
\nonumber \\
&=&  - \frac{G_F}{2}\ (\Lambda_1+\Lambda_2) \ a_2\ f_{K^0}\ 
( m_{D^0}^2 - s_0) \ F_0^{D^0 f_{0}}(m_{K^0}^2)\  \frac{\chi^{n}}{2} \  \Gamma_2^{n*}(s_0). 
\eqa

We now turn to the three isovector-scalar tree amplitudes.  
The isovector-scalar $K^+[{K}^0 K^-] _S $ DCS amplitude, associated to the $a_0(980)^-$ and $a_0(1450)^-$ resonances can be written as (see Fig.~\ref{F3})
\bqa \label{TCFSK0KM}
T^{DCS}_{K^+[{K}^0 K^-]_S^{1} }(s_0,s_-,s_+)
& =& - \frac{G_F}{2}\ \Lambda_2 \ a_1 \  f_{K^+}\ (m_{D^0}^2- s_-) \ \sum_{R_S}F_0^{D^0 R_S[K^0K^-]^{1}}(m_{K}^2)  \nonumber \\
&& G_{R_S[K^0K^-]^{1}}(s_-)\  \langle R_S[K^0K^-]^{1} \vert d\overline{u}\rangle,
\eqa
the $a_0^-$ resonances being built from $\overline{u}d$ pairs and $\langle R_S[K^0K^-]^{1} \vert d\overline{u}\rangle=1$.
In Eq.~(\ref{TCFSK0KM})  $f_{K^+}$ denotes the charged kaon decay constant. Parametrizing, as above, the sum over $R_S$
as
\be \label{IVSK0KM}
\sum_{R_S} F_0^{D^0 R_S[K^0K^-]_u^{1}}(m_{K}^2) \ G_{R_S[K^0K^-]^{1}}(s_-)\ 
\langle R_S[K^0K^-]^{1} \vert d\overline{u}\rangle = G_1(s_-)\ F_0^{D^0 a_0^-}(m_{K}^2),\\
\ee
we get
\be \label{T2S}
T_{2} = T^{DCS}_{K^+[{K}^0 K^-]_S^{1} }(s_0,s_-,s_+) =  - \frac{G_F}{2}\ \Lambda_2 \ a_1 \  f_{K^+}\ (m_{D^0}^2- s_-) \ F_0^{D^0 a_0^-}(m_{K}^2)  \ G_1(s_-).
\ee
The function $G_1(s)$ can be built using the isospin 1 coupled $K \overline K$ and $\pi \eta$ channel description of the $a_0(980)$ and $a_0(1450)$ performed in Ref.~\cite{AFLL}.  As previously for the isoscalar-scalar case, we have assumed here that the variation of the $D^0\to R_S$ transition form factor from one resonance to the other is small and we choose it to be that of the lowest resonance, {\it i.e.}, $R_S \equiv a_0(980)^-$, denoted simply $a_0^-$.

In the case ot the isovector-scalar $[\overline{K}^0 K^+] _S^{1} \ K^-$ \ CF amplitude ($a_1$ term of Eq.~(\ref{TCF}), related to the $a_0(980)^+$ and $a_0(1450)^+$ resonances, one has\footnote{Because of the presence of the very small mass squared difference between the charged and neutral kaons  
 the amplitude~(\ref{IVSK0KP}) will give no constraint on the kaon isovector-scalar form factor.} \bqa \label{IVSK0KP} 
T^{CF}_{K^-  [\overline{K}^0 K^+]_S^{1}}(s_0,s_-,s_+)&=& - \frac{G_F}{2}\ \Lambda_1 \ a_1\  (m_{D^0}^2- m_{K}^2)\ \frac{m_{K}^2-m_{K^0}^2}{s_+}\nonumber \\
&\times&  F_0^{D^0K^-}(s_+) \ F_{0}^{[\overline{K}^0K^+]^1}(s_+),
\eqa
where $F_{0}^{[\overline{K}^0K^+]^1}(s_+)$ is the kaon isovector-scalar form factor and
 denoted also as $F_{0}^{[\overline{K}^0K^+]^1}(s)$ in the second relation of the Eqs.~(\ref{FSKbarK}).
 For the transition form factor $F_0^{D^0K^-}(s_+)$, following Ref.~\cite{Melichow}, we use the parameterization:
\be \label{F0DKm}
F_0^{D^0K^-}(s_+)=\frac{0.78}{1-0.38~ s_+/M_V^2+0.46~ s_+^2/M_V^4},
\ee 
where $M_V=2.11$ GeV. We have then
\be \label{T3S}
T_3= T^{CF}_{[\overline{K}^0 K^+]_S^{1} K^- }(s_0,s_-,s_+) = - \frac{G_F}{2}\ \Lambda_1\  a_1\ (m_{D^0}^2- m_{K}^2)\ \frac{m_{K}^2-m_{K^0}^2}{s_+}\ F_{0}^{[\overline{K}^0K^+]^1}(s_+) \  F_0^{D^0K^-}(s_+) 
\ee

We proceed similarly for the isovector-scalar CF $\overline{K}^0 \ [K^+ K^-]_{S,u}^{1}$ and DCS ${K}^0 \ [K^+ K^-]_{S,u}^{1}$ amplitudes (see Figs.~\ref{F2} and \ref{F4}). It is given by  
\bqa
\label{IVSKPKM}
T^{CF}_{\overline{K}^0 \ [K^+ K^-]_S^{1}} (s_0,s_-,s_+)&=& - \frac{G_F}{2}\ \Lambda_1 \ a_2\  f_{K^0}\ 
( m_{D^0}^2 - s_0) \sum_{R_S} F_0^{D^0 R_S[K^+K^-]_u^{1}}(m_{K^0}^2)\nonumber \\
&\times & G_{R_S[K^+K^-]_u^{1}}(s_0) \ \langle R_S[K^+K^-]_u^{1} \vert u \overline{u}\rangle,
\eqa  
where the sum over $R_S$ runs over the contributing resonances in that channel, i.e., $a_0(980)^0$ and $a_0(1450)^0$ for which   
$ \langle R_S[K^+K^-]_u^{1} \vert  u \overline{u}\rangle = \frac{1}{\sqrt{2}}$.  Then, we get
\be
\label{G1s0}
\sum_{R_S} F_0^{D^0 R_S[K^+K^-]_u^{1}}(m_{K^0}^2) \ G_{R_S[K^+K^-]_u^{I=1}}(s_0) \ \langle R_S[K^+K^-]_u^{1} \vert u \overline{u}\rangle = 
\frac{1}{2} G_1(s_0)\ F_0^{D^0 a_0^0}(m_{K^0}^2),
\ee
where  we assume (isospin invariance) that, to describe the $u \bar u$ transition to the isovector-scalar  $K^+ K^-$ state, it is the same function $G_1(s)$  as that introduced in Eqs.~(\ref{IVSK0KM})  for the $ d \bar u$ transitions to the isovector-scalar $K^0K^-$  state.
In Eq.~(\ref{G1s0}) $a_0^0$ means $a_0(980)^0$.
 We may now rewrite
\be \label{IVSKPKM1}
T^{CF}_{\overline{K}^0 \ [K^+ K^-]_S^{1}} (s_0,s_-,s_+) = - \frac{G_F}{2}\ \Lambda_1 \ a_2\ f_{K^0}\ ( m_{D^0}^2 - s_0)  \ \frac{1}{2} G_1(s_0)  \ F_0^{D^0 a_0^0}(m_{K^0}^2).
\ee

In a similar way, the related isovector-scalar  DCS amplitude reads
 \be  \label{TDCSsc} 
T^{DCS}_{K^0 [K^+K^-]_S^{1}} (s_0,s_-,s_+)= \frac{\Lambda_2}{\Lambda_1}
~T^{CF}_{\overline{K}^0 \ [K^+ K^-]_S^{1}} (s_0,s_-,s_+).
\ee
Recombining Eqs.~(\ref{IVSKPKM1}) and (\ref{TDCSsc}) we get 
\bqa \label{T4S}
T_4  &=& T^{CF}_{\overline{K}^0 [K^+K^-]_S^{1}} (s_0,s_-,s_+) + T^{DCS}_{K^0 [K^+K^-]_S^{1}} (s_0,s_-,s_+) \nonumber \\
& =& - \frac{G_F}{2}\ (\Lambda_1+\Lambda_2) \ a_2\ f_{K^0} \ ( m_{D^0}^2 - s_0)\ \frac{1}{2} G_1(s_0)
  \ F_0^{D^0 a_0^0}(m_{K^0}^2).
\eqa

\subsubsection {\it Vector amplitudes}

Let us now study the vector tree amplitudes associated to the $\overline{K}^0 \ [K^+ K^-]_{P,u}^{I}$ channel. The isoscalar-vector CF amplitude can be  built  from the $\omega$ and $\phi$ resonances (see Fig.~\ref{F2}). It reads
\bqa  \label{T6ISV}
T^{CF}_{\overline{K}^0 \ [K^+ K^-]_P^{0}}(s_0,s_-,s_+) &=&  \frac{G_F}{2}\ \Lambda_1 \ a_2\ f_{K^0} \ (s_--s_+) \  \sum_{R_P} A_0^{D^0 R_P[K^+K^-]_u^{0}}(m_{K^0}^2) \nonumber \\
&\times&  m_{R_P [K^+K^-]}\ G_{R_P [K^+K^-]_u^{0}}(s_0) \ 
 \langle R_P[K^+K^-]_u^{0} \vert u \overline{u}\rangle, 
\eqa
where $m_{R_P [K^+K^-]}$ denotes the mass of the contributing resonances. Now we introduce the following parametrization in terms of the kaon vector form factor  $F_1^{[K^+K^-]_u^{0}}(s_0)  $  and, for the same reasons as introduced in the scalar case [Eqs.~(\ref{ISSKPKM}) and~(\ref{T2S})]
\bqa
 \sum_{R_P} m_{R_P [K^+K^-]}\ A_0^{D^0 R_P[K^+K^-]_u^{0}}(m_{K^0}^2)&&\hspace{-0.6cm}G_{R_P [K^+K^-]_u^{0}}(s_0) \ 
 \langle R_P[K^+K^-]_u^{0} \vert u \overline{u}\rangle \nonumber \\
 &=& \frac{1}{f_\omega} \ A_0^{D^0\omega}(m_{K^0}^2) \ F_1^{[K^+K^-]_u^{0}}(s_0).
\eqa
This approximation relies on the fact that the mixing angle of the vector meson nonet is very close to the ideal mixing angle, $\theta_V = 35.3^{\circ}$, so that  the $\phi$ resonance amplitude gives an almost nul contribution. Note that $f_\omega$ denotes the decay constant for the $\omega(782)$ meson. We have then 
\be \label{TCFom}
T^{CF}_{\overline{K}^0 \ [K^+ K^-]_P^{0}}(s_0,s_-,s_+)  = \frac{G_F}{2}\ \Lambda_1\ {a_2}  \ (s_--s_+)  \frac{f_{K^0} }{f_{\omega} }\ A_0^{D^0 \omega}(m_{K^0}^2)\  F_1^{[K^+K^-]_u^{0}}(s_0).
\ee
The associated  isoscalar-vector  $K^0 [K^+K^-]_{P,u}^{0}$ DCS amplitude  is given by 
 \be  \label{TDCSL2}
T_{K^0 [K^+K^-]_P^{0}}^{DCS}(s_0,s_-,s_+) =\frac{\Lambda_2}{\Lambda_1}
~T^{CF}_{\overline{K}^0 \ [K^+ K^-]_P^{0}}(s_0,s_-,s_+).
\ee
Thus, from Eqs.~(\ref{TCFom}) and~(\ref{TDCSL2}),  the isoscalar-vector amplitude reads 
\bqa \label{T5P}
T_5 &=& T^{CF}_{\overline{K}^0 \ [K^+ K^-]_P^{0}}(s_0,s_-,s_+) + T_{K^0 [K^+K^-]_P^{0}}^{DCS}(s_0,s_-,s_+) \nonumber \\
&=& \frac{G_F}{2}\ (\Lambda_1+\Lambda_2) \ a_2\ (s_--s_+)  \frac{f_{K^0} }{f_{\omega} }\ A_0^{D^0 \omega}(m_{K^0}^2)\  F_1^{[K^+K^-]_u^{0}}(s_0).
\eqa

The isovector-vector $T^{CF}_{\overline{K}^0 \ [K^+ K^-]_P^{1}}$ amplitude related to the $\rho^0$ resonances is given by a similar expression to Eq.~(\ref{T6ISV}) (see Fig.~\ref{F2})
\bqa  \label{T7IVV}
T^{CF}_{\overline{K}^0 \ [K^+ K^-]_P^{1}}(s_0,s_-,s_+) &=&  \frac{G_F}{2}\ \Lambda_1 \ a_2\  f_{K^0} \ (s_--s_+) \ \sum_{R_P} A_0^{D^0 R_P[K^+K^-]_u^{1}}(m_{K^0}^2) \nonumber \\
&\times&  m_{R_P [K^+K^-]}\ G_{R_P [K^+K^-]_u^{1}}(s_0) \  \langle R_P[K^+K^-]^{1} \vert u \overline{u}\rangle,
\eqa
where $ \langle R_P[K^+K^-]^{1} \vert u \overline{u}\rangle = 1/ {\sqrt{2}}$. Again, parametrizing the sum over the vertex functions by
\bqa
\sum_{R_P}  A_0^{D^0 R_P[K^+K^-]_u^{1}}(m_{K^0}^2)  m_{R_P [K^+K^-]}\ f_{R_P[K^+K^-]^{1}}\hspace{-0.5cm}&& G_{R_P [K^+K^-]_u^{1}}(s_0) \ 
 \langle R_P[K^+K^-]^{1} \vert u \overline{u}\rangle \nonumber \\
 &=&  \frac{1}{f_\rho}\ A_0^{D^0\rho^0}(m_{K^0}^2)\   F_1^{[K^+K^-]_u^{1}}(s_0),
\eqa
where $f_{\rho}$ is the charged $\rho(770)$ decay constant, 
\be
T^{CF}_{\overline{K}^0 \ [K^+ K^-]_P^{1}}(s_0,s_-,s_+) = \frac{G_F}{2}\ \Lambda_1 \ {a_2} \ (s_--s_+) \ \frac{ f_{K^0}}{f_\rho}\
A_0^{D^0\rho^0}(m_{K^0}^2)\  F_1^{[K^+K^-]_u^{1}}(s_0). 
\ee
Then comes the contribution of the isovector-vector $K^0 [K^+K^-]_P^{1}$ DCS amplitude
(see Fig.~\ref{F4}). It goes as  
\be
T_{K^0 [K^+K^-]_P^{1}}^{DCS}(s_0,s_-,s_+) = \frac{\Lambda_2}{\Lambda_1}~
T^{CF}_{\overline{K}^0 \ [K^+ K^-]_P^{1}}(s_0,s_-,s_+),
\ee
so that the total isovector-vector amplitude is
\bqa \label{T8P}
T_6 &=& T^{CF}_{\overline{K}^0 \ [K^+ K^-]_P^{1}}(s_0,s_-,s_+) + T_{K^0 [K^+K^-]_P^{1}}^{DCS}(s_0,s_-,s_+) \nonumber \\
& =& \frac{G_F}{2}\ (\Lambda_1+\Lambda_2) \ a_2  \ (s_--s_+) 
 A_0^{D^0 \rho^0}(m_{K^0}^2) \frac{f_{K^0}}{f_\rho}\ F_1^{[K^+K^-]_u^{1}}(s_0).
 \eqa

The isovector-vector $[\overline{K}^0 K^+] _{P}^{1} \ K^- $ CF amplitude has the expression (see Fig.~\ref{F1})
\bqa
T^{CF}_{[\overline{K}^0 K^+]_P^{1} K^- }(s_0,s_-,s_+)&=& - \frac{G_F}{2}\ \Lambda_1 \ a_1 \ \left [s_0 - s_- + (m_{D^0}^2- m_{K}^2)\ \frac{m_{K^0}^2-m_{K}^2}{s_+} \right ] \nonumber \\
&\times& \sum_{R_P} F_1^{D^0K^-}(s_+) \ G_{R_P[\overline{K}^0 K^+] ^{1}}(s_+) \ \langle R_P[\overline{K}^0 K^+]^{1}\vert u \overline{d}\rangle 
\eqa
where  $\langle R_P[\overline{K}^0 K^+] ^{1}\vert u \overline{d}\rangle= 1$ since it is associated to 
 the $\rho(770)^+$, $\rho(1450)^+$ and $\rho(1700)^+$ resonances.  The sum over the vertex functions $G_{R_P[\overline{K}^0 K^+] ^{I=1}}(s_+) $ is expressed in terms of the isovector-vector form factor $ F_1^{\overline{K}^0 K^+}(s_+) $
\be
\sum_{R_P}  F_1^{D^0K^-}(s_+) \ G_{R_P[\overline{K}^0 K^+] ^{1}}(s_+)\ \langle R_P[\overline{K}^0 K^+] ^{1}\vert u \overline{d}\rangle = F_1^{D^0K^-}(s_+)\ F_1^{\overline{K}^0 K^+}(s_+), 
\ee
\bqa \label{T7P}
T_7 &=& T^{CF}_{[\overline{K}^0 K^+]_P^{1} K^- }(s_0,s_-,s_+)=   - \frac{G_F}{2}\ \Lambda_1 \ a_1 \ \nonumber \\
&\times&  \left [s_0 - s_- + (m_{D^0}^2- m_{K}^2)\ \frac{m_{K^0}^2-m_{K}^2}{s_+} \right ] \ 
 F_1^{D^0K^-}(s_+) \ F_1^{[\overline{K}^0 K^+]^1}(s_+).
\eqa
As in Ref.~\cite{Melichow} we parametrize $F_1^{D^0K^-}(s_+)$ as follows
\be\label{F1dKm}
F_1^{D^0K^-}(s_+)=\frac{0.78}{(1-s_+/M_V^2)(1-0.24~ s_+/M_V^2)}
\ee
with, as before in Eq.~(\ref{F0DKm}), $M_V=2.11$ GeV.

The isovector-vector $K^+[{K}^0 K^-] _P $ DCS amplitude is given  by (see Fig.~\ref{F3})
\bqa
T_{K^+[{K}^0 K^-] _P^1 }^{DCS}(s_0,s_-,s_+)&=&  \frac{G_F}{2}\ \Lambda_2 \ a_1 \ \left [s_0 - s_+ + (m_{D^0}^2- m_{K}^2)\ \frac{m_{K^0}^2-m_{K}^2}{s_-} \right ] \ f_{K^+} \  \nonumber \\
&\times&  \sum_{R_P} A_0^{D^0 R_P[K^0K^-]^{1}}(m_{K}^2) \ m_{R_P} \ G_{R_P[K^0K^-]^{1}}( s_-) \ \langle R_P[K^0K^-]\vert d \overline{u}\rangle. 
\eqa
It is linked to the $\rho(770)^-$, $\rho(1450)^-$ and $\rho(1700)^-$ resonances and can be reexpressed  as 
\bqa \label{T6P}
T_8 &=& T_{K^+[{K}^0 K^-] _P^1 }^{DCS}(s_0,s_-,s_+)=  \frac{G_F}{2}\ \Lambda_2 \ a_1 \ \frac{f_{K^+}}{f_{\rho}}
\nonumber \\
&\times& \left [s_0 - s_+ + (m_{D^0}^2- m_{K}^2) \frac{m_{K^0}^2-m_{K}^2}{s_-} \right ] \
 A_0^{D^0 \rho^-}(m_{K}^2) \ F_1^{[K^0 K^-]^1}(s_-).
\eqa
where we have introduced the isovector-vector form factor $ F_1^{[K^0 K^-]^1}(s_-)$ related to the sum over the vertex functions $G_{R_P[K^0K^-]^{1}}( s_-)$ by
\be 
 \sum_{R_P} A_0^{D^0 R_P[K^0K^-]^{1}}(m_{K}^2) \ m_{R_P} \ G_{R_P[K^0K^-]^{1}}( s_-) \  \langle R_P[K^0K^-]_u\vert d \overline{u}\rangle = \frac{1}{f_\rho} \ F_1^{K^0 K^-}(s_-) \  A_0^{D^0 \rho^-}(m_{K}^2)
 \ee
with $ \langle R_P[K^0K^-]_u\vert d \overline{u}\rangle =  1$, and $\rho$ refers to $\rho(770)$.

\subsubsection{\it Tensor amplitudes}

 For the isoscalar-tensor amplitude $\overline{K}^0 \ [K^+ K^-]_{D,u}^{0}$ amplitude, one can write (see Fig.~\ref{F2})
\bqa 
\label{T8IST}
T^{CF}_{\overline{K}^0 \ [K^+ K^-]_D^{0}}(s_0,s_-,s_+) &=&  - \frac{G_F}{2}\ \Lambda_1 \ a_2\ f_{K^0}\ 
\sum_{R_D} F^{D^0  R_D[K^+K^-]_u^{0}}(m_{K^0}^2) \nonumber \\
&\times& G_{R_D[K^+K^-]_u^{0}}(s_0, s_-,s_+) \langle R_D[K^+K^-]_u^{0}\vert u\overline{u}\rangle
\eqa
but it will be dominated by the $f_2(1270)$ resonance with mass $m_{f_2}$; it will be described by a Breit-Wigner representation. Linking the vertex function to the form factor and using
\be
\langle R_D[K^+K^-]_u^{0}\vert u\overline{u}\rangle \approx  \langle f_2(1270) \vert u \overline{u} \rangle = \frac{1}{\sqrt{2}},
 \ee
we have 
\bqa \label{T8IST2}
T^{CF}_{\overline{K}^0 \ [K^+ K^-]_D^{0}}(s_0,s_-,s_+)  &=& - \frac{G_F}{2}\ \Lambda_1 \ a_2\ f_{K^0}\ \frac{1}{\sqrt{2}} \nonumber \\ 
&\times&  F^{D^0f_2}(m_{K^0}^2) \
 g_{f_2 K^+K^-} \ \frac{D({\bf p_2},  {\bf p_0})}{m_{f_2}^2 - s_0 -i m_{f_2}\ \Gamma_{f_2}(s_0)},
\eqa
where $g_{f_2 K^+K^-}$ is the coupling constant of the $f_2 \to K^+K^-$ transition
and the function ${D({\bf p_2},  {\bf p_0})}$ is defined by
\be \label{Dpipj}
{D({\bf p_2},  {\bf p_0})} = \frac{1}{3} \ {(\vert{\bf p_2}\vert \ \vert {\bf p_0}\vert)^2 - ({\bf p_2}\cdot{\bf p_0})^2}.
\ee

The three-momenta ${\bf p_2},  {\bf p_0}$ are defined in the $[K^+K^-]$ center-of-mass (c.m.) system. One has 
\be \label{p2p+}
{{\bf p_2}= {\bf p_+} = - \bf{p_-} },\hspace{2cm}\hspace{2.cm}
 {\vert {\bf p_2} \vert } = {\frac{1}{2}}\ \sqrt{s_0 - 4 m_{K}^2}
\ee
and 
\be \label{p2p0}
 \vert{\bf p_0}\vert = \frac{\sqrt{[m_{D^0}^2-(\sqrt{s_0}+m_{K^0})^2] \  [m_{D^0}^2-(\sqrt{s_0}-m_{K^0})^2] }} {2\ \sqrt{s_0}}.
\ee
The scalar product ${\bf p_2}\cdot{\bf p_0}$ which enters the function  ${D({\bf p_2},  {\bf p_0})}$ is given by the relation 
\be \label{4p2p0}
4 \ {\bf p_2}\cdot{\bf p_0} = s_- - s_+.
\ee 

 One has  
\be
 g_{f_2 K^+K^-}= m_{f_2}\ \sqrt{\frac{60\pi\ \Gamma_{f_2 K^+K^-}}{q_{f_2}^5}},
  \ee
where $\Gamma_{f_2 K^+K^-}$ is the $f_2(1270)$ decay constant into $K^+K^-$ and
the momentum $q_{f_2}=\frac{1}{2}\ \sqrt{m_{f_2}^2- 4 m_{K}^2}$.
In Eq.~(\ref{T8IST2}), because of the large width of the $f_2$ meson, an energy dependent total width  $\Gamma_{f_2}(s_0)$ has been introduced (see Eqs.~(121) and~(122) in Ref.~\cite{JPD_PRD89}) such that 
\be
 \Gamma_{f_2}(s_0)= \left (\frac{q}{q_{f_2}} \right )^5 \ \frac{m_{f_2}}{\sqrt{s_0}}\ 
 \frac{(q_{f_2}r)^4 + 3 (q_{f_2}r)^2 +9}{(qr)^4 + 3 (qr)^2 +9} \ \Gamma_{f_2},
\ee
where $r=4.0$ GeV$^{-1}$ and $q \equiv {\vert {\bf p_2} \vert }$.
In Eq.~(\ref{T8IST2}) the $D^0\to f_2$ effective transition form factor $F^{D^0 f_2}(m_{K^0}^2)$
will be treated as a free complex parameter.

To this amplitude one has to add the  isoscalar-tensor $K^0 [K^+K^-]_D^{0}$ DCS amplitude given by (see Fig.~\ref{F4})
\be
T_{K^0 [K^+K^-]_D^{0}}^{DCS}(s_0,s_-,s_+)=   \frac{\Lambda_2}{\Lambda_1}~
T^{CF}_{\overline{K}^0 \ [K^+ K^-]_D^{0}}(s_0,s_-,s_+).
\ee
The total  isoscalar-tensor amplitude then reads 
\bqa \label{T9D}
T_{9}& =& T^{CF}_{\overline{K}^0 \ [K^+ K^-]_D^{0}}(s_0,s_-,s_+)  + T_{K^0 [K^+K^-]_D^{0}}^{DCS}(s_0,s_-,s_+) \nonumber \\
&=&   - \frac{G_F}{2}\ (\Lambda_1+\Lambda_2) \ a_2\ f_{K^0}\ \frac{1}{\sqrt{2}}\ F^{D^0f_2}(m_{K^0}^2)\  
g_{f_2 K^+K^-} \ \frac{D({\bf p_2},  {\bf p_0})}{m_{f_2}^2 - s_0 -i m_{f_2}\ \Gamma_{f_2}(s_0)}.
\eqa

\subsection{Annihilation amplitudes} \label{SAA}

\subsubsection{Scalar amplitudes}
The isoscalar-scalar CF annihilation amplitude corresponding to $\overline{K}^0 [K^+K^-]_{S,s}^0$  final states (Fig.~\ref{F5}) is given by
\bqa
A_{\overline{K}^0 [K^+K^-]_S^0}^{CF}(s_0,s_-,s_+)  &=& - \frac{G_F}{2}\ \Lambda_1 \ a_2 \ f_{D^0} \ (m_{K^0}^2 - s_0) \
 \sum_{R_S} F_0^{\overline{K}^0 R_S[K^+K^-]_s^0}(m_{D^0}^2) \ \nonumber \\
&\times & G_{R_S[K^+K^-]_s^0}(s_0) \ 
\langle R_S[K^+K^-]\vert \overline{s}s\rangle 
\eqa
while the isoscalar-scalar DCS amplitude  corresponding to ${K}^0 [K^+K^-]_{S,s}^0$ final states  (Fig.~\ref{F6W}) is 
\be
A_{K^0 [K^+K^-]_S^0}^{DCS}(s_0,s_-,s_+) = \frac{\Lambda_2}{\Lambda_1}~
A_{\overline{K}^0 [K^+K^-]_S^0}^{CF}(s_0,s_-,s_+),
\ee
where we have used the equality 
\be
F_0^{{K}^0 R_S[K^+K^-]_s^0}(m_{D^0}^2) =  F_0^{\overline{K}^0 R_S[K^+K^-]_s^0}(m_{D^0}^2).
\ee
Thus the total isoscalar-scalar annihilation amplitude reads 
\bqa
A_1 &=& A_{\overline{K}^0 [K^+K^-]_S^0}^{CF}(s_0,s_-,s_+) + A_{K^0 [K^+K^-]_S^0}^{DCS}(s_0,s_-,s_+) \nonumber \\
 &=& - \frac{G_F}{2}\ (\Lambda_1+\Lambda_2) \ a_2 \ f_{D^0} \ (m_{K^0}^2 - s_0) 
 \nonumber \\ 
&\times &  \sum_{R_S} F_0^{{K}^0 R_S[K^+K^-]_s^0}(m_{D^0}^2) \ G_{R_S[K^+K^-]}(s_0) \ 
\langle R_S[K^+K^-]\vert \overline{s}s\rangle. 
\eqa

Following the steps in Sec.\ref{STA} it leads to 
 \be  \label{A1S}
 A_1 =  - \frac{G_F}{2}\ (\Lambda_1+\Lambda_2) \ a_2 \ f_{D^0} \ (m_{K^0}^2 - s_0) \ 
 F_0^{K^0f_0}(m_{D^0}^2) \ \frac{1}{\sqrt{2}}\chi^s \ \Gamma_2^{s*}(s_0),
 \ee
where $\chi^s$ is a complex constant and $\Gamma_2^{s*}(s_0)$ is the kaon strange scalar form factor.

The isovector-scalar annihilation DCS amplitude, associated to the $K^+[K^0K^-]_S^1$ final states containing the $a_0(980)^-$ and $a_0(1450)^-$, is given by 
 \bqa 
 A_2 = A_{K^+[K^0K^-]_S^1}^{DCS}(s_0,s_-,s_+) &=&  - \frac{G_F}{2}\ \Lambda_2 \ a_2 \ f_{D^0} \ (m_{K}^2 - s_-) 
 \sum_{R_S} F_0^{K^+ R_S[{K}^0K^-]^1}(m_{D^0}^2) \  \nonumber \\
 &\times &  G_{R_S[{K}^0K^-]^1}(s_-) \ 
 \langle R_S[{K}^0K^-]^1\vert d\overline{u} \rangle 
 \eqa 
and hence with,
  \bqa
  \label{G1sm}
  \sum_{R_S} F_0^{K^+ R_S[{K}^0K^-]^1}(m_{D^0}^2) \   G_{R_S[{K}^0K^-]^1}(s_-) \ 
 \langle R_S[{K}^0K^-]^1\vert d\overline{u}\rangle =  F_0^{K^+a_0^-}(m_{D^0}^2) \ G_1(s_-),  
 \eqa
 reads
  \be \label{A2S}
 A_2  =  - \frac{G_F}{2}\ \Lambda_2 \ a_2 \ f_{D^0} \ (m_{K}^2 - s_-) \ F_0^{K^+a_0^-}(m_{D^0}^2) \ G_1(s_-).
\ee

The corresponding  isovector-scalar annihilation CF amplitude associated to the $K^- [\overline{K}^0 K^+]_S^1$ reads 
\bqa
A_3= A_{K^-[\overline{K}^0K^+]_S^1}^{CF}(s_0,s_-,s_+) &=& - \frac{G_F}{2}\ \Lambda_1 \ a_2 \ f_{D^0} \ (m_{K}^2 - s_+) 
 \sum_{R_S} F_0^{K^- R_S[\overline{K}^0K^+]^1}(m_{D^0}^2) \ \nonumber \\
&\times & G_{R_S[\overline{K}^0K^+]^1}(s_+) \ \langle R_S[\overline{K}^0K^+]^1\vert \overline{d}u \rangle  
\eqa 
 and contains the $a_0(980)^+$ and $a_0(1450)^+$. With the approximation 
 \bqa
 \label{G1sp}
  \sum_{R_S} F_0^{K^- R_S[\overline{K}^0K^+]^1}(m_{D^0}^2) \ 
G_{R_S[\overline{K}^0K^+]^1}(s_+) \ \langle R_S[\overline{K}^0K^+]^1\vert \overline{d}u \rangle 
= F_0^{K^-a_0^+}(m_{D^0}^2) \ G_1(s_+)  
 \eqa 
 we reach
\be  \label{A3S}
A_3 = - \frac{G_F}{2}\ \Lambda_1 \ a_2 \ f_{D^0} \ (m_{K}^2 - s_+)  \ F_0^{K^-a_0^+}(m_{D^0}^2) \ G_1(s_+).
\ee 
The  $[K^+K^-]_S^1$ final states which would contain the $a_0(980)^0$ and $a_0(1450)^0$ mesons cannot be formed from a $s\overline{s}$ pair and thus the corresponding  ${K}^0 [K^+K^-]_S^1$  isovector-scalar DCS amplitude  is zero.
 
\subsubsection{Vector amplitudes}

We now turn to the vector-annihilation amplitudes. The isoscalar-vector CF amplitude corresponding to  $\overline{K}^0[K^+K^-]_{P,s}^0$ final states read (see Fig.~\ref{F5})
\bqa
A_{\overline{K}^0[K^+K^-]_P^0}^{CF}(s_0,s_-,s_+) &=&   \frac{G_F}{2}\ \Lambda_1 \ a_2 \ f_{D^0} \ (s_- - s_+) 
\sum_{R_P} A_0^{\overline{K}^0 R_P[K^+K^-]_s^0}(m_{D^0}^2) \ m_{R_P} \nonumber \\
&\times & G_{R_P[K^+K^-]_s^0}(s_0) \ 
\langle R_P[K^+K^-]\vert \overline{s}s\rangle,
\eqa
and is associated to the $\phi$ mesons. It may be reexpressed as
\be
A_{\overline{K}^0[K^+K^-]_P^0}^{CF}(s_0,s_-,s_+) = \frac{G_F}{2}\ \Lambda_1 \ a_2 \ \frac{f_{D^0}}{f_{\phi}} \ (s_- - s_+)  \
A_0^{\overline{K}^0 \phi}(m_{D^0}^2)\  F_1^{[K^+K^-]_s^0}(s_0).
\ee

One has to add the associated DCS amplitude  corresponding to ${K}^0[K^+K^-]_{P,s}^0$ final states (see Fig.~\ref{F6W}); 
since $A_0^{\overline{K}^0 \phi}(m_{D^0}^2) =  A_0^{K^0 \phi}(m_{D^0}^2)$ we have 
\bqa \label{A5P}
 A_4 = \frac{G_F}{2}\ (\Lambda_1+ \Lambda_2) \ a_2 \ \frac{f_{D^0}}{f_{\phi}} \ (s_- - s_+)  \
A_0^{{K}^0 \phi}(m_{D^0}^2)\  F_1^{[K^+K^-]_s^0}(s_0).
\eqa

The isovector amplitude corresponding to ${K}^- [\overline{K}^0K^+]_P^1$ final states,  
which contains  the $\rho(770)^+$, $\rho(1450)^+$ and $\rho(1700)^+$ mesons,
\bqa \label{A7P}
A_5 &=& - \frac{G_F}{2}\ \Lambda_1 \ a_2 \ \frac{f_{D^0}}{f_{\rho}} \ \left [s_0-s_- + \frac{(m_{D^0}^2-m_{K}^2)
( m_{K^0}^2-m_{K}^2)}{s_+}\right ] 
\nonumber \\
&\times & \sum_{R_P} A_0^{K^- R_P[\overline{K}^0 K^+]^1}(m_{D^0}^2) \ m_{R_P}\ G_{R_P[\overline{K}^0 K^+]^1}(s_+) \ 
\langle R_P[\overline{K}^0 K^+]\vert \overline{d} u \rangle
\eqa 
may be written as
\be
A_5 = - \frac{G_F}{2}\ \Lambda_1 \ a_2 \ \frac{f_{D^0}}{f_{\rho}} \ \left [s_0-s_- + \frac{(m_{D^0}^2-m_{K}^2)
( m_{K^0}^2-m_{K}^2)}{s_+}\right ] \ A_0^{K^-\rho^+}(m_{D^0}^2)\ F_1^{[\overline{K}^0K^+]^1}(s_+).
\ee

 Similarly, the isovector-DCS amplitude corresponding to ${K}^+ [K^0K^-]_P^1$ final states  reads 
\be \label{A6P}
A_6 =  - \frac{G_F}{2}\ \Lambda_2 \ a_2 \ \frac{f_{D^0}}{f_\rho} \ \left [s_0-s_+ + \frac{(m_{D^0}^2-m_{K}^2)(m_{K^0}^2-m_{K}^2)}{s_-}\right ] \ A_0^{K^+\rho^-}(m_{D^0}^2)\ F_1^{[K^0K^-]^1}(s_-).
\ee
It contains the $\rho(770)^-$, $\rho(1450)^-$ and $\rho(1700)^-$ mesons. 

\subsubsection{\it Tensor amplitudes}
Finally we present the tensor amplitudes. The two isoscalar CF and DCS amplitudes associated to the $\overline{K}^0[K^+K^-]_{D,s}^0$ and $K^0[K^+K^-]_{D,s}^0$  final states read respectively 
\bqa \label{TCD2SVT}
A_{\overline{K}^0[K^+K^-]_D^0}^{CF}(s_0,s_-,s_+)  &=& \frac{G_F}{2}\ \Lambda_1 \ a_2 \ f_{D^0} \ {D({\bf p_2},{\bf p_0})} 
 \sum_{R_D} F^{\overline{K}^0 R_D[K^+K^-]_s^0}(m_{D^0}^2) \nonumber \\
&\times& G_{R_D[K^+K^-]_s^0}(s_0)\ 
\langle R_D[K^+K^-]\vert \overline{s}s\rangle 
\eqa
and
\bqa \label{TFD2SVT}
A_{K^0[K^+K^-]_D^0}^{DCS}(s_0,s_-,s_+)= \frac{\Lambda_2}{\Lambda_1}~
A_{\overline{K}^0[K^+K^-]_D^0}^{CF}(s_0,s_-,s_+)
\eqa
They contain the  $f_2(1270)$ meson. In the last equation we have used the relation 
\be
F^{\overline{K}^0 R_D[K^+K^-]_s^0}(m_{D^0}^2)=F^{{K}^0 R_D[K^+K^-]_s^0}(m_{D^0}^2).
\ee
 Hence the total isoscalar-tensor amplitude reads 
\bqa \label{A9D}
\hspace{-3cm} A_7&=&  A_{\overline{K}^0[K^+K^-]_D^0}^{CF}(s_0,s_-,s_+)  + A_{K^0[K^+K^-]_D^0}^{DCS}(s_0,s_-,s_+)  \nonumber \\
 &=&  \frac{G_F}{2}\ (\Lambda_1+\Lambda_2) \ a_2 \ f_{D^0} \frac{1}{\sqrt{2}}
 F^{K^0f_2}(m_{D^0}^2)
~ g_{f_2 K^+K^-} \ \frac{D({\bf p_2},  {\bf p_0})}{m_{f_2}^2 - s_0 -i m_{f_2}\ \Gamma_{f_2}(s_0)}.
\eqa

\subsection{Combination of amplitudes}  
\label{scaamp}

The full  scalar amplitude ${\cal M}_1(s_0)$ is built up by the isoscalar and isovector amplitudes associated to the channel $[K^+K^-]_SK^0_S$ with the $f_0$ and $a_0^0$ resonances [Eqs.~(\ref{T1S}),~(\ref{T4S}) and~(\ref{A1S})]
\be
{\cal M}_1(s_0) = T_1 + A_1 + T_4={\cal M}_1^{n,I=0}(s_0) + {\cal M}_1^{s,I=0}(s_0) + {\cal M}_1^{I=1}(s_0), \label{M1T}
\ee
In Eq.~(\ref{M1T})  the ${\cal M}_1^{n,I=0}(s_0) $ and ${\cal M}_1^{s,I=0}(s_0)$ amplitudes are associated with the isoscalars  $f_0$,
\be \label{M11}
{\cal M}_1^{n,I=0}(s_0) = - \frac{G_F}{2}\ (\Lambda_1+\Lambda_2) \ a_2\ f_{K^0}\ ( m_{D^0}^2 - s_0)\  
F_0^{D^0 f_{0}}(m_{K^0}^2)\ \frac{\chi^{n}}{2} \Gamma_2^{n*}(s_0),
\ee
\be \label{M12}
{\cal M}_1^{s,I=0}(s_0) = -\frac{G_F}{2}\ (\Lambda_1+\Lambda_2) \ a_2\ f_{D^0}\ (m_{K^0}^2 - s_0)\ F_0^{K^0 f_{0}}(m_{D^0}^2) \frac{\chi^{s}}{\sqrt{2}} \Gamma_2^{s*}(s_0),
\ee
 while
the ${\cal M}_1^{I=1}(s_0)$ amplitude is associated with the isovectors $a_0^0$
\be \label{M13}
{\cal M}_1^{I=1}(s_0)= - \frac{G_F}{2}\ (\Lambda_1+\Lambda_2) \ a_2\ f_{K^0}\ ( m_{D^0}^2 - s_0)\  
F_0^{D^0 a_{0}^0}(m_{K^0}^2) \ \frac{1}{2} G_1(s_0).
\ee

The isoscalar amplitudes corresponding to the $\omega$ mesons [Eqs.~(\ref{T5P}) and~(\ref{A5P})] can be recombined with the isovector amplitudes  [Eq.~(\ref{T8P})] related to the  $\rho^0$ mesons
\bqa \label{M2}
{\cal M}_2 &=& T_5+T_6+A_4\nonumber \\
&=& \frac{G_F}{2}\ (\Lambda_1+\Lambda_2) \ a_2 \ (s_--s_+)
\left [ \frac{f_{K^0}}{f_{\rho}} \ A_0^{D^0 \rho}(m_{K^0}^2) \ F_u^{[K^+K^-]}(s_0) \right .\nonumber \\
&& \left .  \hspace{5cm} +\frac{f_{D^0}}{f_{\phi}} \ A_0^{K^0 \phi}(m_{D^0}^2) \ F_1^{[K^+K^-]^0_s}(s_0) \right ].
\eqa
Here we have used $A_0^{D^0 \omega}(m_{K^0}^2)/f_{\omega} \approx A_0^{D^0 \rho}(m_{K^0}^2)/f_{\rho}$  and defined
\be\label{Fu}
F_u^{K^+K^-}(s_0)=F_1^{[K^+K^-]^0_u}+F_1^{[K^+K^-]^1_u}.
\ee
The form factors $F_1^{[K^+K^-]^0_s}(s_0)\equiv F_s^{K^+K^-}(s_+)$ (in Eq.~\ref{M2})
and $F_u^{K^+K^-}(s_0)$ have been written in the forms given by Eqs.~(23) and~(25) of Ref.~\cite{PLB699_102}, respectively. The first form factor takes contributions from the $\phi(1020)$ and $\phi(1680)$ resonances while the second one from eight vector meson resonances $\rho(770)$, $\rho(1450)$, $\rho(1700)$ for the isovector part and $\omega(782)$, $\omega(1420)$, $\omega(1680)$ for the isoscalar part, as determined in Ref.~\cite{Bruch2005} for the constrained fit to kaon form factors.

Since the isovector-scalar form factor  $F_{0}^{[\overline{K}^0K^+]^1}(s_+)$ is related to the function $G_1(s_+)$ by the relation 
\be \label{G1formf}
F_{0}^{[\overline{K}^0K^+]^1}(s_+)=\frac{G_1(s_+)}{G_1(0)},
\ee
the isovector amplitude associated to the $a_0^+$ resonances in the channel $[K^0_SK^+]_SK^-$ [Eqs.~(\ref{T3S}) and~(\ref{A3S})] can be expressed as 
\bqa \label{M3Tb} 
{\cal M}_3(s_+) = T_3+A_3 =  \frac{G_F}{2}\ \Lambda_1&&\left[a_1 \ (m_{D^0}^2- m_{K}^2)\ \frac{m_{K}^2-m_{K^0}^2}{s_+}\ F_0^{D^0K^-}(s_+)\ \frac{1}{G_1(0)} \right. \nonumber\\
&& \left.  + a_2 \ f_{D^0} \ (m_{K}^2 - s_+)  F_0^{K^-a_0^+}(m_{D^0}^2)\right ]  G_1(s_+).
\eqa

The amplitude associated to the $\rho^+$ resonances [Eqs.~(\ref{T7P}) and~(\ref{A7P})] reads 
\bqa \label{M6Tb}
{\cal M}_4 &=& T_7 +A_5  = - \frac{G_F}{2}\ \Lambda_1 \  \left [s_0 - s_- + (m_{D^0}^2- m_{K}^2)\ \frac{(m_{K^0}^2-m_{K}^2)}{s_+} \right ] \nonumber \\
&& \hspace{3cm} \left \{ a_1\  F_1^{D^0K^-}(s_+) + a_2\ \frac{f_{D^0}}{f_{\rho}}  A_0^{K^-\rho^+}(m_{D^0}^2) \right \}
\ F_1^{[\overline{K}^0 K^+]^1}(s_+).
\eqa
The form factor $F_1^{[\overline{K}^0 K^+]^1}(s_+)=2~F_1^{[K^+K^-]^1_u}(s_+)$ gets contributions from the three $\rho$ resonances as explained below Eq.~(\ref{Fu}).

The isovector amplitude associated to the $a_0^-$ resonances 
in the channel $[K^0_SK^-]_SK^+$ [Eqs.~(\ref{T2S}) and~(\ref{A2S})] is
\bqa \label{M2Tb}
{\cal M}_5(s_-) &=& T_2 +A_2 
= - \frac{G_F}{2}\ \Lambda_2  \left [ a_1  \  f_{K^+}\ (m_{D^0}^2- s_-) \ F_0^{D^0 a_0^-}(m_{K}^2)\right . \nonumber\\
&& \hspace{4cm}\left .+  a_2 \ f_{D^0} \ (m_{K}^2 - s_-) \ F_0^{K^+a_0^-}(m_{D^0}^2)  \right ]  G_1(s_-) .
\eqa

The isovector amplitude associated to the $\rho^-$ resonances is given by Eqs.~(\ref{T6P}) and~(\ref{A6P})
\bqa \label{M6}
{\cal M}_6= T_8 + A_6 &=& \frac{G_F}{2}\ \Lambda_2  \ \left [s_0 - s_+ + (m_{D^0}^2- m_{K}^2)\ \frac{(m_{K^0}^2-m_{K}^2}{s_-}) \right ] 
\nonumber \\
&\times& \ \left \{ \ a_1\frac{f_{K^+}}{f_{\rho}} A_0^{D^0 \rho^-}(m_{K}^2) - a_2 \frac{f_{D^0}}{f_{\rho}} \ A_0^{K^+\rho^-}(m_{D^0}^2) \right \}
 F_1^{[\overline{K}^0 K^+]^1}(s_-),
\eqa
where we have applied the relation
$F_1^{[K^0 K^-]^1}(s_-)=F_1^{[\overline{K}^0 K^+]^1}(s_-)$.

Finally, the isoscalar-tensor amplitudes related to the $f_2$ [Eqs.~(\ref{T9D}) and~(\ref{A9D})] can be recombined to give 
\bqa \label{M7Tb}
{\cal M}_7 &=&T_9 +A_7 \nonumber \\
&=&-  \frac{G_F}{2}\ (\Lambda_1+ \Lambda_2)\ a_2 
\frac{1}{\sqrt{2}}f_{K^0} P_D
~g_{f_2 K^+K^-} \ \frac{D({\bf p_2},  {\bf p_0})}{m_{f_2}^2 - s_0 -i m_{f_2}\ \Gamma_{f_2}(s_0)}, 
\eqa
where 
\be \label{PD}
P_D= F^{D^0 f_2}(m_{K^0}^2)-\frac{f_{D^0}}{f_{K^0}} F^{K^0 f_2}(m_{D^0}^2)
\ee
can be treated as a complex constant parameter fitted to data.

\section{\bf Near  threshold comparison of the $S$-wave $K^+ K^-$ and $K^0_S K^+$ effective mass projections}
\label{comp}
   
Our study  can provide information on the $S$-wave content of the $\overline K K$ effective mass densities. In the region of low effective masses, near the $\overline  K K$ thresholds, one expects  dominant contributions of the
$S$- and $P$-wave amplitudes which simplifies the partial wave analysis of the experimental Dalitz plot distribution. This analysis has been performed by the \textit{BABAR} Collaboration for the following three decay reactions: $D^0 \to \overline  K^0 K^+ K^-$\cite{B5}, $D^0 \to K^- K^+ \pi^0$~\cite{B7} and $D_s^+ \to K^+ K^- \pi^+$~\cite{B11}. In the $K^+K^-$ $S$-wave effective mass distributions both scalar resonances $f_0(980)$ and $a_0(980)$ contribute while in the $\overline  K^0 K^+$ case only the $a_0(980)^+$ resonance is present. 

In the analyses of Refs.~\cite{B10} and~\cite{B5} the $f_0(980)$ contribution has not been introduced. A possible argument in favor of this choice has been formulated in Ref.~\cite{B5}, namely the authors have expected that the presence of the $f_0(980)$ resonance would lead to an {\it excess} in the $K^+K^-$ mass spectrum with respect to $\overline K^0K^+$. However, based on the limited statistics of 12540 events they have observed that both $\overline K K$ spectra are approximately equal. Below, using a much larger sample of about 80000 signal events~\cite{B10}, we show that the  $K^+K^-$ and $\overline K K$ spectra in $D^0$ decays into $K^0_S K^+ K^-$ are significantly different for low $\overline K K$ effective masses. Thus, the contribution of the $f_0(980)$ resonance is required to obtain a good description of the data of Ref.~\cite{B10}.
  
To proceed further, the definitions of the $\overline K K$ effective $S$-wave mass distributions corrected for phase space are needed. If we denote by $N(s_0,s_+)$ the number of events of the $D^0~\to~\overline K^0 K^+ K^-$ reaction, the corresponding Dalitz-plot density distribution is given by $d^2N(s_0,s_+)/ds_0ds_+$. 
 The $K^0_S K^+$ effective mass squared distribution corrected for phase space can be then defined as
\be \label{e1}
\frac{dn_{K^0_S K^+}(s_+)}{ds_+}= \frac{1}{s_{0max}-s_{0min}} \int_{s_{0min}}^{s_{0max}} \frac{d^2N(s_0,s_+)}{ds_0ds_+}ds_0,
\ee
\begin{figure}[t!]  \begin{center}
\includegraphics[width=0.85\textwidth]{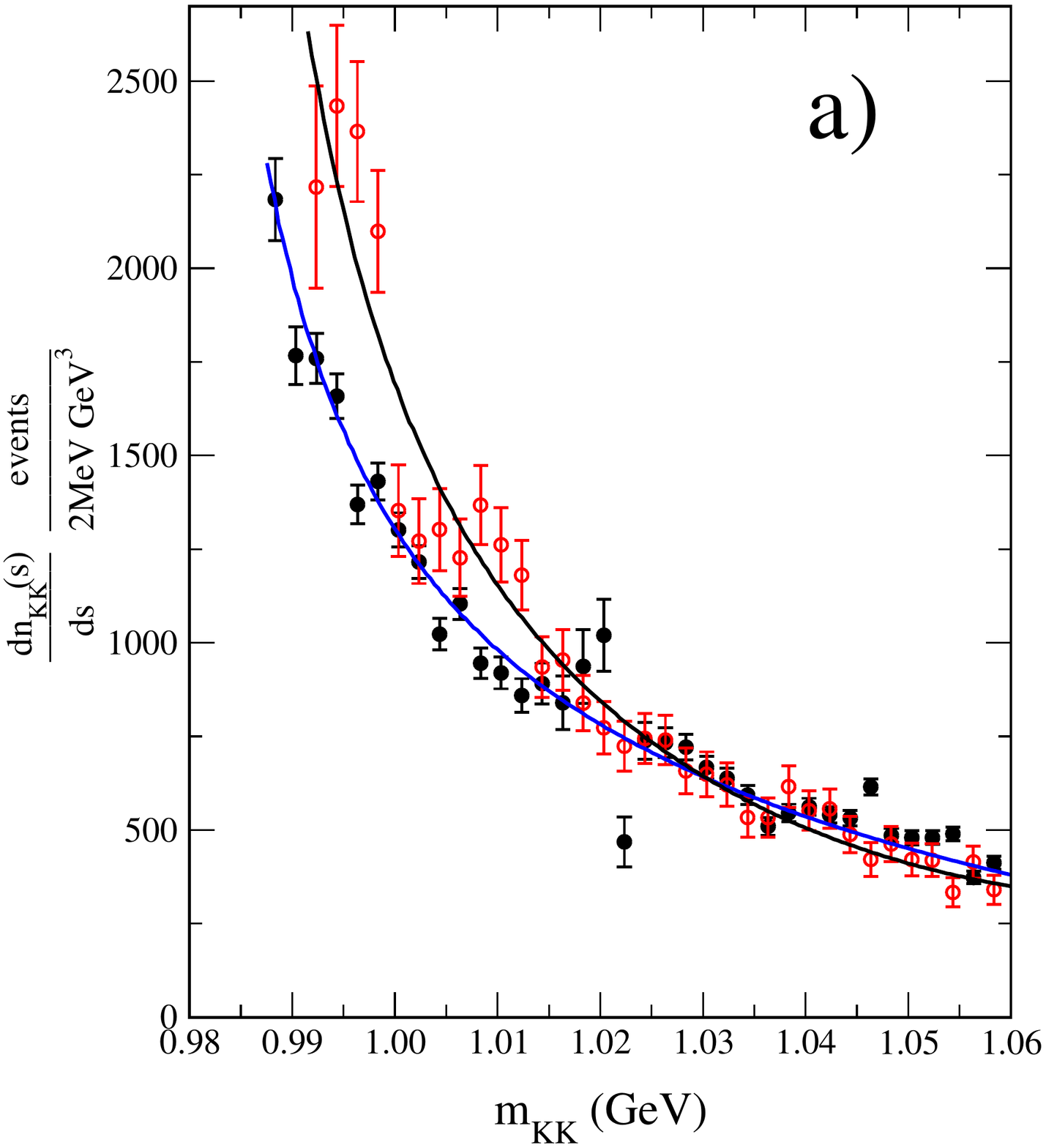}~
\hspace{-6.cm}
\includegraphics[width=0.85\textwidth]{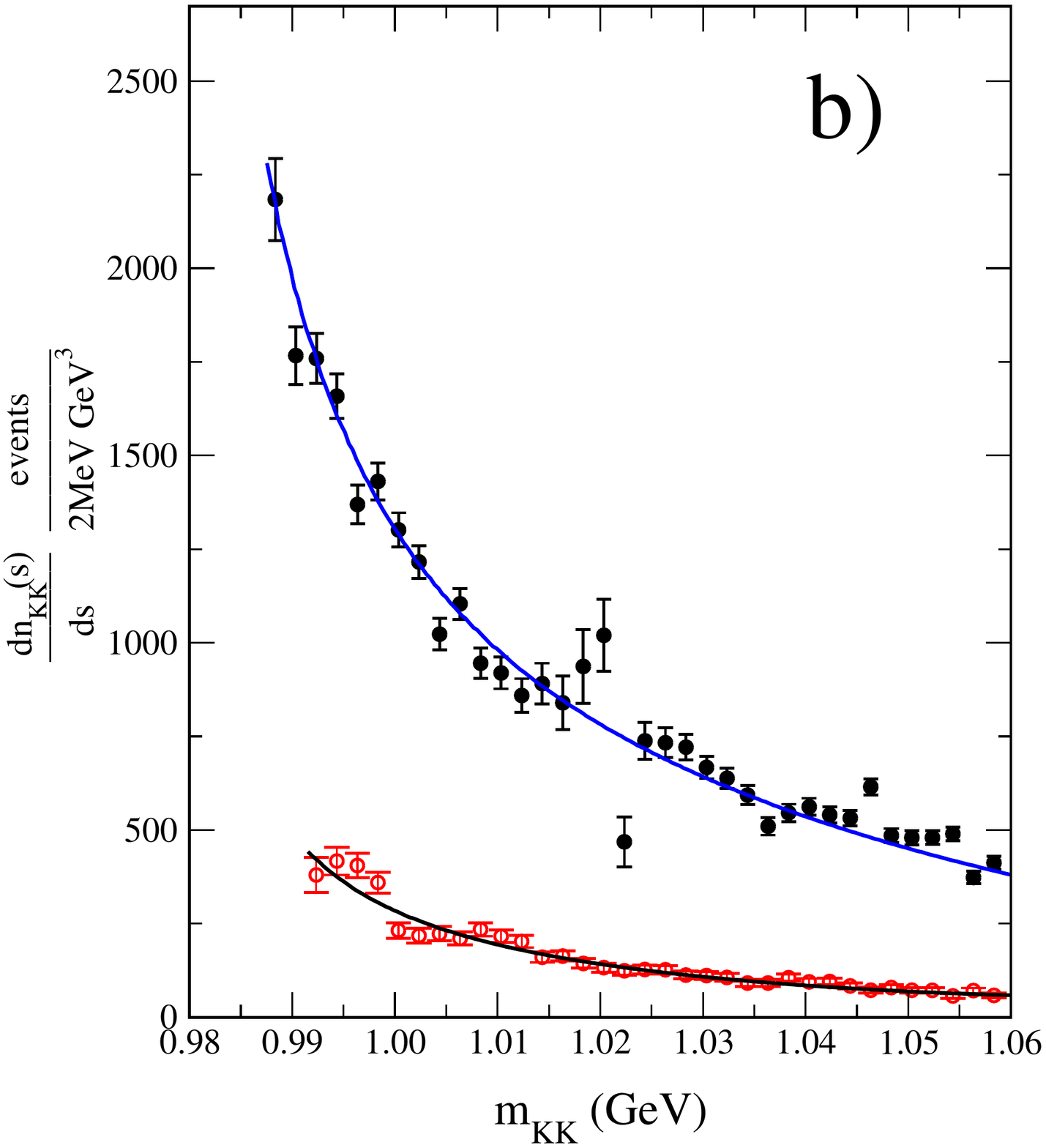}
\vspace*{6pt}
\caption{  Comparison of $K^+K^-$ and $\overline K^0 K^+$ $S$-wave effective mass squared distributions corrected for phase space as functions of the variable $m_{KK}$ in bins of 2 MeV [see Eqs.~(\ref{e1}) and~(\ref{e5})].
The variable $m_{KK}$ is equal to $m_{K^0_S K^+}$ for the $\overline K^0 K^+$ data points (open red circles) or equal to $m_0$ for the $K^+K^-$ data points (filled black circles). 
In the left panel a) the $\overline  K^0 K^+$ distribution is normalized to the number of events of the $K^+K^-$  distribution when integrated over $m_{KK}$ from the $ K^+ K^-$ threshold up to~1.05 GeV.
In the right panel b) the $\overline  K^0 K^+$ distribution has not been  renormalized.
The curves correspond to the theoretical model calculations (see Sec.~\ref{results}). 
}
\label{Projab} \end{center}
\end{figure}
where $s_{0max}$ and $s_{0min}$ are the maximum and minimum $s_0$ values at fixed $s_+$.
If we limit ourselves to the low $s_+$ values (for example, up to about 1.05 GeV$^2$) then to a good accuracy the above distribution corresponds to the $S$-wave part of the total decay amplitude related to the isovector-scalar$a_0(980)^+$ resonance. The reason is that the dominant  $P$-wave contribution, related to the $\phi(1020)$ resonance, is only present in the $K^+K^-$ decay channel. Moreover, the low-mass $K^0_S K^+$ and  $K^+K^-$ distributions are very well separated on the Dalitz plot~\cite{B10}.

For the low effective $K^+K^-$ masses one has to subtract the $P$-wave contribution. Following Ref.~\cite{B5},
this can be done by calculating the spherical harmonic moments
\be \label{e2}
\sqrt{4\pi}\langle Y^0_0(s_0) \rangle=\int_{s_{+min}}^{s_{+max}} \frac{d^2N}{ds_0ds_+} ds_+
\ee
and
\be \label{e3}
\sqrt{4\pi}\langle Y^2_0(s_0) \rangle=\sqrt{5}\int_{s_{+min}}^{s_{+max}} P_2(cos\theta)\frac{d^2N}{ds_0ds_+} ds_+,
\ee
with 
\be \label{e4}
P_2(cos\theta)=\frac{1}{2}(3~ cos^2\theta-1), 
\ee
and where $\theta$ is the helicity angle of the $K_S^0$ meson defined with respect to the $K^+$ direction in the $K^+K^-$ center-of-mass frame, $s_{+max}$ and $s_{+min}$ being the maximum and minimum $s_+$ values at fixed $s_0$.
The $S$-wave $K^+ K^-$ effective mass squared distribution corrected for phase space is then defined as
\be \label{e5}
\frac{dn_{K^+K^-}(s_0)}{ds_0}= \frac{\sqrt{4\pi}}{s_{+max}-s_{+min}}  [\langle Y^0_0(s_0) \rangle-\frac{\sqrt{5}}{2}\langle Y^2_0(s_0) \rangle].
\ee
For completeness we give below the kinematical relation for the cosine of the helicity angle  
\be \label{e6}
cos\ \theta=\frac{s_--s_+}{s_{+max}-s_{+min}},
\hspace{1cm}{\rm where} \hspace{1cm}
s_{+max}-s_{+min}=4 \lvert{\bf p}_{+}\rvert  \lvert {\bf p}_{0}\rvert,
\ee
with $|{\bf p}_{+}|$ and  $|{\bf p}_{0}|$  defined by Eqs.~(\ref{p2p+}) and (\ref{p2p0}), respectively.

  We have performed a simplified partial wave analysis of the \textit{BABAR} data published in Ref.~\cite{B10}. As described above, only the $S$- and $P$-waves have been included and the effective $K^0_S K^+$ and  $K^+K^-$ masses were smaller than 1.05 GeV$^2$.  The number of signal events of the $D^0 \to \bar K^0 K^+ K^-$ decays was 79900$\pm$300. Based on the Dalitz plot density distributions corrected for reconstruction efficiency and background,  the $S$-wave $K^0_S K^+$ and $K^+ K^-$ effective mass distribution corrected for phase space are calculated using Eqs.~(\ref{e1}) and~(\ref{e5}). 

The  comparison of the calculated $S$-wave $K^+K^-$ and $\overline K^0 K^+$ distributions is shown in Fig.~\ref{Projab}. 
In the left panel a)  a clear surplus of the $\bar K^0 K^+$ distribution over the $K^+K^-$ one is
seen below 1.02~GeV.  Above $m_{KK}=1.02$ GeV the open circles corresponding to $\overline K^0 K^+$ spectrum are in majority located below the closed circles ($K^+K^-$ events), so we observe a crossing of the two distributions. This effect is statistically significant. It was not so clear in 2005 when the first set of the \textit{BABAR} data was published. But even then, in Fig. 8 of Ref.~\cite{B5}, one can see the same cross-over tendency as in Fig.~\ref{Projab} although the statistics was lower by a factor  larger than 6.
In the right panel b), one sees that unrenormalized $\overline K^0 K^+$ distribution is lower than the   $K^+K^-$ distribution by a factor of about 4.
The lines show the corresponding theoretical distributions based on the best fit to the Dalitz plot density distributions described in the next section\footnote{These distributions are also relatively well described by the two alternative models given in the Appendix~\ref{fitswMOBM} except for the two first data point of the $m_{KK}$ distribution.}.
 
   In conclusion, the shape of the $K^+K^-$ and $\overline K^0 K^+$ $S$-wave effective mass squared distributions, corrected for phase space, is significantly different, so in the phenomenological analysis of the $D^0 \to \overline K^0 K^+ K^-$ data one cannot neglect the $f_0(980)$ contribution in the decay amplitude. 

\section{Results and discussion} \label{results}
The differential branching fraction or the Dalitz plot density distribution is defined as
\be \label{d2Br}
\frac{d^2{\rm Br}}{ds_+ds_0}=\frac{|{\cal M}|^2}{32(2\pi)^3m_{D^0}^3 \Gamma_{D^0}},
\ee
where ${\cal M}=\sum_{i=1}^7{\cal M}_i$ is the decay amplitude for the process studied and $\Gamma_{D^0}$ is the $D^0$ width. The decay amplitudes ${\cal M}_i$ have been derived in Sec.~\ref{amplitudes}.
In Table~\ref{fixed} one can find  some constant parameters which appear in these amplitudes.

\begin{table*}[h]
\caption{Values  of coupling constants (in GeV) and the fixed form factors.}
\label{fixed}
\begin{center} \begin{tabular}{lcc}
\hline \hline
\hspace{0.0cm}Parameter &Value & Reference\\ 
\hline
$f_{K^0}$=$f_{K^+}$          &   0.1561  &  \cite{JPD_PRD89}   \\ 
$f_{\rho}$                 &   0.209   &  \cite{Beneke2003}      \\
$f_{\phi}$                  &   0.22    &  \cite{Ball2007}        \\
$f_{D^0}$                  &   0.2067  &  \cite{JPD_PRD89}   \\
$F_0^{D^0 f_0}(m_{K^0}^2)$=$F_0^{D^0 a_0^0}(m_{K^0}^2)$  & 0.18 & \cite{El-Bennich_PRD79}    \\
$A_0^{D^0 \rho^0}(m_{K^0}^2)$  &   $0.7$   & \cite{JPD_PRD89} \\
\hline \hline
\end{tabular} \end{center}
\end{table*}

To make a comparison of experimental data with model predictions 
the Dalitz diagram has been divided into five regions as shown in Fig.~\ref{Figure:Cells}. The dimensions in different regions have been adjusted to the density of experimental events. This has been done in two steps. In the first step
the units $u_0 = 8.94 \times10^{-4}$~GeV$^2$ and $u_+ = 8.97 \times 10^{-4}$~GeV$^2$ 
corresponding to the one thousand of the full kinematic range of the variables $s_0$ and $s_+$ have been chosen. A small difference between $u_0$ and $u_+$ comes form the difference between the $K^0_S$ and $K^+$ masses. The cells in the ranges I, III and IV have the dimensions $11u_0\times 11u_+$ while in the range II the cells are larger having the dimensions $41u_0 \times 41u_+$. Because of the high density of experimental events around the position of the $\phi(1020)$ resonance, the cells in the narrow range V have the dimensions $1u_0 \times 35u_+$. In the second step we have checked whether the experimental number of events in a given cell was higher than ten. When this was not the case the adjacent cells have been combined together to group a sufficient number of events in an enlarged cell.
Altogether the total number of cells was equal to $N_{cells} = 1196$ (including 164 enlarged cells). The cell numbers in the regions I, II, III, IV and V were equal to 135, 282, 242, 187 and 350, respectively. 

The fit of the model parameters to the experimental data has been performed using the $\chi^2_{tot}$ function defined as a sum of two components:
\be \label{Chi2Tot}
\chi^2_{tot} = \sum_{i=1}^{N_{cells}} \chi^2_i + \chi^2_{{\rm Br}}.
\ee

The value of $\chi^2_i$ for each cell $i$ has been defined as in Ref.~\cite{chi2}:
\be \label{chi2Log}
\chi^2_i = 2\Big [ N^{exp}_i-N^{th}_i+N^{th}_i ln(\frac{N^{th}_i}{N^{exp}_i})\Big ],
\ee
where $N^{exp}_i$ is a number of experimental signal events in the cell $i$ corrected for the reconstruction efficiency and $N^{th}_i$ is the theoretical number of events in the same cell\footnote{The efficiency and the signal distributions on the Dalitz diagram have been provided to us by Fernando Martinez-Vidal ~\cite{B10}. The samples of the $D^0$ and $\bar D^0$ decays into $K^0_S K^+ K^-$ have been combined.}.
Including the above corrections one gets the total number of experimental events equal to
$N^{exp}= 80379$. The total number of theoretical events is then taken equal to $N^{exp}$. 

The second component of the $\chi^2$ function is given by
\be \label{Chi2Br}
\chi^2_{{\rm Br}} = w ~ \Big ( \frac{{\rm Br}^{exp} - {\rm Br}^{th}}{\Delta {\rm Br}^{exp}} \Big )^2.
\ee
In our fit the experimental branching ratio for the decay $D^0 \to K_S^0 K^+ K^-$ 
has been taken equal to ${\rm Br}^{exp} = 4.45  \times 10^{-3}$ and its error $\Delta {\rm Br}^{exp} =0.34 \times 10^{-3}$. These values agree well with recent values of the Particle Data Group~\cite{PDG2020}. The theoretical branching fraction ${\rm Br}^{th}$ is obtained from Eq.~(\ref{d2Br}) after integrations of $\frac{d^2{\rm Br}}{ds_+ds_0}$ over the variables $s_+$ and $s_0$.
The weight factor $w$ in our fit has been set to 100 in order to obtain a good agreement of the theoretical branching fraction with its experimental value.

\begin{figure}[h]  \begin{center}
\hspace{2.cm}
\includegraphics[scale = 0.5]{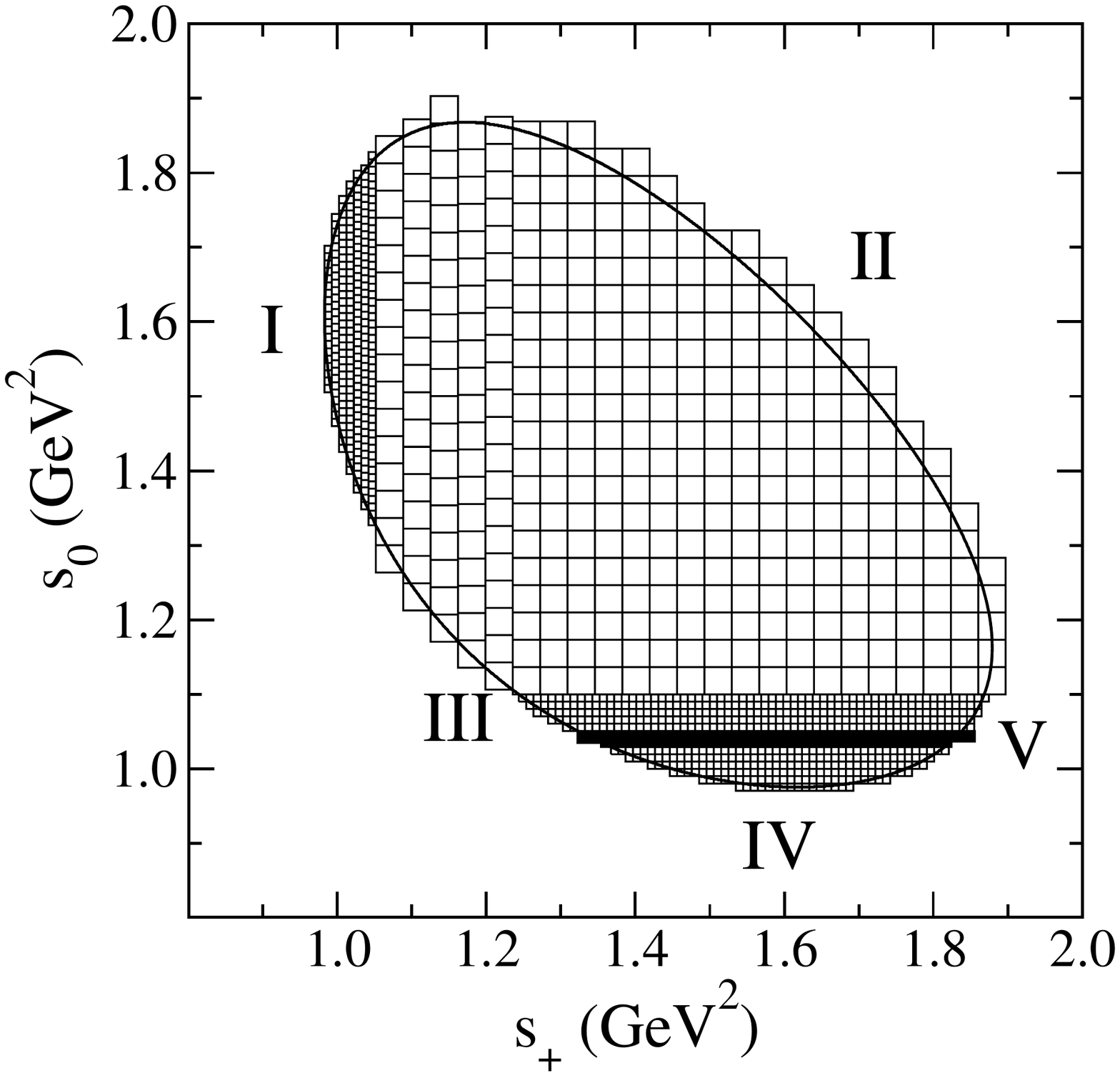}
\caption{Partition of the Dalitz contour into five regions. Different sizes of the ($s_+,s_0$) cells are shown.}
\label{Figure:Cells}
\end{center}
\end{figure}

We have performed many fits with our model using different parameter sets. The best fit $\chi^2~=~1474.4$ has been obtained with the nineteen free parameters which are displayed in Table~\ref{parameters}. Since the number of degrees of freedom is $ndf=1196-19=1177$, the $\chi^2$ per degree of freedom is  $\chi^2/ndf=1.25$.

\begin{table}
\caption{Parameters of our model amplitudes and their errors. Phases are given in radians.}
\begin{center} \begin{tabular}{ccc}
\hline \hline
Parameter & modulus & phase \vspace*{0.2cm} \\ 
\hline
$\chi_n$ &   22.5 GeV$^{-1}$  fixed & $2.22^{+0.82}_{-0.98}$ \vspace*{0.2cm} \\   
$F_0^{K^0f_0}(m_{D^0}^2)$ & $2.22^{+0.26}_{-0.17}$ & $2.21 \pm 0.10$ \vspace*{0.2cm} \\ 
$r_2$ & $(5634^{+509}_{-560})$ GeV$^{3/2}$ & \vspace*{0.2cm} \\ 
$r_1/r_2$  &  0.88  $\pm 0.01$  & \vspace*{0.1cm} \\ 
$s'$ & $(1.558^{+0.016}_{-0.014})$ GeV$^2$ & \vspace*{0.2cm} \\  
$p_1$ & (-1.84 $\pm 0.01)$  GeV$^{-2}$ & \vspace*{0.1cm} \\ 
$p_2$ &  (1.09  $\pm 0.01)$ GeV$^{-4}$ & \vspace*{0.1cm} \\ 
$p_3$ & -(0.212  $\pm 0.004)$  GeV$^{-6}$ & \vspace*{0.1cm} \\ 
$F_0^{K^-a_0^+}(m_{D^0}^2)$ & $0.25^{+0.02}_{-0.03}$ & $5.33^{+0.12}_{-0.08}$
 \vspace*{0.1cm} \\
$A_0^{K^0\phi}(m_{D^0}^2)$  & 0.985  $\pm 0.007$ &  $3.67^{+0.12}_{-0.09}$ \vspace*{0.1cm} \\    
$M_{\phi}$  &  (1019.58   $\pm 0.02$) MeV & \vspace*{0.1cm} \\ 
$\Gamma_{\phi}$ & (4.72   $\pm 0.04$) MeV & \vspace*{0.2cm} \\ 
$A_0^{K^-\rho^+}(m_{D^0}^2)$  & $9.38^{+0.63}_{-0.58}$ & $5.01^{+0.06}_{-0.05}$ \vspace*{0.2cm} \\ 
$P_D$ &   $5.52^{+1.25}_{-1.24}$ & $3.97^{+0.23}_{-0.25}$ \vspace*{0.2cm} \\ 
\hline \hline
\end{tabular} \end{center}
\label{parameters}
\end{table}

The value of the constant $\vert \chi^n \vert $  has been estimated using a relation derived similarly as Eq.~(18) in Ref.~\cite{AF} in which the coupling constants of the $f_0(980)$ resonance to the $K^+K^-$ pair are taken into account instead of the $f_0(980)$ coupling to the $\pi\pi$ system. However, in the present case one has to include two close $f_0(980)$ poles sitting on the sheets (-+-) and (-++) (for their complex energy positions, $E_{R_1}$ and $E_{R_2}$, see Table~\ref{Poles} in Appendix~\ref{UpdatedTpipi}). One can generalize Eq.(18) from Ref.~\cite{AF}, valid for the pole position of a single resonance, to the case of two close resonances:
\be \label{chin}
\vert \chi^n \vert \approx\frac{1}{|\Gamma_2^n(s_0)|}\Big |\frac{g_{R_1K^+K^-}}{E_{R_1}^2-s_0}+\frac{g_{R_2K^+K^-}}{E_{R_2}^2-s_0}\Big |,
\ee
where $g_{R_1K^+K^-}$ and $g_{R_2K^+K^-}$ are the coupling constants of the two $f_0$ resonances to $K^+K^-$.
If one takes the effective $K^+K^-$ mass in the range between 960 MeV and 990 MeV then using 
Eq.~(\ref{chin}) the averaged value of $\vert \chi^n \vert $ calculated in this range is 22.5 GeV$^{-1}$.

The magnitude of the $\chi^s$ parameter is taken equal to that of $\vert \chi^n \vert $ and its phase is set to zero. The reason for this choice is the presence of the undetermined complex value of the form factor $F_0^{K^0 f_0}(m^2_{D^0})$ which is multiplied by $\chi^s$ in the ${\cal M}_1$ amplitude. The $F_0^{K^0 f_0}(m^2_{D^0})$ value results from the fit to data.

The form factors $\Gamma_2^n(s_0)$ and $\Gamma_2^s(s_0)$ have been calculated in a three-channel model of meson-meson interactions ($\pi\pi$, $K \bar{K}$ and an effective $2\pi\ 2\pi$), introduced in Ref.~\cite{DedonderPol}.
These form factors depend not only on the values of the meson-meson parameters listed in Table~\ref{fitparamet} in Appendix~\ref{UpdatedTpipi} but also on two other parameters $\kappa$ and $c$ defined by Eqs. (28) and (39) in Ref.~\cite{DedonderPol}, respectively. Their values $\kappa=2807.3$ MeV and $c=0.109$ GeV$^{-4}$ have been fitted to the 
$B^{\pm}\to K^{\pm} K^+K^-$ decay data analyzed in Ref.~\cite{KKK}.

In Fig.~\ref{Gamns} we show the effective energy dependence $E$ of moduli and phases of the $K \overline K$ isoscalar-scalar nonstrange $\Gamma_2^n(E)$ and strange $\Gamma_2^s(E)$ form factors. The energy $E$ is equal to the square root of $s$. 

\begin{figure}[t!] \begin{center}
\includegraphics[width=0.6\textwidth]{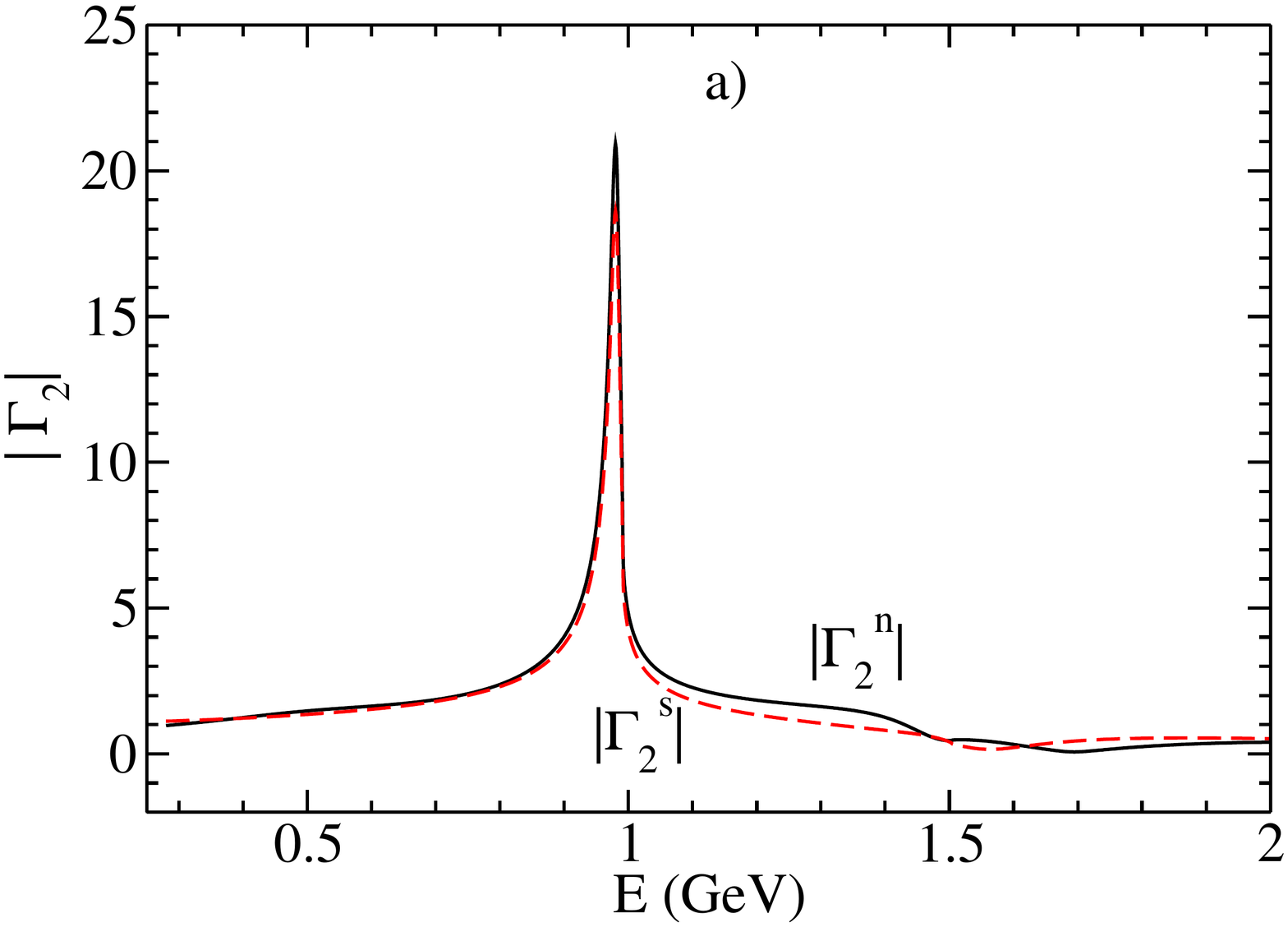}~
\hspace{-1.8cm}
\includegraphics[width=0.6\textwidth]{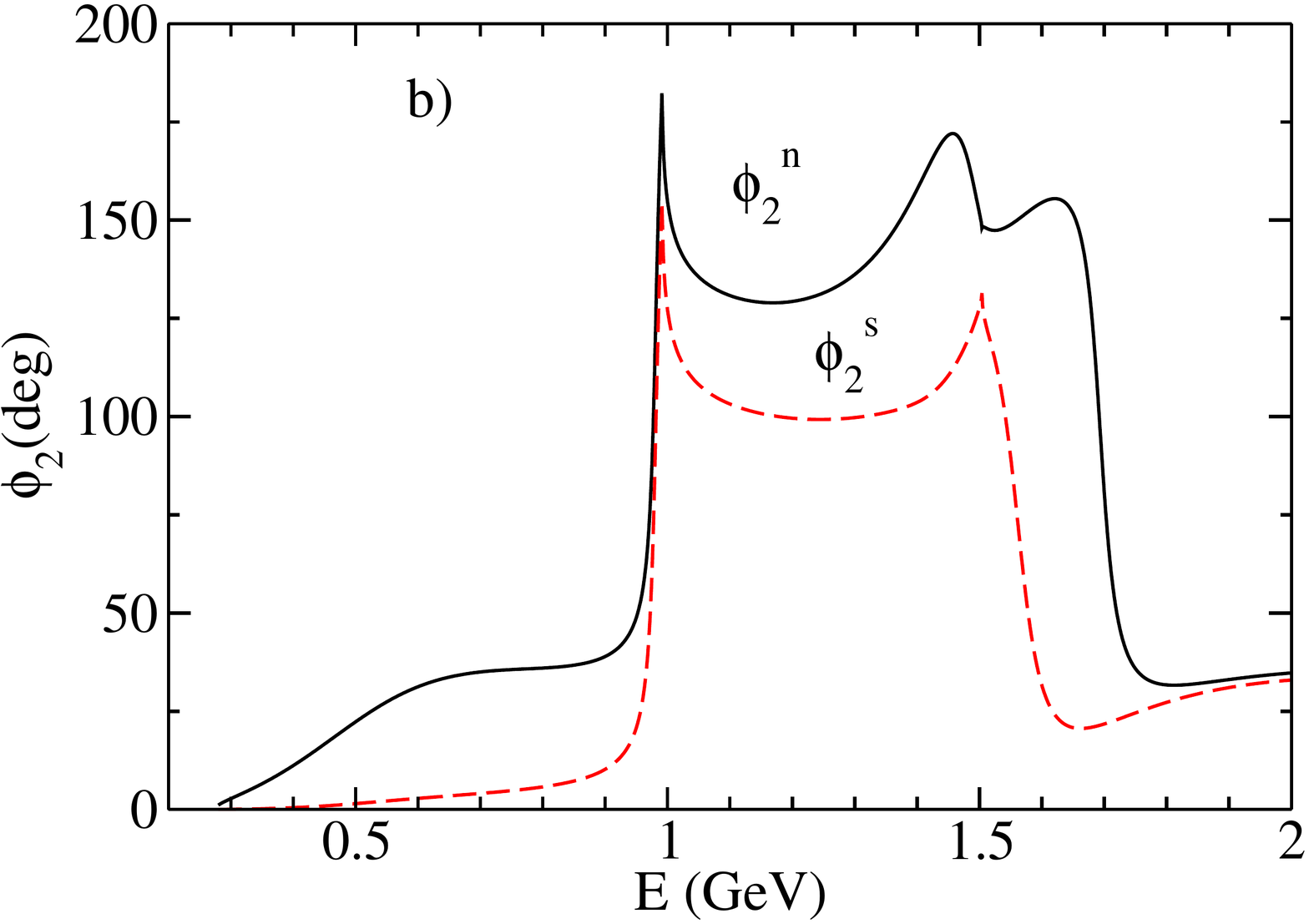}
\vspace{-1cm}
\caption{ a) moduli of the isoscalar-scalar  kaon nonstrange  $\Gamma_2^n(E)$ (solid line) and strange $\Gamma_2^s(E)$ (dashed line) form factors; b) the corresponding phases.}
\label{Gamns} \end{center} 
\end{figure}

In the above model the kaon threshold energy was set equal to the sum of the charged and neutral kaon masses. However, the ${\cal M}_1$ amplitude corresponds to the isoscalar $K^+K^-$ $S$-wave state with a threshold lower by about 3.9 MeV in comparison with the $K^+K^0_S$ threshold energy. In order to take this effect into account in an approximate way,  we introduce the variable 
$$\bar s_0= s_0 \ \frac{s_{av}}{s_{th}}$$
with $s_{th}=4 m_{K}^2$ and the correction factor  is $s_{av}/s_{th}= 1.008172$. The kaon form factors have to be evaluated at this argument, i.e.,  $\Gamma_2^n(\bar s_0)$ and $\Gamma_2^s(\bar s_0)$. To improve the quality of the data fit the form factors $\Gamma_2^n(\bar s_0)$ and $\Gamma_2^s( \bar s_0)$ have been multiplied by the function
\be \label{polygamma}
P(s_0)=\frac{1-\frac{s_0-s_{th}}{s'-s_{th}}}{1 + b s_0^3},
\ee
where $s'$ is a new parameter which is fitted to the data (see Table~\ref{parameters}). It corresponds to a zero of $P(s_0)$. The third order polynomial in the denominator, with the constant $b$ fixed to  0.0154 GeV$^{-6}$, is introduced in order to control asymptotically the high energy  behavior of the ${\cal M}_1$ amplitude. This denominator replaces the denominator $(1+c E^4)$ with $E=\sqrt{s_0}$ in Eq. (39) of ~\cite{DedonderPol}. 
A plot of the function~(\ref{polygamma}) used in our fit is shown as the continuous black line denoted R$_{BF}$ in Fig.~\ref{PlotPandPFs0}~(a) where it is compared to the corresponding functions used in the two alternative fits  MO$_{P1(P2)}$ described in  Appendix~\ref{fitswMOBM}. This function  reduces
the moduli of the amplitudes which depend on the isoscalar-scalar form factors.

The masses and widths of the isovector-scalar resonances $a_0(980)$ and $a_0(1450)$ are presented in 
Table~\ref{Table_a0}. They have been fixed during the minimization of the $\chi^2$ function. The parameters of the $a_0(1450)$ on sheet $(-~-)$ were taken from Ref.~\cite{PDG2020}. However, we have studied the influence of the position of the $a_0(980)$ pole on sheet $(-~+)$ in the complex energy domain on the $\chi^2$ minimum curve. In this way the $a_0(980)$ mass and width on sheet $(-~+)$ have been determined together with an estimation of their errors. The masses and widths of other two associated $a_0$ poles are also given in Table~\ref{Table_a0}.

\begin{table} \begin{center}
\caption{Parameters of resonances $a_0(980)$ and $a_0(1450)$.}
\begin{tabular}{cccc}
\hline
\hline
& mass (MeV) &  width (MeV) & Riemann sheet \\
\hline
$a_0(980)$ & $979^{+3}_{-2}$  & $25^{+8}_{-6}$ & $-~~+$\\
$a_0(980)$ & $959$& $34$&$-~~-$\\
$a_0(1450)$ & $1474$ & $132$& $-~~-$\\
$a_0(1450)$ & $1470$ & $91$& $-~~+$\\
\hline
\hline
\label{Table_a0}
\end{tabular} \end{center}
\end{table}

\begin{table}
\caption{Potential parameters of the $K \overline K$ $S$-wave isospin one interaction.}
\begin{center} 
\begin{tabular}{cc}
\hline \hline
$\beta_1$ & 21.662 GeV  \\
$\beta_2$ & 21.831 GeV  \\
$\lambda_1$ & $-2.9850 \times 10^{-2}$  \\
$\lambda_2$ & $-6.7977 \times 10^{-2}$ \\
$\lambda_{12}^2$ & $2.2142 \times 10^{-7}$ \\
\hline \hline
\label{pot_param}
\end{tabular} \end{center}
\end{table}

The coupled channel model of the $a_0(980)$ and $a_0(1450)$ resonances described in Ref.~\cite{AFLL} has been implemented. There, the separable $\pi \eta$ and  $K \overline K$ interactions have been used in the calculation of the $S$-wave isospin one scattering amplitudes. Altogether the model has five parameters: two range parameters $\beta_1$ and
$\beta_2$, two channel coupling constants $\lambda_1$ and $\lambda_2$, and the interchannel coupling constant $\lambda_{12}$ (here the channel $\pi \eta$ is labeled by $1$ and the channel  $K \overline K$ by $2$). The potential parameters are given in Table~\ref{pot_param}.
There exist direct numerical relations between the four parameters describing the positions of the $a_0(980)$ and $a_0(1450)$ resonances in the complex energy plane (Table~\ref{Table_a0}) and the four potential parameters $\beta_1$, $\lambda_1$, $\lambda_2$ and $\lambda_{12}$ at fixed value of the fifth parameter $\beta_2$. These relations are given in Ref.~\cite{LL}.

The function $G_1(s)$ is introduced to describe a transition from the $u \bar{u}$ pair to the  $K \overline K$ spin zero isospin one state. Two isovector-scalar resonances $a_0(980)$ and $a_0(1450)$ can be formed during that transition. Both resonances are also coupled to the $\pi \eta$ state. Therefore it is natural to consider three cases for the transition from the $u \bar{u}$ pair to the  $K \overline K$ state. In the first case the  $K \overline K$ pair is directly formed from the $u \bar{u}$ pair. In the second case the $K \overline K$ pair undergoes the elastic rescattering in the final state. In the third case the intermediate $\pi \eta$ pair is formed and then the inelastic transition to the  $K \overline K$ state takes place. 
The interaction between the meson-meson pairs is treated in the framework of the separable potential model fully described in Ref.~\cite{LL} and used to study the properties of the $a_0$ resonances (Refs.~\cite{AFLL},~\cite{AFLL2}). 

Below we briefly derive the dependence of the $G_1(s)$ function on the meson-meson transition amplitudes.  
Labelling by 1 the $\pi \eta$ channel and by 2 the  $K \overline K$ channel, one can express $G_1(s)$ as a superposition of three terms:
\be \label{G123}
G_1(s)= R_2(s) + I_{22}(s)+ I_{12}(s),
\ee
where 
\bqa
R_2(s)&=&r_2 \ W(s) \ g_2(k_2), \label{R2} \\
I_{22}(s)&=&r_2\  W(s)\  \frac{T_{22}(s)}{g_2(k_2)} \ C_2(s), \label{I22} \\
I_{12}(s)&=&r_1 \ W(s) \  \frac{T_{12}(s)}{g_1(k_1)} \ C_1(s). \label{I12}
\eqa
Here $r_1$ and $r_2$ are the coupling constants corresponding to the $u \bar{u}$ transitions to the $\pi \eta$ and $K \overline K$ states, respectively.  The function $W(s)$ is the third-degree polynomial 
\be \label{polyG1}
W(s)=1 + p_1 s + p_2 s^2 + p_3 s^3.
\ee
where $p_1$, $p_2$ and  $p_3$ are the real parameters included in the list of  the model free parameters (see Table~\ref{parameters}). We keep the same $p_j, j=1,2,3$, parameters for both channels. The fitted polynomial $W(s_0)$ is plotted as the continuous black line in  Fig.~\ref{PlotPandPFs0}(b) where it is compared to the polynomial $P_F(s_0)$ defined by Eq.~(\ref{PFs0}) and used in the  alternative MO$_{P1(P2)}$ fits discussed in Appendix~\ref{fitswMOBM}. The introduction of these polynomials improves the quality of the $\chi^2$ fit, in particular in the region~II where the density of events is small. The functions $g_i(k_i), i=1, 2$, are the vertex functions 
\be \label{gi}
g_i(k_i)=\sqrt{\frac{2\pi}{m_i}}\frac{1}{k_i^2+\beta_i^2},
\ee
where $m_i$ are the channel reduced masses, $k_i$ are the channel momenta and  $\beta_i$ are the range parameters.
In the $\pi \eta$ channel $m_1=m_{\pi}m_{\eta}/(m_{\pi}+m_{\eta})$, in the  $K \overline K$
channel $m_2=m_K/2$. We take the neutral $\pi$ mass $m_{\pi}=134.977$ MeV and $m_{\eta}=547.862$ MeV. 
The function $T_{22}(s)$ in Eq.~(\ref{I22}) is the elastic  $K \overline K$scattering amplitude
and $T_{12}(s)$ in Eq.~(\ref{I12}) denotes the transition amplitude from the $\pi \eta$ channel to the  $K \overline K$ channel.
In Eq.~(\ref{I22}) one finds the integral
\be \label{C2}
C_2(s)=\int \frac{d^3p}{(2\pi)^3} \frac{g_2^2(p)}{E+i\epsilon -2 E_K(p)}
\ee
and  in Eq.~(\ref{I12}) we have
\be \label{C1}
C_1(s)=\int \frac{d^3p}{(2\pi)^3} \frac{g_1^2(p)}{E+i\epsilon -E_{\pi}(p)-
E_{\eta}(p)},
\ee
where the energies are defined as 
$E=\sqrt{s}$, $E_K(p)=\sqrt{p^2+m_K^2}$, 
$E_{\pi}(p)=\sqrt{p^2+m_{\pi}^2}$
and $E_{\eta}(p)=\sqrt{p^2+m_{\eta}^2}$.
The modulus and the phase of the resulting $G_1(s)$ function are plotted in
Figs.~\ref{F12}~(a) and \ref{F12}~(b), respectively.

\begin{figure}[h]  \begin{center}
\includegraphics[width=0.6\textwidth]{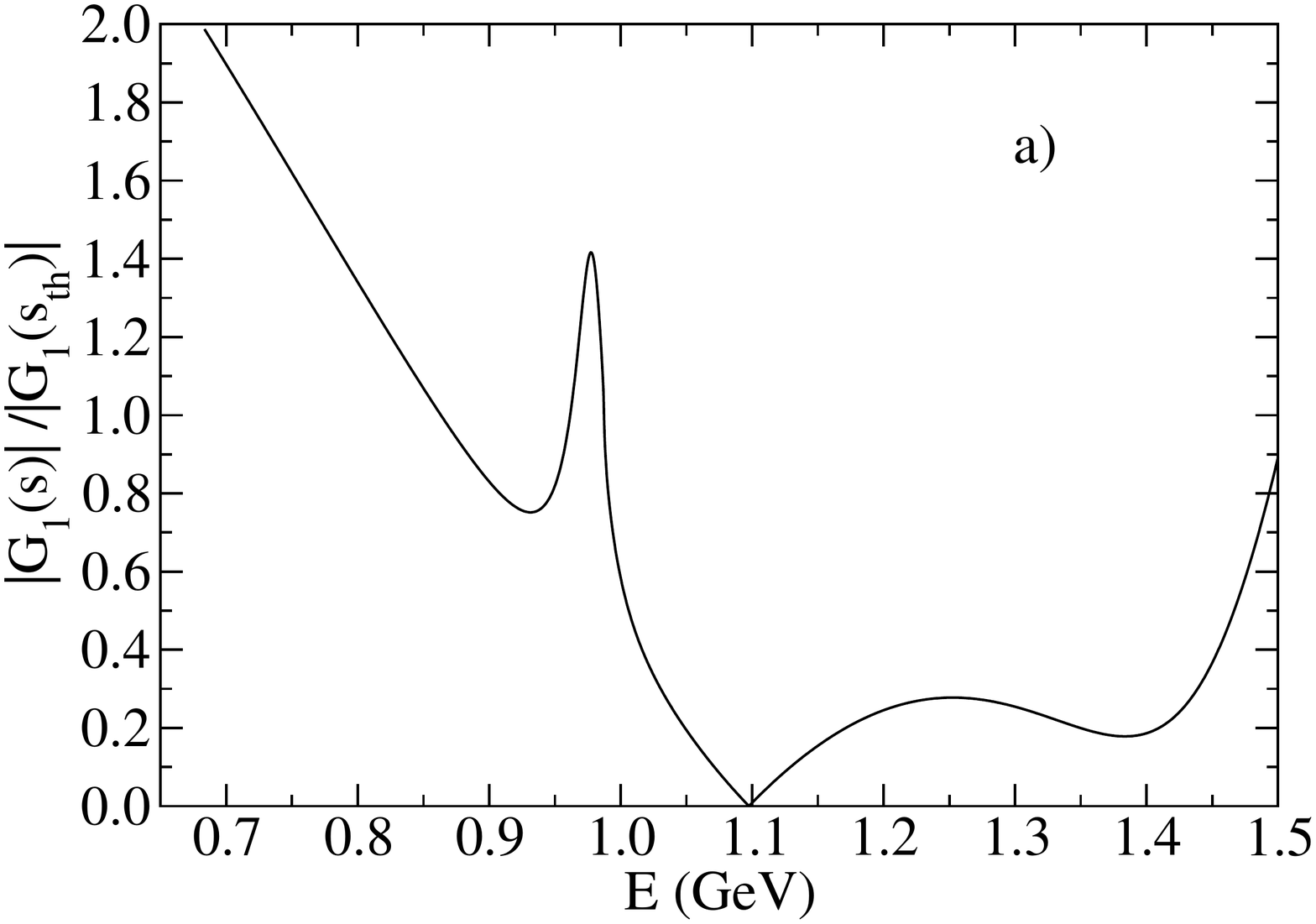}~
\hspace{-1.8cm}
\includegraphics[width=0.6\textwidth]{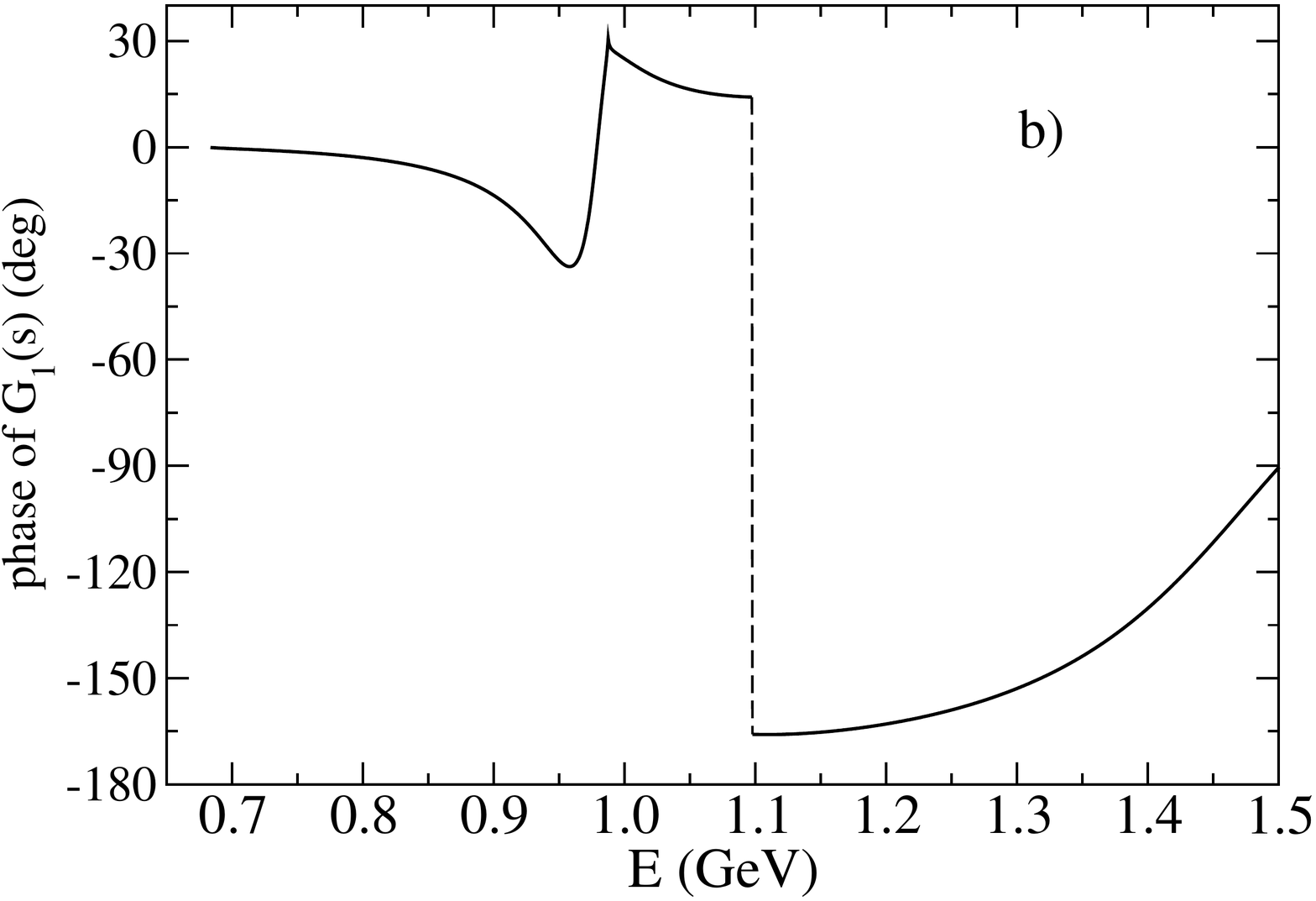}
\vspace{-0.8cm}
\caption{a) modulus of the $G_1(s)$ function normalized to 1 at the $K^+K^-$ threshold; b) phase of the $G_1(s)$ function ($E={\sqrt s}$). At threshold $G_1(s_{th})$ = 304.69~GeV$^{-1}$.}\label{F12}
\end{center} \end{figure}

The importance of the annihilation diagrams in the description of the experimental data should be here underlined. The annihilation terms, proportional to $f_{D^0}$, are present in all the decay amplitudes ${\cal M}_i$ and their magnitudes strongly dominate over other amplitudes which contribute to the total decay amplitude. The annihilation amplitudes depend on the appropriate form factors calculated for the momentum transfer squared $m_{D^0}^2$. These fitted form factors are given in the second, ninth, tenth and thirteenth rows of Table~\ref{parameters}.

The mass and the width of the $\phi(1020)$ resonance seen in Table~\ref{parameters} are in agreement with the corresponding \textit{BABAR} values of ($1019.55\pm0.02$) MeV and ($4.60\pm0.04$) MeV, respectively~\cite{B10}. The obtained width is higher, by about 0.5 MeV,  than the averaged value of ($4.249\pm0.013$) MeV given by the Particle Data Group in Ref.~\cite{PDG2020}. This can be explained by a finite experimental energy resolution.

The branching fraction distributions $\frac{d^2{\rm Br}_i}{ds_+ds_0}$ corresponding to the amplitudes ${\cal M}_i$, $i=1,...,7$ are obtained if in Eq.~(\ref{d2Br}) the amplitude ${\cal M}$ is replaced by ${\cal M}_i$. One can also define the off-diagonal elements $\frac{d^2{\rm Br}_{ij}}{ds_+ds_0}$, $i\neq j$,
\be \label{Brij}
\frac{d^2{\rm Br}_{ij}}{ds_+ds_0}=\frac{Re[{\cal M}_i^*{\cal M}_j]}{32(2\pi)^3m_{D^0}^3 \Gamma_{D^0}}.
\ee
If we integrate over $s_+$ and $s_0$ the differential branching fractions $\frac{d^2{\rm Br}}{ds_+ds_0}$, $\frac{d^2{\rm Br}_i}{ds_+ds_0}$ and $\frac{d^2{\rm Br}_{ij}}{ds_+ds_0}$ then we get the
corresponding branching fractions ${\rm Br}$, ${\rm Br}_i$, $i=1$~to~7 or the off-diagonal elements ${\rm Br}_{ij}$, where $i\neq j$. The matrix ${\rm Br}_{ij}$ is symmetric: ${\rm Br}_{ij}= {\rm Br}_{ji}$.
 
\begin{table}[h]
\caption{Branching fractions (Br) for different quasi-two-body channels in the best 
fit to the \textit{BABAR} data~\cite{B10}.}
\label{TabBr}
\begin{center}
\begin{tabular}{clc}
\hline
\hline
Amplitude & \hspace{0.3cm}channel & \multicolumn{1}{c}{Br$_i$ ($\%$)}\\
\hline 
${\cal{M}}_1$ &  $[K^+\, K^-]_S \,K^0_S$         & $60.9^{+24.4}_{-10.6}$\\
${\cal{M}}_2$ &  $[K^+\, K^-]_P \,K^0_S$         & $45.5\pm{0.7}$\\
${\cal{M}}_3$ &  $[K^0_S \,K^+]_S \,K^-$         & $20.7^{+9.4}_{-6.0}$ \\
${\cal{M}}_4$ &  $[K^0_S \,K^+]_P \,K^-$         & $21.5^{+3.1}_{-2.8}$ \\
${\cal{M}}_5$ &  $[K^0_S \,K^-]_S \,K^+$         & \hspace{0.2cm} $0.76^{+0.18}_{-0.15}$\\
${\cal{M}}_6$ &  $[K^0_S \,K^-]_P \,K^+$         & \hspace{0.2cm} $0.08\pm 0.01$\\
${\cal{M}}_7$ &  $[K^+\, K^-]_D \,K^0_S$         & \hspace{0.2cm} $0.05\pm 0.02$\\
\hline
$\sum_{i=1,7}$ Br$_i$           &                              &  \hspace{-0.2cm} $149.5^{+26.9}_{-12.3}$\\
\hline
\hline
\end{tabular}
\end{center}
\end{table}

In Table~\ref{TabBr} we give uncertainties of the branching fractions. 
They have been obtained  by choosing 10 000 different combinations of the 19 model parameters. 
The parameters values have been generated from the Gaussian distributions taking into account the parameter 
uncertainties written in Table~\ref{parameters} and some correlations between the parameters in the amplitudes ${\cal{M}}_1$ and ${\cal{M}}_3$.
Then the branching fraction uncertainties have been obtained from the distributions of the 10 000 values of each branching fraction and of their sum. 

Let us notice the particularly large uncertainties of the branching fraction ${\rm Br}_1 = 60.9^{+24.4}_{-10.6} \%$.
This is due to the fact that the amplitude ${\cal{M}}_1$ consists of three components and contains 9 free parameters. 
}

As seen in Table~\ref{TabBr} the largest contribution (near 61 $\%$) to the summed branching fraction ${\rm Br}~=~\sum_{i=1}^7 {\rm Br}_{i}$ comes from the first amplitude ${\cal M}_1$. It corresponds to the quasi-two body channel consisting of $K^0_S$ and the $K^+K^-$ pair in the $S$-wave.
The second contribution (near 46 $\%$) to the integrated branching fraction ${\rm Br}$ is due to the 
amplitude ${\cal M}_2$. In this case the $K^+K^-$ pair is in the $P$-wave and its major part is
related to the $\phi(1020)$ resonance. This resonance largely dominates in the region V of the Dalitz diagram.

There are two almost equal contributions of about 21$\%$ from the channels 
$[K^0_S \,K^+]_S \,K^-$ and $[K^0_S \,K^+]_P \,K^-$  (amplitudes ${\cal M}_3$ and ${\cal M}_4$, respectively).
The ${\cal M}_3$ amplitude can be related to a presence of the two isovector-scalar resonances $a_0(980)$ and $a_0(1450)$. 
As seen in Table~\ref{Table_a0} the mass of the resonance $a_0(980)$ equal to $979^{+3}_{-2}$ on sheet $-+$ is lower than the 
$K^0_S K^+$ threshold mass of about 991.3 MeV. However, due its finite width of $25^{+8}_{-6}$~MeV, this resonance, together with the second $a_0(980)$ resonance on sheet $--$ at ($959 - i~ 34$) MeV, can strongly influence the near threshold $s_+$ range of the Dalitz plot density distribution.
 On the other hand, the mass of the $a_0(1450)$ resonance
lies above the upper range of the $K^0_S K^+$ effective mass close to 1371 MeV. However, the $a_0(1450)$ resonances are wide and they can also affect the distribution of the $D^0 \to K^+ K^- K^0_S$ events on the Dalitz plot. 

The contribution of the quasi-two body channel $[K^0_S \,K^+]_P \,K^-$ 
 is related to nonzero couplings of the $P$-wave resonances $\rho(770)^+$, $\rho(1450)^+$ and $\rho(1700)^+$ to $K^0_S K^+$. Although the $\rho(770)^+$ mass lies below the $K^0_S \,K^+$ threshold its width is
 sufficiently large to influence the differential density distribution of the Dalitz plot for $s_+$ values above the threshold. The $\rho(1450)^+$ width is even larger than that of $\rho(770)^+$, so the
 whole $s_+$ range on the Dalitz plot is sensitive to the strength of its coupling to $K^0_S \,K^+$. The above three $\rho$ resonances, being wide, cannot create a clear structure or a well distinguished band on the Dalitz plot. This could be a reason why they have not been included in the isobar model analyses~\cite{B10} and ~\cite{B5}.
 
These results can be compared to the results of the experimental analysis which finds a summed branching fraction of 163.4~\%, mainly with 71.1~\% coming from the $a_0(980)^0$ and $a_0(1450)^0$, 44.1~\% from the $\phi(1020)$ resonance and 45.1~\% from the $a_0(980)^+$ and $a_0(1450)^+$.

\begin{table*}[h]
\caption{Matrix of the branching fractions components Br$_{ij}$ (Eq. (\ref{Brij})) for the best fit to the \textit{BABAR} data~\cite{B10}.
All numbers are in per cent. }
\label{tabBrij}
\begin{center}
\begin{tabular}{c|ccccccc}
\hline
\hline
  & ${\cal{M}}_1$ & ${\cal{M}}_2$ & ${\cal{M}}_3$ & ${\cal{M}}_4$ & ${\cal{M}}_5$ & ${\cal{M}}_6$ &  ${\cal{M}}_7$ \\ 
\hline
${\cal{M}}_1$ & 60.92  & 0.00 & 2.99 & -20.76 & -2.49 & -0.69 & 0.00\\
${\cal{M}}_2$ & & 45.52  & -3.37 & -1.29 & -0.66 & -0.06 & 0.00 \\
${\cal{M}}_3$ & & & 20.73  & 0.00 & -0.21 & 0.52 & 0.13\\
${\cal{M}}_4$ & & & & 21.47  & 0.61 & 0.58 & -0.06 \\
${\cal{M}}_5$ & & & & & 0.76  & 0.00 & 0.01\\
${\cal{M}}_6$ & & & & & & 0.08  & 0.00 \\
${\cal{M}}_7$ & & & & & & & 0.05  \\
\hline
\hline
\end{tabular}
\end{center}
\end{table*}

In Table~\ref{tabBrij} the diagonal branching fraction terms already shown in Table~\ref{TabBr} are given together with the off-diagonal terms ${\rm Br}_{ij}$.
The sum of the off-diagonal terms equals to -$49.53~\%$. One should remark here that
some off-diagonal terms are exactly equal to zero. This is due to the orthogonality of certain wave functions. For example, the interference term ${\rm Br}_{12}$ vanishes since the wave functions of the $S$- and $P$-states of the $K^+K^-$ system are
orthogonal. Due to the matrix symmetry the elements of the branching fractions below the diagonal are not written.

The  amplitude ${\cal M}_1$ is a sum of three terms [see Eqs.~(\ref{M1T})-(\ref{M13})].
The first isoscalar term is proportional to the conjugated kaon nonstrange scalar form factor $\Gamma_2^{*n}(s_0)$ and the second one to the conjugated kaon strange scalar form factor $\Gamma_2^{*s}(s_0)$. 
The third term is proportional to the function $G_1(s_0)$ describing the transition from the $u \bar{u}$ pair of quarks into the $K^+K^-$ pair of mesons in the isospin one and spin zero state. 

\begin{table*}[h]
\caption{Matrix of the branching fraction components Br of the ${\cal{M}}_1$  amplitude calculated for the best fit to the \textit{BABAR} data~\cite{B10}.
All numbers are in per cent.}
\begin{center}
\begin{tabular}{c|ccc}
\hline
\hline
  &  ${\cal{M}}^{n,I=0}_1$ & ${\cal{M}}^{s,I=0}_1$ & ${\cal{M}}^{I=1}_1$ \\ 
\hline
${\cal{M}}^{n,I=0}_1$         & $\,\,\,$1.19  & -7.32  & -0.36 \\
${\cal{M}}^{s,I=0}_1$         & & 59.82  & 5.40 \\
${\cal{M}}^{I=1}_1$           & & & $\,\,\,$4.48  \\
\hline
\hline
\label{TabBr123}
\end{tabular} \end{center}
\end{table*}

In Table~\ref{TabBr123} the diagonal and the off-diagonal components of the branching fraction related to the ${\cal M}_1$ amplitude are given. They are defined in a similar way as the ${\rm Br}_{ij}$ components in Eq.~(\ref{Brij}).
From this Table we see that the major contribution close to 60\% is related to the strange scalar isospin zero component of the annihilation ($W$-exchange) amplitude ${\cal M}_1^{s, I=0}$. Here the isoscalar-scalar resonances like $f_0(980)$ are formed from the strange-antistrange pair of quarks. Following the result of the fit shown in the above Table the formation of the isoscalar-scalar resonances from the $u \bar{u}$ quarks is suppressed (the diagonal ${\cal M}_1^{n, I=0}$ branching fraction 
is equal only to 1.19~\%). Also the branching fraction equal to 4.48\%, corresponding to the isovector-scalar amplitude ${\cal M}_1^{I=1}$,  is much smaller than that related to the ${\cal M}_1^{s, I=0}$ amplitude. The sum of all the off-diagonal components equals to -4.57~\%.
A comparison of the results for this best fit model with those for the  alternative MO$_{P1(P2)}$ ones can be found in the Appendix~\ref{fitswMOBM}.

\begin{figure}[h]  \begin{center}
\includegraphics[scale = 0.3]{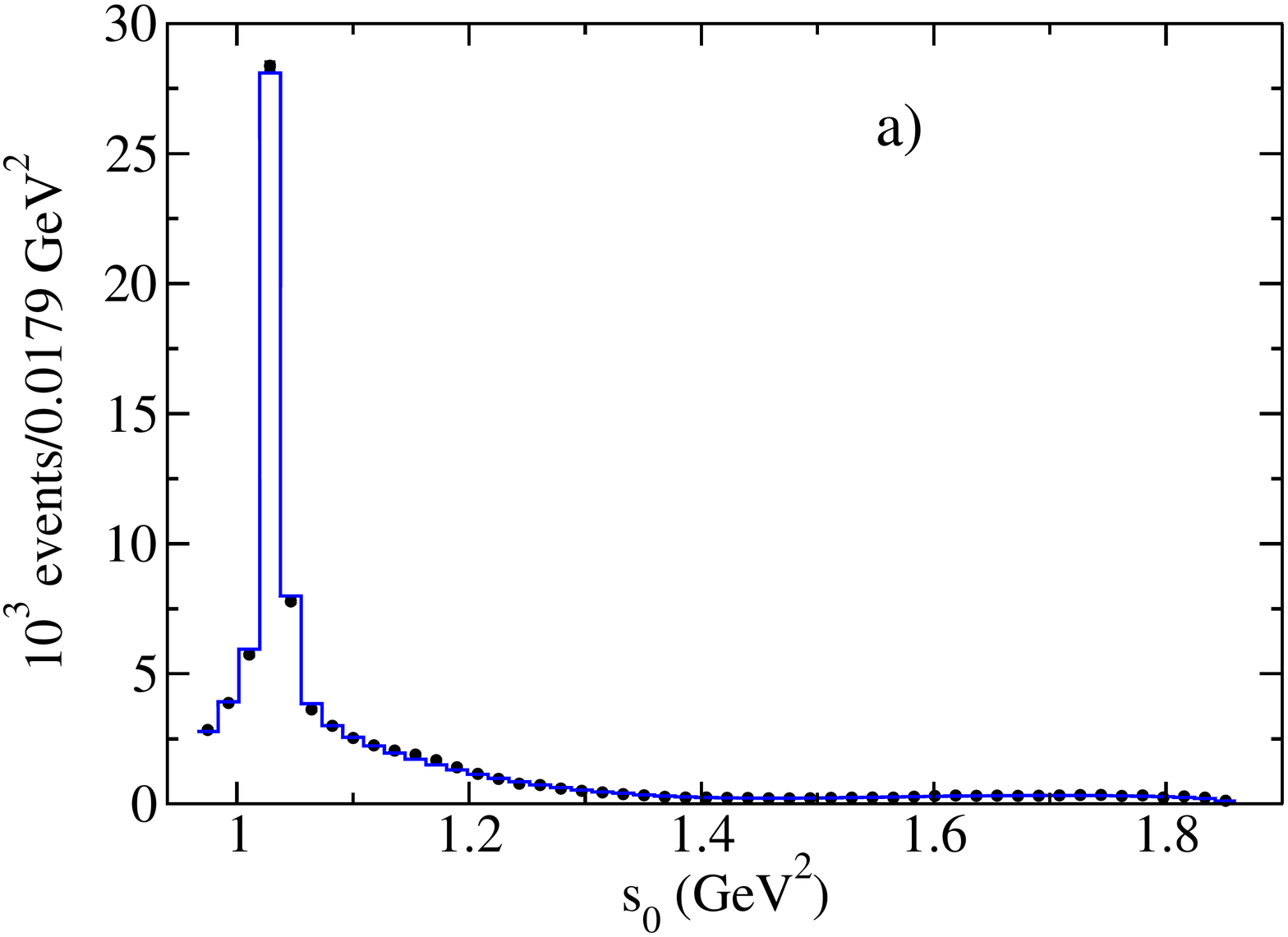}
\hspace{-1.3cm}
\includegraphics[scale = 0.3]{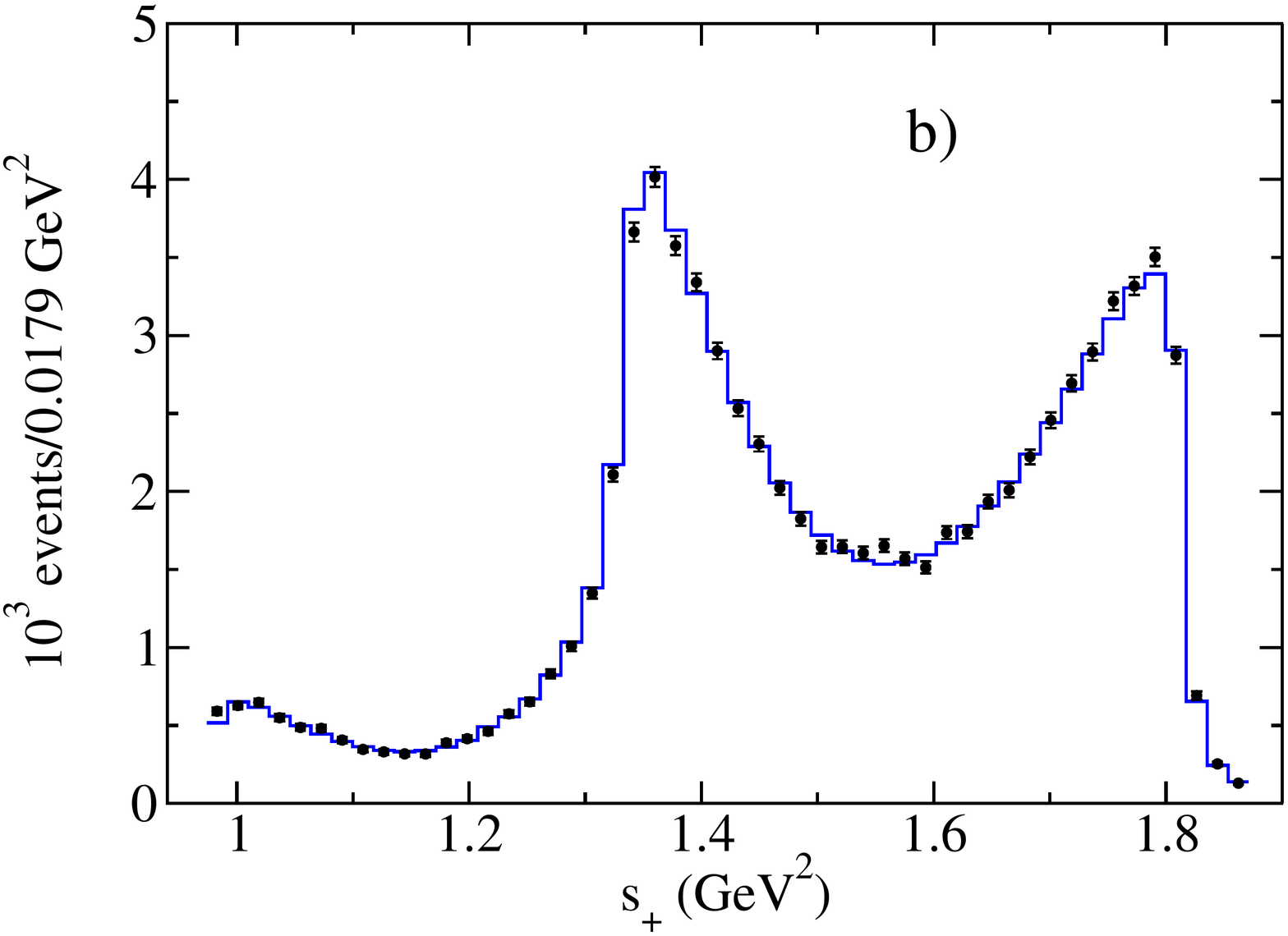}
\includegraphics[scale = 0.3]{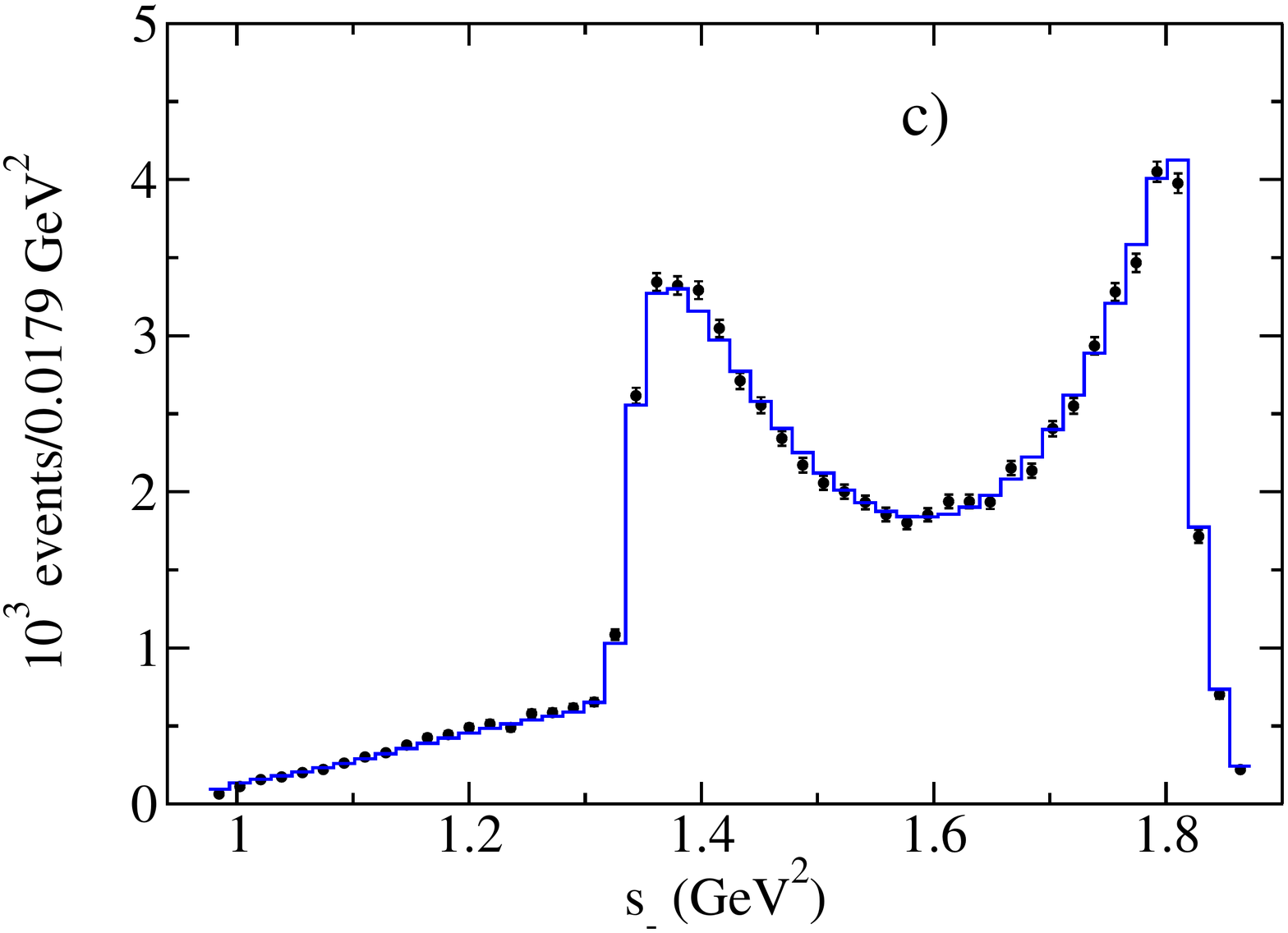}
\caption{ Dalitz plot projections for $D^0~\to K^0_S K^+ K^-$. The points indicate the \textit{BABAR} data~\cite{B10} together with their statistical errors. Histograms  represent the best fit theoretical distributions.}
\label{Projections}
\end{center} \end{figure}
 
Dalitz plot projections or one-dimensional effective mass squared distributions
of events are calculated by a proper integration of the two-dimensional density distributions. They are shown in Fig.~\ref{Projections}.
The errors of the experimental signal weighted event number distributions are the statistical ones. The histograms correspond to the theoretical
distributions normalized to the same total number of events.

The distribution in Fig.~\ref{Projections}(a) is strongly dominated by the maximum corresponding to the $\phi(1020)$ resonance decaying to the $K^+K^-$ pair.
This decay is in the $P$-wave and leads to  a characteristic two-maximum shape of the Dalitz plot distribution as a function of $s_+$ - the square of the $K^+K^-$ effective mass. Since the branching fraction for the channel $[K^+\, K^-]_P \,K^0_S$ is large
(45.5~\%) the two Dalitz projections in Figs.~\ref{Projections}(b) and(c) have a two-maximum character. There are also two other important  contributions
related to the amplitudes ${\cal M}_1$ and  ${\cal M}_3$. However, they do not produce
any pronounced structures on the Dalitz plot since both are due to the $S$-wave in the 
$K^+K^-$ or in the $K^0_S \,K^+$ configuration. 

Finally let us discuss the low effective mass parts of the $K^+ K^-$ and 
$K^0_S \,K^+$ distributions ($m_{KK} < 1.06$ GeV).
Since the differential branching fraction [Eq.~(\ref{d2Br})] is proportional to the Dalitz plot density distribution of events $\frac{d^2N}{ds_+ds_0}$, one can calculate the theoretical one-dimensional distributions of the event numbers using Eqs.~(\ref{e1})-(\ref{e3}) of Sec.~\ref{comp}. 
They are displayed in Fig.~\ref{Projections} as solid histograms. One can see that the \textit{BABAR} data agree well with the corresponding lines. 
This agreement enforces the statement about the significant difference between the 
$K^+ K^-$ and $K^0_S \,K^+$ effective mass distributions which is due to the dominant $f_0(980)$ resonance contribution to the $K\bar{K}$ final state interaction amplitude.

\section{\bf CONCLUSIONS} \label{conclusions}

A theoretical model of the $D^0 \to K_S^0 K^+ K^-$ decay amplitude has been constructed  within a quasi-two-body QCD factorization approach  introducing scalar kaon form factors to describe the $S$-wave kaon-kaon final  state interaction.
In doing so, the contribution of isoscalar-scalar $f_0$ resonance family, viz. $f_0(980), f_0(1370)$ and  that of the isovector-scalar $a_0$ one, viz. $a_0^0(980), a_0^\pm(980), a_0^0(1450), a_0^\pm(1450)$ are taken into account. 
The isospin zero and one kaon-kaon  $S$-wave interactions have been treated in a unitary way using either  coupled channel relativistic equations, or a dispersion relation framework. The  $P$- and $D$- waves of the final state kaon-kaon interactions have also been taken into account.

Independently of any model assumptions, we have shown that the $K^+K^-$ and $\bar K^0 K^+$ $S$-wave effective mass squared distributions, corrected for phase space, are significantly different.
This means that, in the analyses of the $D^0 \to \bar K^0 K^+ K^-$ data,  one cannot neglect the contribution of the $f_0(980)$ resonance and retain only the $a_0(980)$ contribution. 

In Appendix~\ref{UpdatedTpipi}, we have updated the meson-meson $S$-wave isospin zero scattering amplitudes.
These include the three coupled, $\pi \pi$, $\bar K K$ and  an effective $2\pi \ 2\pi$ channels. 
Using the above amplitudes the new kaon nonstrange and strange form factors $\Gamma_2^n(s_0)$ and $\Gamma_2^s(s_0)$ have been calculated following Ref.~\cite{KKK} and introduced in the data analysis.
 As seen  Fig.~\ref{ModGamma2}, these form factors are quite similar to those derived using the Muskhelishvili-Omn\`es dispersion relation approach~\cite{Moussallam_2000, Moussallam_2019}.

In the factorization framework, for the $D^0 \to K^0_S K^+ K^-$ process one has to evaluate the matrix elements of the $D^0$ transitions to two-kaons or the transitions between one kaon and two kaons.
The knowledge of these transitions requires that of the three-body strong interaction between the  $D^0, K^0_S$ and  $K^\pm$ mesons and that between the $K^0_S, K^+$ and $K^-$ mesons. 
Here, to describe these transitions with the two final kaons in $S$-wave state, we had to go beyond the simple multiplication of the scalar kaon form factors by a complex constant. 
 And to obtain good fits we have multiplied  the  isoscalar-scalar kaon form factor by a one free parameter energy-dependent  function and  introduced into the isovector-scalar function an energy-dependent phenomenological polynomial involving three free parameters.

The undetermined free parameters of our seven $D^0 \to K^0_S K^+ K^-$ amplitudes are then related to the strength of the isoscalar-scalar kaon form factor, to the function proportional to the isovector-scalar kaon form factor and to the unknown meson to meson transition form factors. They are obtained through a $\chi^2$ minimization to the \textit{BABAR} Dalitz plot distribution~\cite{B10}. It should be pointed out that the low density of events in the central region of this Dalitz plot distributions (see Fig.~\ref{Figure:Cells})  is difficult to reproduce. 
Using unitary relativistic equations to built the isoscalar-scalar form factor and a function proportional to the isovector-scalar one, we obtain a best fit (denoted R$_{\rm 
BF}$), with a $\chi^2/ndf$ of 1.25 with 19 free parameters to be compared to that of 1.28 for Ref.~\cite{B10} which uses 17 free parameters.

In Appendix~\ref{fitswMOBM}, we have studied two alternative fits with scalar-kaon form factors derived in the Muskhelishvili-Omn\`es dispersion relation framework. All other amplitudes are parametrized as in the best fit model.
If the scalar form factors are multiplied by energy dependent phenomenological functions, we obtain two good fits, one, denoted MO$_{P1}$ with a $\chi^2/ndf $ of 1.32 and 16 free parameters and another  one, MO$_{P2}$, with a $\chi^2/ndf $ of 1.31 and 16 free parameters. 
 
Our fits indicate the dominance of  the annihilation amplitudes and for the best fit a large dominance of the 
$[K^+K^-]_S$ isospin 0 $S$-wave contribution and a sizable branching fraction to the 
$[K^0_S K^+]_P K^-$ final state with the $[K^0_S K^+]$  pair coupled to  $\rho(770)^+$, $\rho(1450)^+$ and $\rho(1700)^+$.
 The alternative fits show important contributions from both the $f_0$ and $a_0^0$ mesons and a weaker $\rho^+$ mesons role. For all our models, the one-dimensional distributions agree well with that of the \textit{BABAR} data.

One can estimate the strength of the contributions of the different amplitudes by looking at their branching ratio
compared to the sum of their branching ratios.
As can be seen in Table~\ref{TabBr45Req} for the best fit model this sum\footnote{The numbers in  brackets are the corresponding values of the MO$_{P1}$ and MO$_{P2}$ fits, respectively} is 149.5~[126.3,~164.1]~\% (163.4\% in Ref.~\cite{B10}), which points to sizable interference contributions. 
The  kaon-kaon S-wave interactions, related to the $f_0$ and  $a_0^0$ resonances, gives a large branching of $\sim$61~[45,~63]~\%  with a large value (for BF,  MO$_{P1}$  and MO$_{P2}$, see Table~\ref{TabBr45Reqa}) of
$\sim$60~[23,~46]~\%, for the amplitude proportional to the  strange isoscalar-scalar form factor ($f_0$ contributions) and  smaller branching $\sim$5~[16,16]~\% for the amplitude proportional to the  isovector-scalar form factor ($a_0^0$ contributions).
Corresponding figures in the isobar \textit{BABAR} analysis \cite{B10} are $\sim$71~\%, dominated by the $a_0(980)^0$ and $a_0(1450)^0$ with no $f_0(980)$ and a $f_0(1370)$ $\sim$~
 2\% .

The branching fraction of the isospin~0 $P$-wave  $\sim$46~[45,~45]~\%, dominated by the $\phi(1020)$ resonance, is similar to that found, $\sim$44~\%, in Ref.~\cite{B10}.
The branching of the isovector amplitude associated to the $a_0^+$ resonances is $\sim$21~[26, 40]~\% to be compared to $\sim$~45~\% in Ref.~\cite{B10}.
The branching fraction of the amplitude related to the  $[\rho(770)^+ + \rho(1450)^+ + \rho(1700)^+] K_S^0$ final state, not introduced in Ref.~\cite{B10}, has a value of $\sim$22~[8,~13]~\%.
One could say that, this  contribution with no bumps in the Dalitz plot distribution, is in Ref.\cite{B10} taken into account by a part of that of the  $a_0^+$.

 The charmless hadronic $B^0 \to K^+ K^- K_S^0$  studied by the Belle~\cite{Be2010} and \textit{BABAR}~\cite{BA2012} Collaborations 
   has the same meson final states  as the $D^0$ decay we have been studied here.
 A quasi-two-body  QCD factorization analysis of this $B^0$ decay process should allow, to constrain, not only the weak interaction observables but also the scalar kaon form factors, the transitions between one kaon and two kaons and to learn about the $B^0$ transition to two kaons.

\section*{Acknowledgements}
\label{ACKN}
  We are grateful to Fran\c{c}ois Le Diberder who, at the early stage of this work, has helped us in getting access to the \textit{BABAR} data. We are deeply indebted to Fernando Martinez-Vidal from the \textit{BABAR} Collaboration who provided us with experimental information for this study. We thank him for many fruitful exchanges. We are also indebted to Bachir Moussallam for very profitable correspondence and for the communication of the results of his calculation of scalar form factors. We also acknowledge helpful discussions with Piotr \.{Z}enczykowski.

This work has been partially supported by a grant from the French-Polish exchange program COPIN/CNRS-IN2P3, collaboration 08-127.\\

\appendix

\section{Updated $\pi \pi$, $\bar K K$ and effective $2\pi \ 2\pi $ $S$-wave amplitudes}
 \label{UpdatedTpipi}
  
   Here we present updated results for the meson-meson $S$-wave isospin zero scattering amplitudes. They include the following three coupled channels: $\pi \pi$, channel~1, $\bar K K$, channel~2 and effective $2\pi \ 2\pi $, channel~3. Our previous fits to the meson-meson scattering data were obtained in the late nineties~\cite{KLL, EPJ}. Since, new precise 
low energy  $\pi \pi$  data have appeared~\cite{NA48}. Moreover, as noticed by Bachir Moussallam~\cite{Moussallam_Nov2018}, we  used  
an assumption valid only below the opening of the third channel, namely the phase of the $\pi \pi~\to~\bar K K$ transition amplitude was set equal to the sum of the elastic $\pi \pi$ and $\bar K K$ phaseshifts. The derivation of the kaon isoscalar-scalar form factors  $\Gamma_2^n(s_0)$ and $\Gamma_2^s(s_0)$, used in the present analysis for $s_{0min}=0.98$~GeV$^2~\le~s_0~\le~s_{0max}=1.87$~GeV$^2$, requires the knowledge of the meson-meson amplitudes at energies above~$s_{0max}$.

Thus, dropping the above mentioned assumption, we have performed a new analysis based on an enlarged set of data.
Using the same three coupled-channel  separable potential model as developed in Refs.~\cite{KLL} and~\cite{EPJ}, we fit the following data: 
    
\noindent
a) for the effective $\pi\pi$ mass  $E$ between 286 and 390 MeV,  the 10 values of the elastic $\pi \pi$ phase shifts from the NA48 data~\cite{NA48},
 
\noindent
b) for $610$~MeV$\le~E~\le 1580$~MeV, the  50 values of the  $\pi \pi$ phase shifts $\delta_{\pi\pi}$ and for $1010$~MeV$\le~E~\le 1580$~MeV the 30 values of the $\pi \pi$ inelasticities $\eta_{\pi\pi}$, both quantities obtained in the experimental analysis of Ref.~\cite{KLRyb}, 

\noindent
c) for $995$~MeV$\le~E~\le 1580$~MeV, the 23 values of the moduli of the transition $\pi\pi \to \overline K K$ amplitude $T_{12}$ extracted from Fig. 27 of Ref.~\cite{Cohen},

\noindent
d) for   the $\delta_{\pi\pi\to K \overline K}$ phases of the $T_{12}$ amplitude, the 21 values extracted in the analysis of Ref.~\cite{Cohen}  for $1016$~MeV$\le~E~\le 1530$~MeV,  
 
 \noindent
e) plus the  6 data points for these phases between 1538 and 1741 MeV  determined in Ref.~\cite{Etkin}. 

The total number of fitted data is then equal to 140.  As in Ref.~\cite{KLL}, the fitting method is based on the $\chi^2$ function being a sum of five components related to the five data sets enumerated above. The resulting $\chi^2$ is equal to 135.04 which, for 14 free model parameters, gives the value $\chi^2/ndf= 1.07$ when divided by $ndf=140-14=126$ degrees of freedom. 

\begin{table}[ht] 
\caption{Model parameters fitted to data.}
\begin{center} {\begin{tabular}{cc} 
\hline \hline
 Parameter&value \rule{0mm}{6mm}\vspace{4pt}\\
\hline
 $\Lambda_{11,1}$&-0.14434 $\cdot 10^{-3}$ $\vphantom{\frac{1}{fg}}$\rule{0mm}{6mm}\\
 $\Lambda_{11,2}$&-0.21102\\
 $\Lambda_{22}$&-0.62730\\
 $\Lambda_{33}$&-0.81318$\cdot 10^{-3}$\\
 $\Lambda_{12,1}$&0.25184$\cdot 10^{-4}$\\
 $\Lambda_{12,2}$&0.033294\\
 $\Lambda_{13,1}$&0.25063$\cdot 10^{-4}$\\
 $\Lambda_{13,2}$&-0.34913\\
 $\Lambda_{23}$&-5.4206\\
 $\beta_{1,1}$&3.0366$\cdot 10^{3}$~~GeV\\
 $\beta_{1,2}$&1.1019~~GeV\\
 $\beta_2$&0.98412~~GeV\\
 $\beta_3$&0.047940~~GeV\\
 $m_3$&0.75200~~GeV\\
\hline \hline
\end{tabular}}
\end{center}\label{fitparamet}
\end{table}

  \begin{figure}[t!]
\centering
\includegraphics[width=0.54\textwidth]{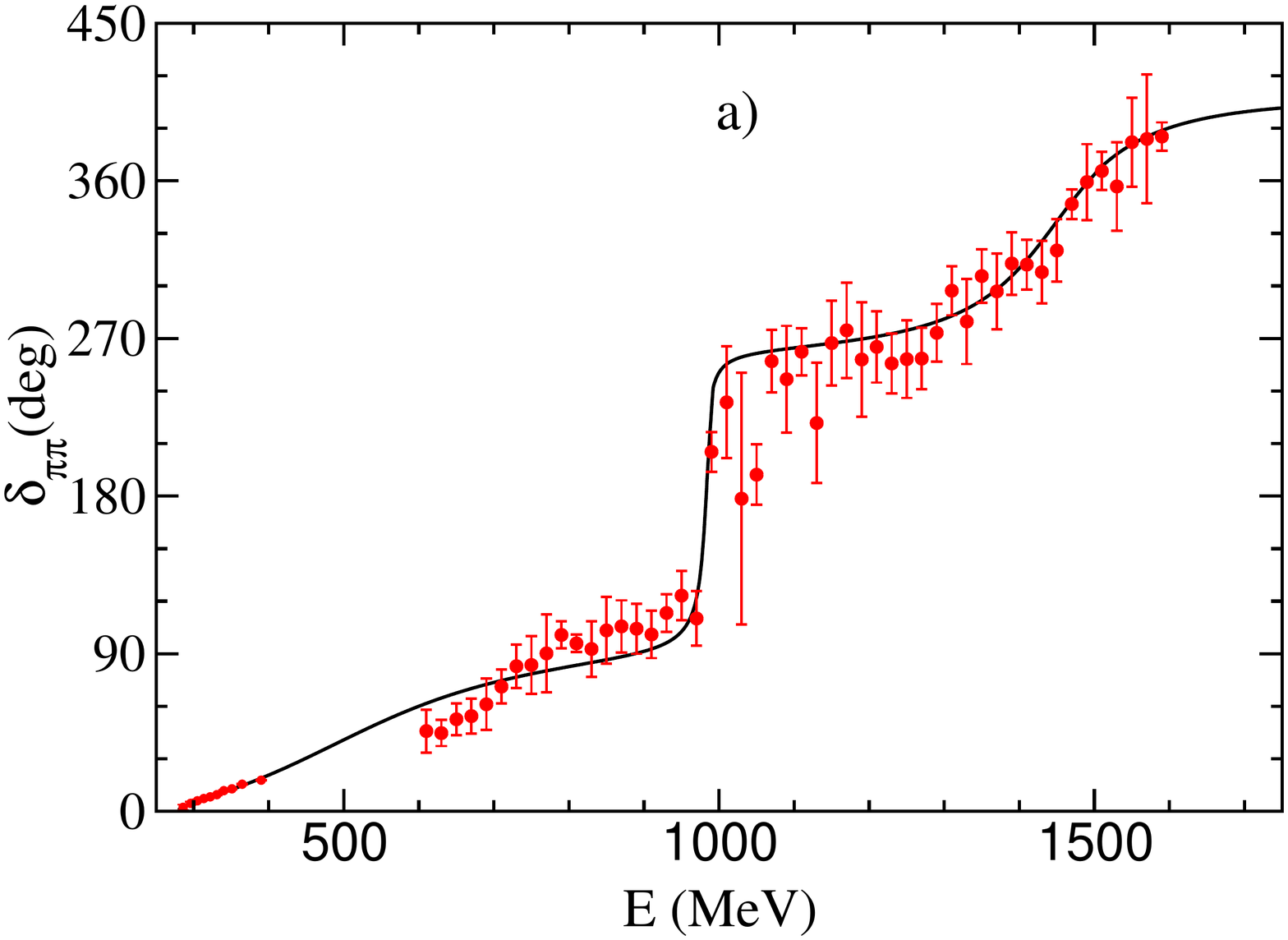}
\hspace{-1.7cm}
\includegraphics[width=0.54\textwidth]{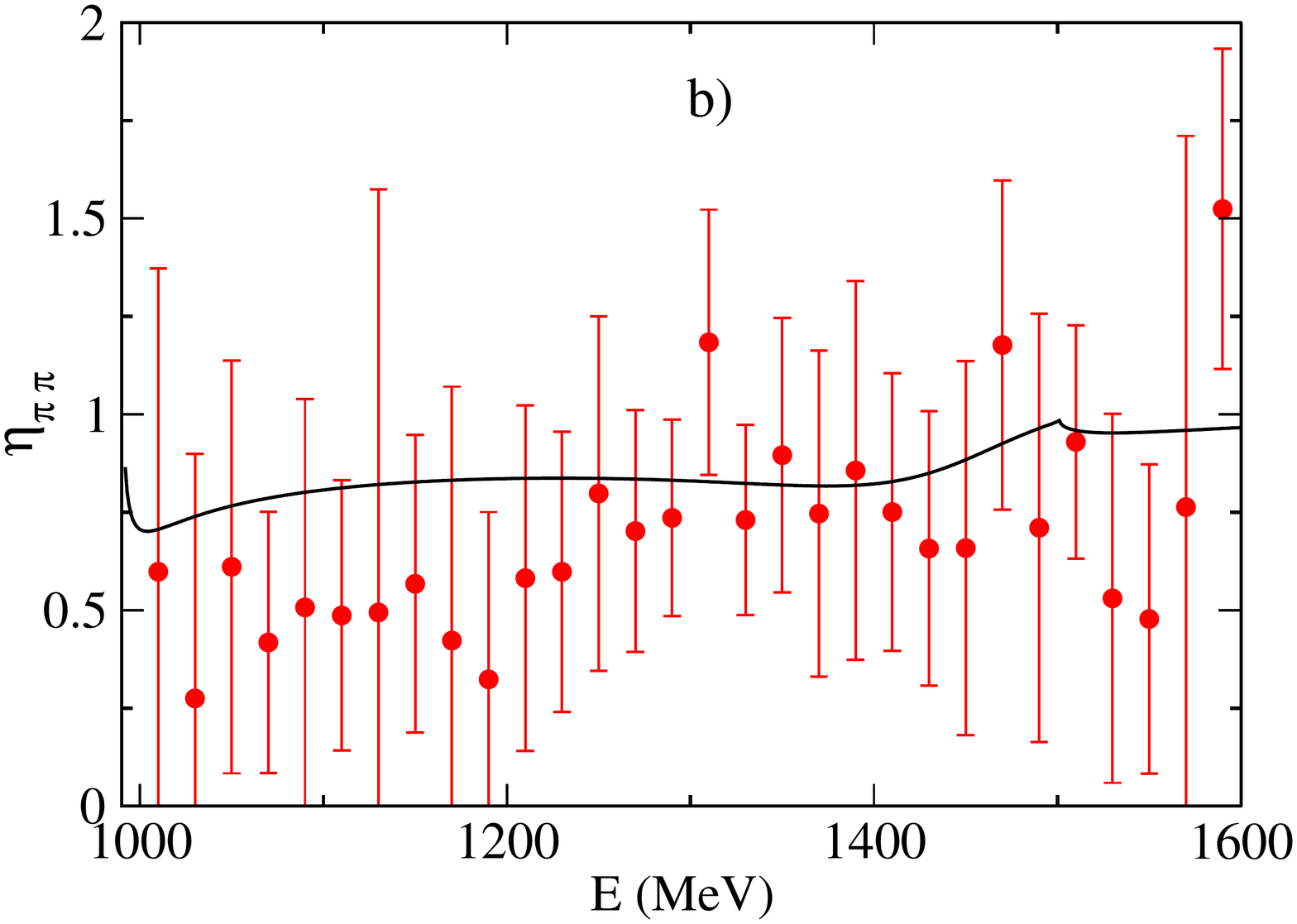}
\vspace{-0.8cm}
\caption{  Comparison to the data of our fit (solid line): a) the $\pi\pi$  elastic phase shifts versus the $\pi\pi$ center-of-mass energy $E$, b) the  $\pi\pi$ inelasticities  and for energies $E$  higher than the $\overline K K$ mass threshold.
The data below $E$~=~400 MeV are taken from Ref.~\cite{NA48} and those above 600 MeV from Ref.~\cite{KLRyb} for the ``down-flat" solution. 
}
\label{Delta} 
\end{figure}

\begin{table}[ht]
\caption{Positions of $S$ matrix poles ($E$ in MeV).}
\begin{center}
{\begin{tabular}{cccc}
\hline \hline
  & & Sign of& \\ 
 Re~$E$&Im~$E$&Im~$k_1$~Im $k_2$~Im $k_3$&~~n$^\circ$ \\
\hline
227&0&$-$~~~~~$-$~~~~~$-$& I\\
230&0&$-$~~~~~$-$~~~~~$+$& II\\
230&0&$+$~~~~~$-$~~~~~$-$& III\\
232&0&$+$~~~~~$-$~~~~~$+$& IV\\
485&-233&$-$~~~~~$+$~~~~~$-$& V\\
485&-233&$-$~~~~~$+$~~~~~$+$& VI\\
506&-262&$-$~~~~~$-$~~~~~$-$& VII\\
507&-265&$-$~~~~~$-$~~~~~$+$& VIII\\
967&-10&$-$~~~~~$+$~~~~~$-$& IX\\
982&-8&$-$~~~~~$+$~~~~~$+$&X\\ 
1442&-100&$-$~~~~~$+$~~~~~$-$&XI\\
1444&-93&$-$~~~~~$+$~~~~~$+$&XII\\
1448&-97&$-$~~~~~$-$~~~~~$+$& XIII\\ 
1465&-98&$-$~~~~~$-$~~~~~$-$& XIV\\ 
1553&-211&$+$~~~~~$-$~~~~~$-$&XV\\
1559&-213&$-$~~~~~$-$~~~~~$-$&XVI\\
1581&-138&$-$~~~~~$-$~~~~~$+$&XVII\\
1584&-134&$+$~~~~~$-$~~~~~$+$&XVIII\\
\hline \hline
\end{tabular}}
\end{center} \label{Poles}
\end{table}

\begin{table}[ht]
\caption{Coupling constants $g_i$ in GeV for a few representative $S$-matrix poles ($E$ in MeV).}
\begin{center}
{\begin{tabular}{cccccc} 
\hline \hline
 Re~$E$&Im~$E$&$|g_1|^2/4 \pi$&$|g_2|^2/4 \pi$&$|g_3|^2/4 \pi|$&~~n$^\circ$ \\
\hline
506&-262&0.86&0.03&0.00& VII\\
967&-10&0.08&1.68&0.05& IX\\
982&-8&0.07&1.27&0.04& X\\
1448&-97&1.02&0.08&0.16& XIII\\ 
1559&-213&0.07&2.24&0.77&XVI\\ 
\hline \hline
\end{tabular}}
\end{center} \label{Couplings}
\end{table}

\begin{figure}[t!]
\centering
\includegraphics[width=0.53\textwidth]{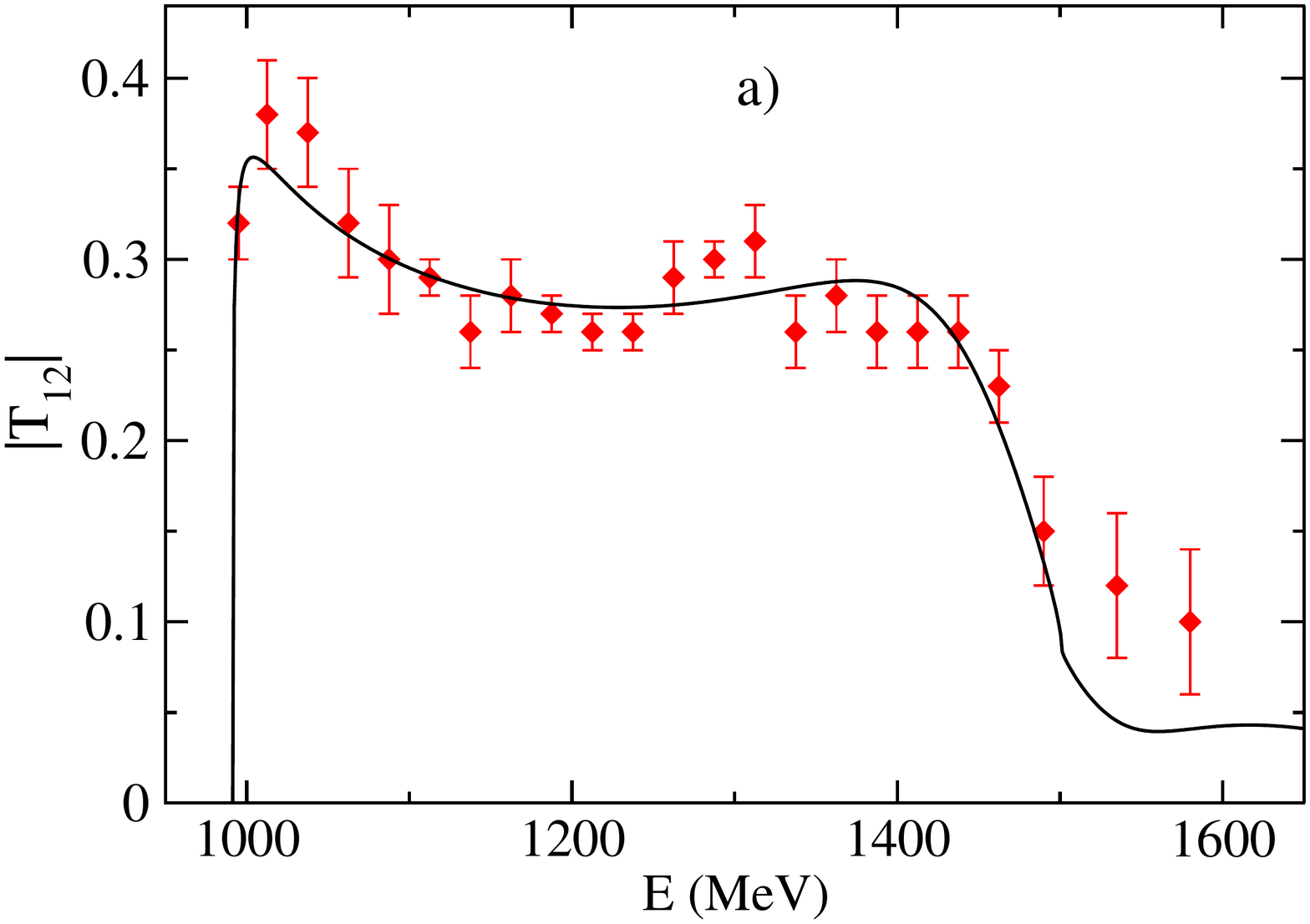}
\hspace{-1.3cm}
\includegraphics[width=0.53\textwidth]{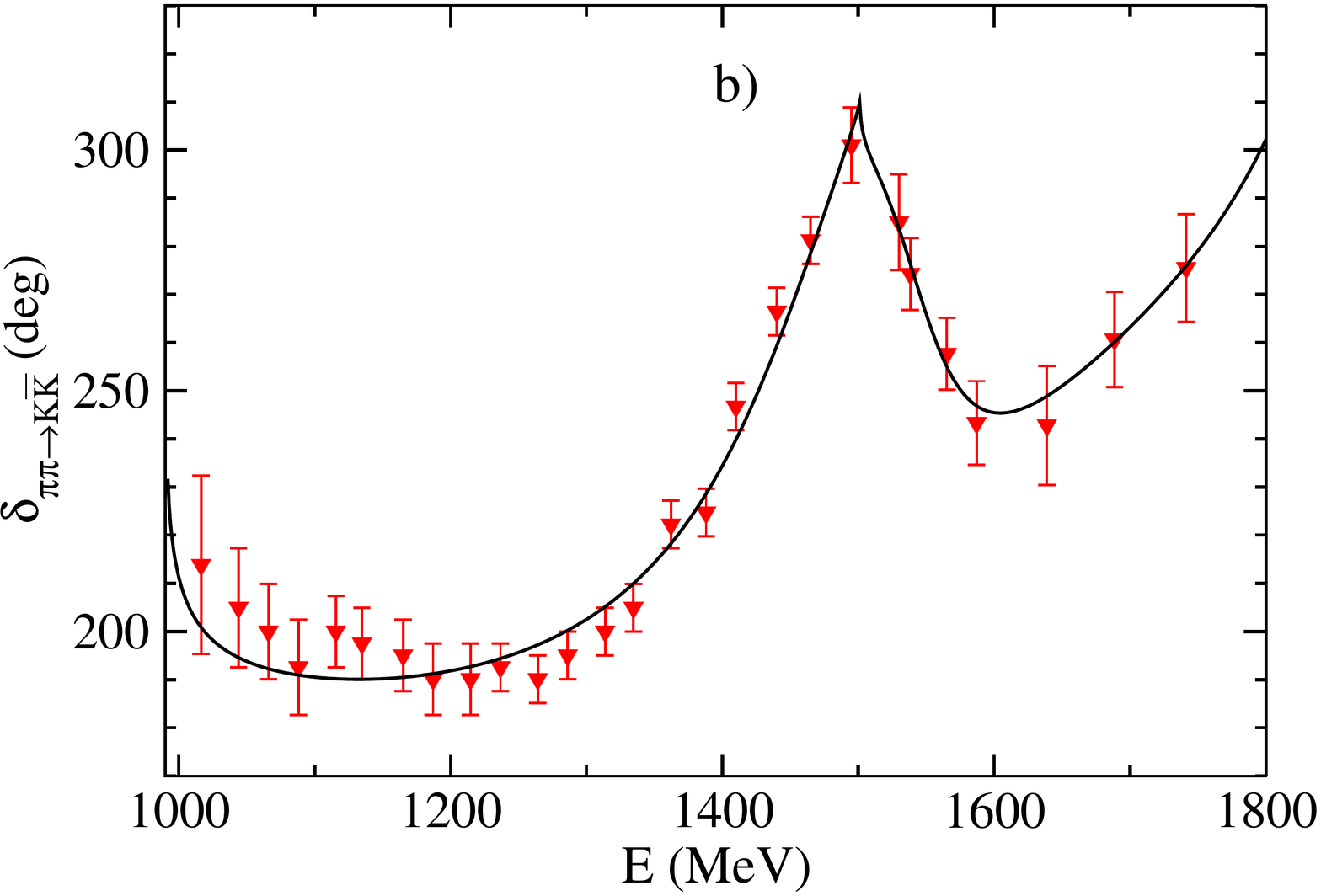}
\vspace{-0.8cm}
\caption{ Comparison to the data~\cite{Cohen} of our fit (solid line): a) Modulus of the $\pi\pi \to \bar K K$ transition amplitude $T_{12}(E)$ normalized as in Ref.~\cite{Cohen}, b) $T_{12}(E)$ phase. 
\label{T12}
}
\end{figure}

The quality of our fit for the $\pi\pi$ phase shifts and inelasticities is illustrated in Fig.~\ref{Delta}. 
As seen in Fig.~\ref{T12} a good fit is achieved for the moduli and the phases of the $T_{12}$  amplitude. All 
experimental data sets are well reproduced by our phenomenological  model. The resulting separable interaction parameters
 are listed in Table~\ref{fitparamet}, their notation being identical to that of Ref.~\cite{KLL}.
 
Positions of the $S$-matrix poles in the complex energy $E$ plane are given in Table~\ref{Poles}. The signs of the imaginary parts of the channel complex momenta $k_i$, i=1,2,3 are indicated in order to mark the corresponding pole position on different Riemann sheets. The total width $\Gamma$ of a given pole equals to twice $\lvert$Im$E\rvert$.

As in the case of solution A (see Table 3 of Ref.~\cite{EPJ}) one finds 18 poles. The first four (I to IV), lying on the real axis below  the $\pi \pi$ threshold, are related to the $S$-matrix poles in the absence of interchannel couplings. The next four (V to VIII), located on different sheets, correspond to the wide resonance $f_0(500)$. There are two close poles (IX and X) related to the narrow resonance $f_0(980)$ and four poles (XI to XIV) attributed to the wider resonance $f_0(1400)$.  The four poles (XV to XVIII) located between 1553 and 1584 MeV are responsible for the structure in the phase of the transition $\pi \pi \to K \overline K$ amplitude as can be seen in the right panel of Fig.~\ref{T12}  and in Fig.~6 of Ref.~\cite{Etkin}; there is a maximum near 1500 MeV, close to the opening of the third channel,  followed by a dip at about 1600 MeV. These latter poles could be related to the $f_0(1370)$ and $f_0(1500)$ resonances.

In Table~\ref{Couplings} we present values of the moduli of the channel coupling constants calculated for five typical $S$ matrix poles (for their definitions see Eq.~(34) of Ref.~\cite{EPJ}). The $f_0(500)$ poles like that with n$^\circ$ VII are mainly coupled to the $\pi\pi$ channel ($i=1$).  Also the four poles close to Re $E$ = 1450 MeV have a strong coupling only to the $\pi\pi$ channel. The $f_0(980)$ poles n$^{\circ s}$ IX and X are preferentially coupled to the $\bar K K$ channel ($i=2$) like the four other poles n$^{\circ s}$ XV to XVIII. This last group of poles has also a substantial coupling to the $(2\pi)(2\pi)$ channel ($i=3$) in addition to the strong coupling to the $\bar K K$ (i=3) one. All these poles lie above the opening of the third channel taking place at $2 m_3=1504$~MeV.

\section{Fits using kaon scalar form factors derived from dispersion relation approach}
 \label{fitswMOBM}

\begin{figure}[h]  \begin{center}
\includegraphics[scale=0.357]{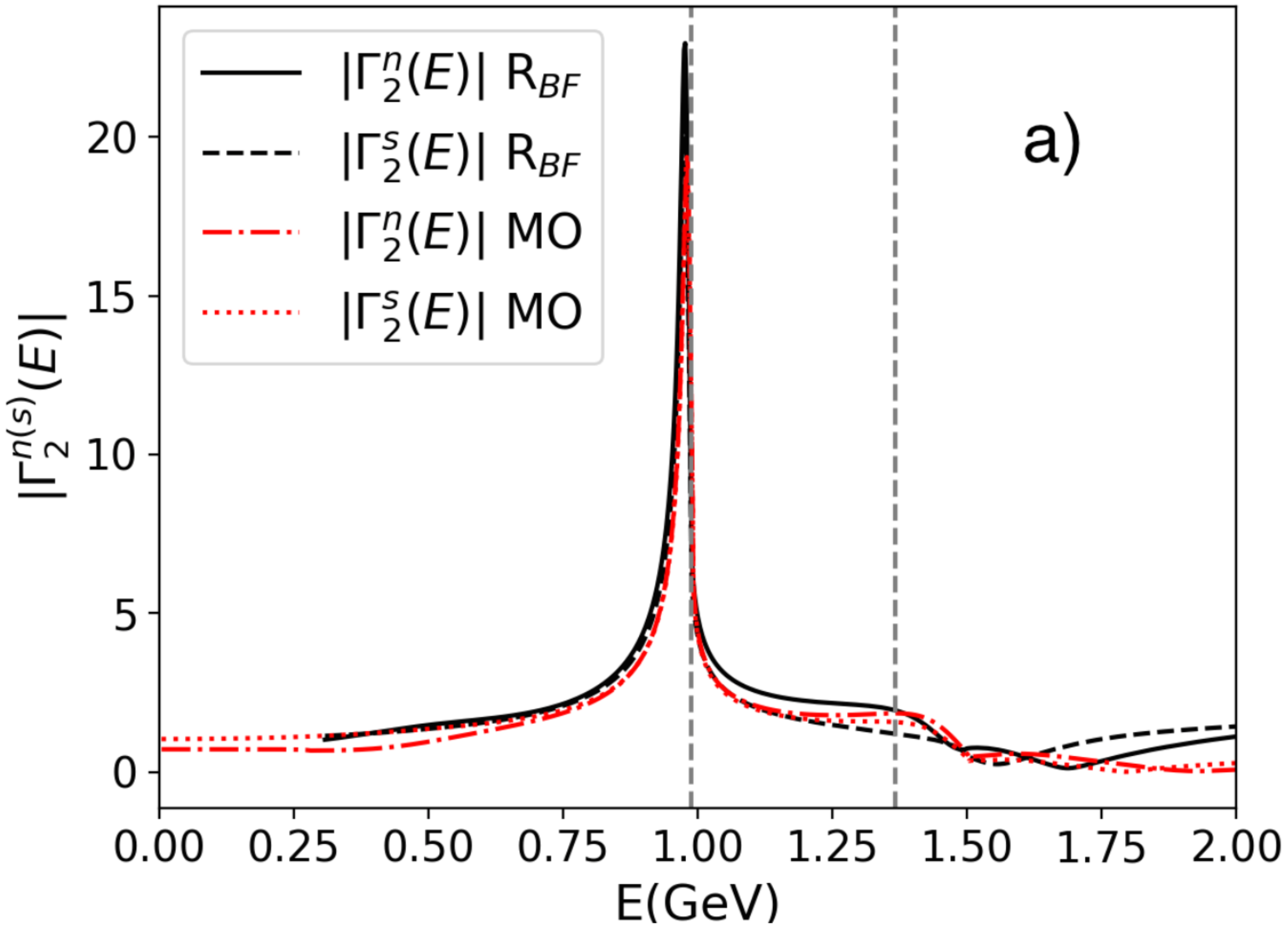}
\includegraphics[scale=0.357]{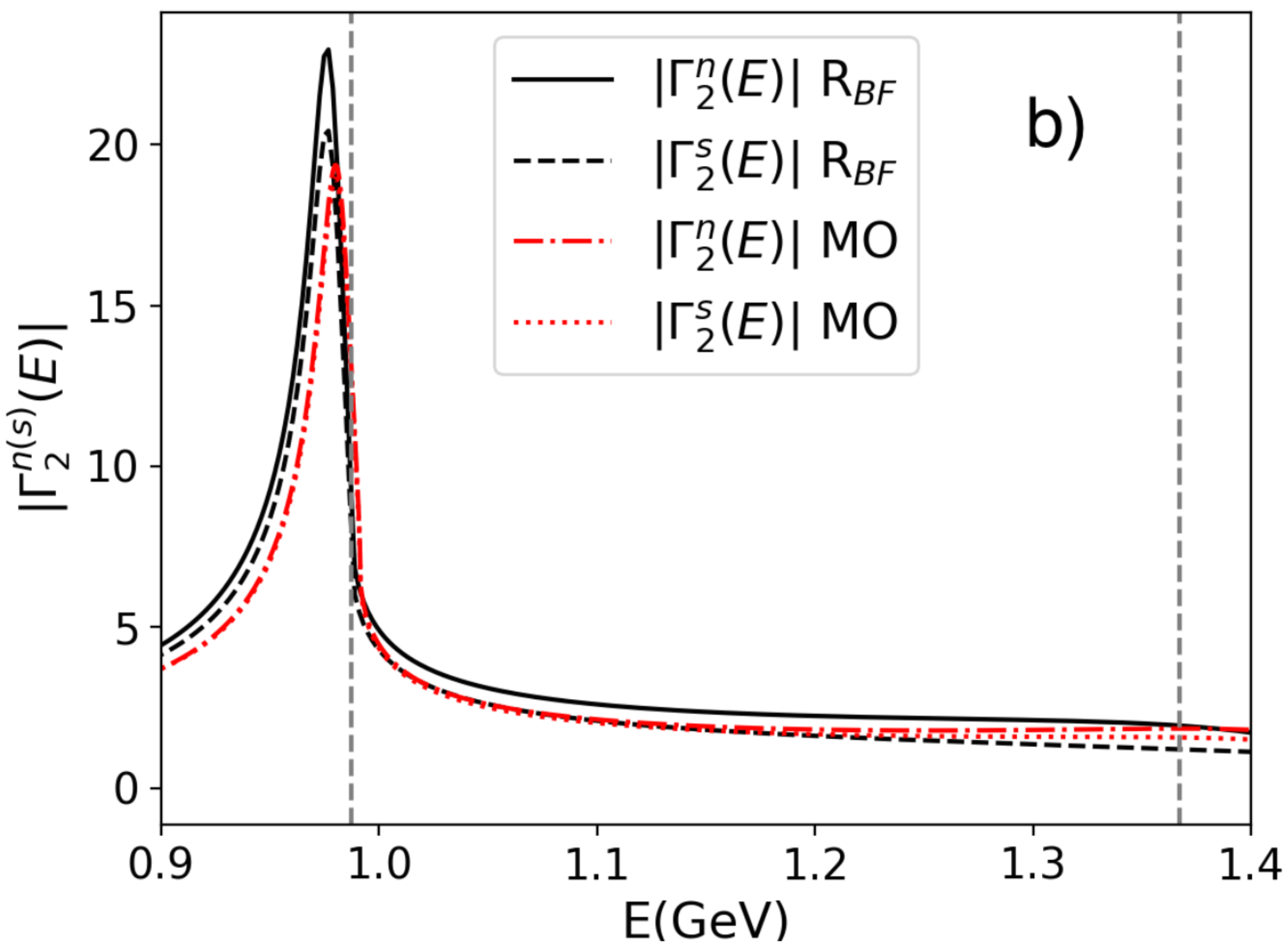}
\caption{  Moduli,  a) for 0~GeV $\leqslant E \leqslant 2$~GeV and b)  for 0.9~GeV $\leqslant E \leqslant ~1.4$ GeV,  of the isoscalar-scalar kaon form factors calculated using the Muskhelishvili-Omn\`es dispersion relation approach~\cite{Moussallam_2019}. 
They are compared to those (continuous black lines) derived in the best fit in Sec.~\ref{results} from a unitary relativistic three coupled-channel model. Both approaches use the updated $T_{\pi \pi}$ matrix derived in Appendix~\ref{UpdatedTpipi}.
The physical $E$ region, 0.987~GeV~$\lesssim E \lesssim $1.367~GeV, is delimited by the two  vertical dashed lines.
}
\label{ModGamma2}
\end{center} \end{figure}

In this appendix we complete our study by describing the results of two fits of the  \textit{BABAR}-Collaboration Dalitz-plot  distribution~\cite{B10} taking, in the amplitudes with final kaon-kaon states in  $S$ wave and isospin 0, the scalar $K \overline K$ form factors derived from the Muskhelishvili-Omn\`es (MO) approach~\cite{MO}.
The same parametrizations as those described in Sec.~\ref{amplitudes} are used for all other amplitudes. In the MO dispersion-relation framework the  isoscalar-scalar $\Gamma_2^{n,s}(s)$ form factors have been calculated by B.~Moussallam~\cite{Moussallam_2000, Moussallam_2019}  from the MO equation using the updated  $\pi \pi$-$T$ matrix  of the  $\pi \pi$, $K \overline K $ and effective $(2\pi)(2\pi)$ coupled-channel model of Ref.~\cite{EPJ}  (see Appendix~\ref{UpdatedTpipi}).
 In Fig.~\ref{ModGamma2} the moduli of these MO form factors are compared to those derived in Sec.~\ref{results} from a relativistic coupled-channel model.
  In Sec.~\ref{amplitudes} one has introduced for the form factors  $\Gamma_2^{n,s}(s)$  complex phenomenological coefficients of proportionality  $\chi^{n,s}$ and in Sec.~\ref{results}, to achieve good fits, notably to reproduce the low density of events in the central region of the Dalitz distribution (see Fig.~\ref{Figure:Cells}, region II), we have been led to multiply them by the energy-dependent phenomenological  functions $P_i(s_0)$ defined below in Eqs.~(\ref{P1}) and~(\ref{P2}).

\begin{figure}[h]  \begin{center}
\includegraphics[scale=0.6]{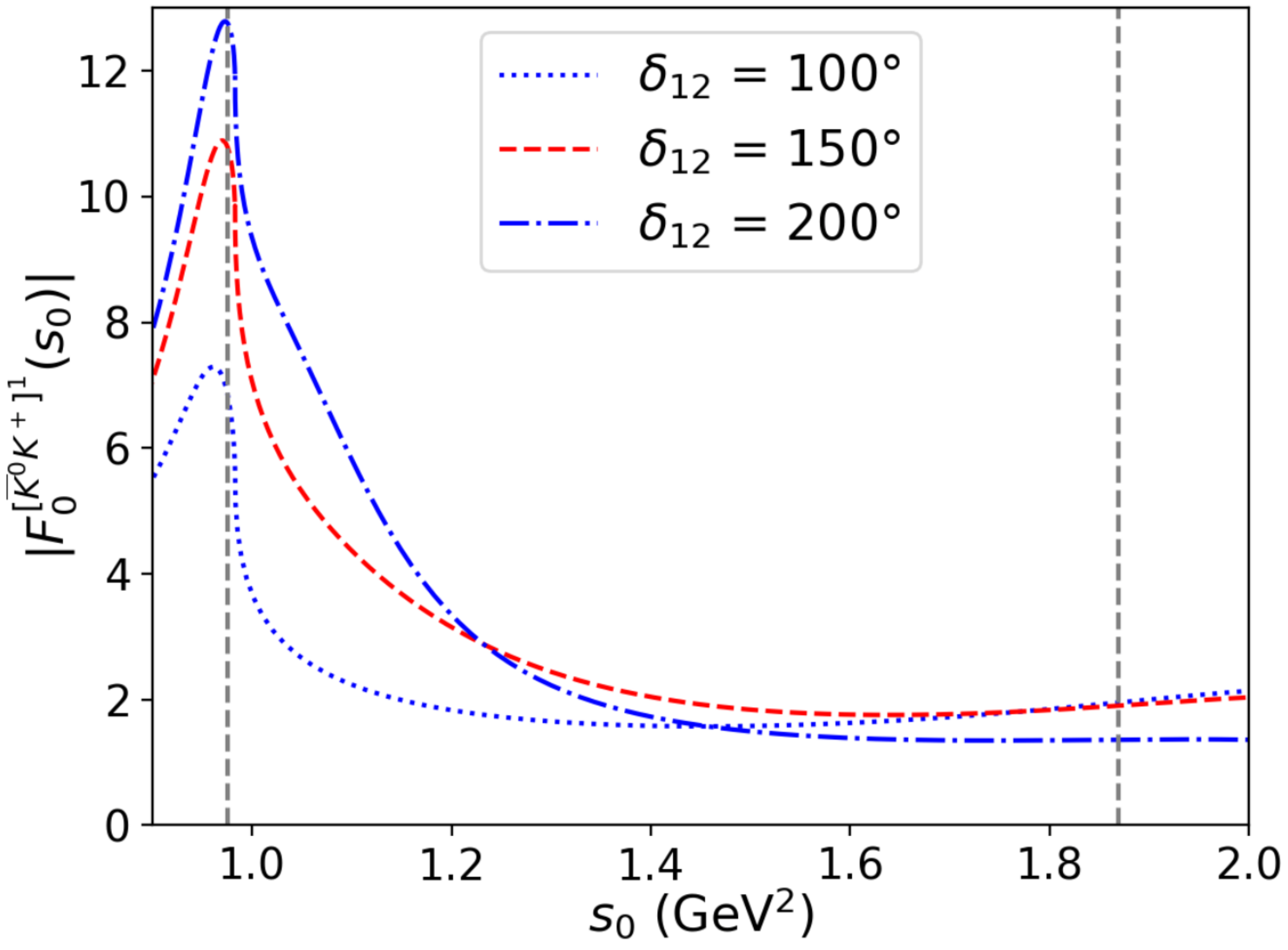}
\caption{  Moduli of the isovector-scalar kaon form factors calculated from MO equations in Ref.~\cite{Bachir2015} for different $\delta_{12}$  parameters which correspond to the sum of the $\eta \pi \to \eta \pi$ and $K \bar K \to K \bar K$ phase shifts at $\sqrt{s}= m_{a_0 (1450)}$. The two vertical dashed lines delimit the physical $s_0$ region, 0.975~GeV$^2$ $\lesssim s \lesssim $1.87~GeV$^2$.
}\label{ModFs}
\end{center} \end{figure}

The isovector-scalar  $F_{0}^{[\overline{K}^0K^+]^1}(s)$  form factor has been calculated in Ref.~\cite{Bachir2015} from coupled MO equations for $\pi \eta$ and $ K \overline K$  channels.
Its modulus, for the parameters $\delta_{12}$= 100$^\circ$, 150$^\circ$ and 200$^\circ$ which are equal to the sum of the $\eta \pi \to \eta \pi$ and $K \bar K \to K \bar K$ phase shifts at $\sqrt{s}= m_{a_0 (1450)}$ , is plotted in Fig.~\ref{ModFs}. 
The isovector amplitudes associated to the isospin-1 $a_0^0$ and $a_0^+$  resonances can  be expressed  in terms of this form factor by using,  in the  Eqs.~(\ref{M13})  and~(\ref{M3Tb}), the relation~(\ref{G1formf}) with $G_1(0)=\chi^1$. 
The strength $\chi^1$ is real  and to obtain good fits, it was necessary to multiply it by the phenomenological polynomial
\be \label{PFs0}
P_F(s_+)\equiv 1 +c_{1} (s_+-s'') + c_{2}( s_+-s'')^2+c_{3} (s_+-s'')^3,
\ee
where the free parameters  $c_i, \ i=1, 2, 3$ and $s''$ are real.

 \begin{figure}[h] \begin{center}
\includegraphics[scale=0.37]{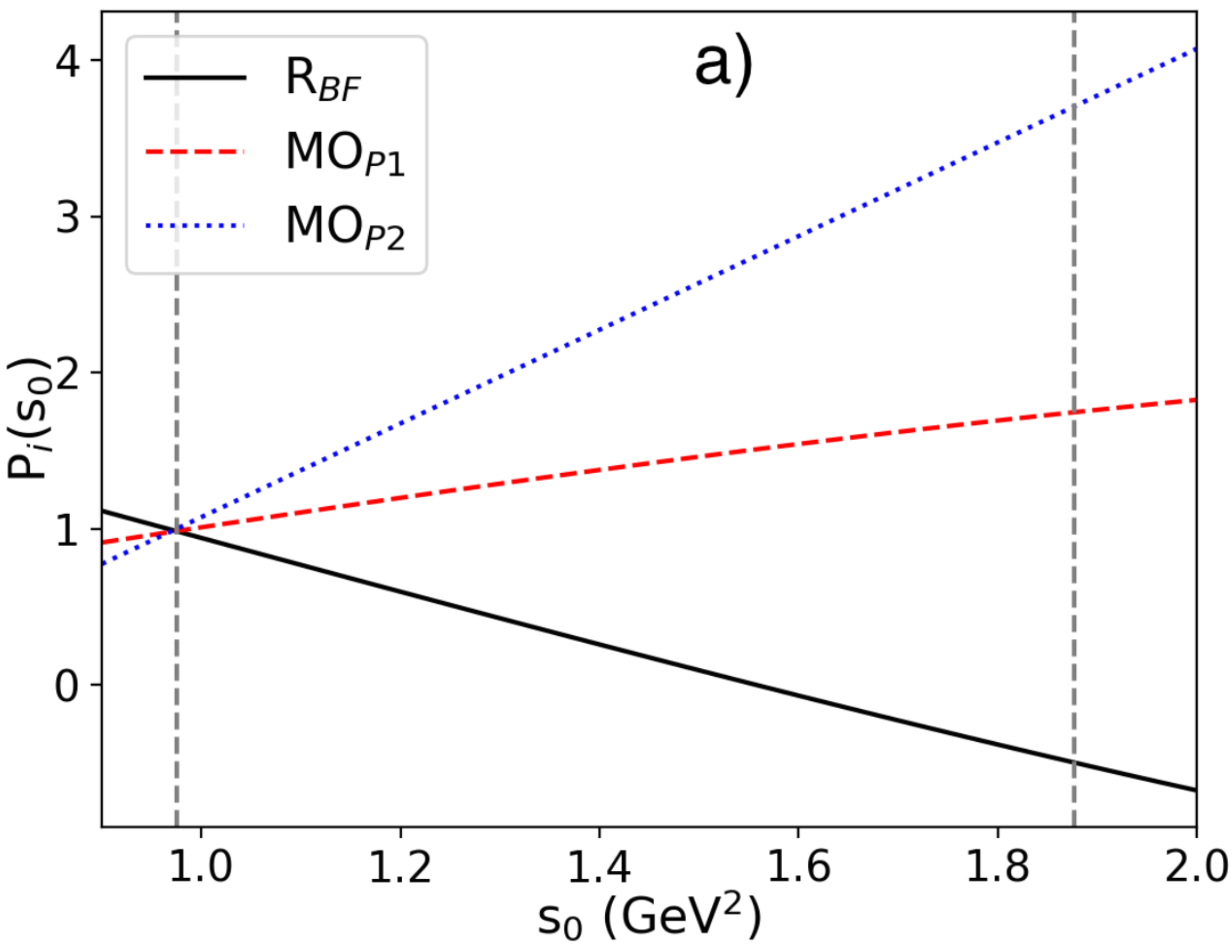}
\includegraphics[scale=0.37]{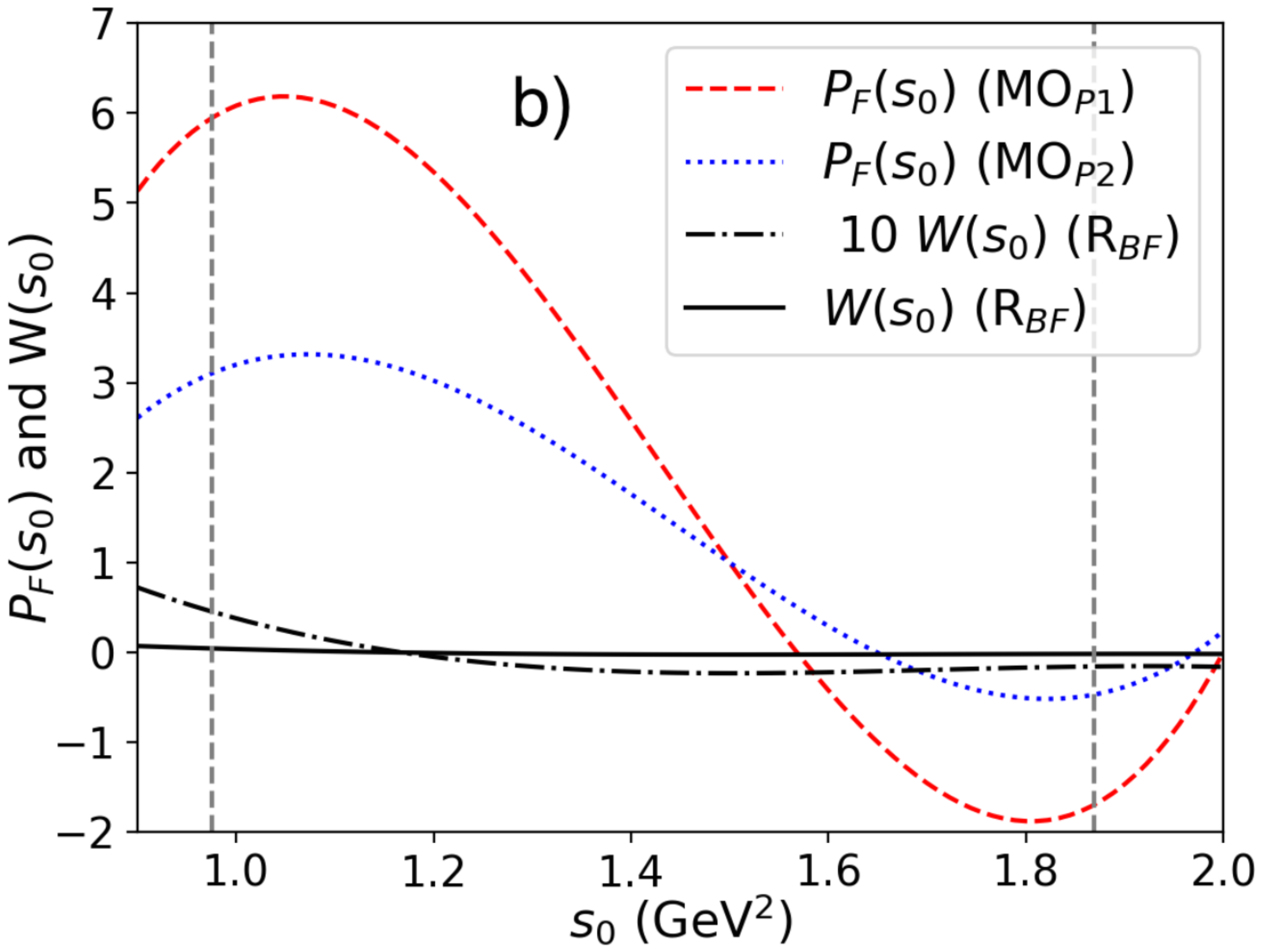}
\caption{ Comparison  of: a) the  functions $P_i(s_0 )$ multiplying,  for the three models MO$_{P1}$, MO$_{P2}$ and R$_{\rm BF}$, the isoscalar-scalar form factors $\Gamma_2^{n(s)}(s_0)$; b) the polynomials $P_F(s_0)$ multiplying, for the  two models MO$_{P1(P2)}$, the isovector-scalar form factor $F_{0}^{[\overline{K}^0K^+]^1}(s_0)$ and $W(s_0)$ introduced in~Eq.~(\ref{polyG1}) for the $G_1(s_0)$ function of the best fit R$_{BF}$. Vertical dashed lines as in Fig.~\ref{ModFs}.
}
\label{PlotPandPFs0}
\end{center} \end{figure}

An estimation of  the phenomenological  strength parameters $\chi^{n,s}$  using  Eq.~(\ref{chin}) with  $|\Gamma_{2}^{n, s}(m^2_{f_0})|\simeq~19$ (see Fig.~\ref{ModGamma2}) leads to $\vert \chi^{n,s}\vert \simeq 26\ \rm{GeV}^{-1}$.
For the kaon isovector-scalar form factor  with $\chi^1 \vert F_{0}^{[\overline{K}^0K^+]^1}(m^2_{a_0})\vert = {g_{a_0 K^+ K^-}}/[{m_{a_0}\,\Gamma_{tot}(a_0)}]=89.66~\rm{GeV}^{-1}$, using (see Ref.~\cite{AFLL}), $\Gamma_{tot}(a_0)=71\pm14$ MeV, $\vert g_{a_0 K^+ K^-} \vert^2/(4\pi)= 0.275$ GeV$^2$, $m_{a_0}=980$ MeV   and for  $\delta_{12}=150^\circ$ $\vert F_{0}^{[\overline{K}^0K^+]^1}(m^2_{a_0})\vert\simeq 10.89$ (see Fig.~\ref{ModFs}),  one obtains $ \chi^1 \simeq 8.2 \rm\ {GeV}^{-1}$.
As expected, from our study in Sec.~\ref{comp} of the near  threshold comparison  between  the  $K^+ K^-$  $S$-wave effective mass projection with that of the $\overline K^0 K^+$, a good fit without the contribution associated with the isospin 0 $f_0$ resonances ($\chi^{n,s}\equiv 0$) cannot be obtained.

\begin{table}[ht] 
\caption{Comparison between the  MO$_{P1}$, MO$_{P2}$ parameters from the two fits using kaon scalar-form factors derived from dispersion relations (see the text) and  the corresponding R$_{{\rm BF}}$  ones, from the best fit with  kaon scalar-form factors calculated from unitary relativistic equation (see Table~\ref{parameters}).
The number of free parameters is denoted by $N_p$.  Parameters without uncertainties are kept fixed during the minimization procedure. Phases $\varphi$ are in radians and detailed  definitions of all parameters are given in the text.}
\begin{center} \begin{tabular}{c c c c c c } \hline\hline
Fit &MO$_{P1}$&MO$_{P2}$&R$_{{\rm BF}}$\\
\hline
$\chi^2(N_p, \chi^2/ndf)$ & 1559.7(16, 1.32)& 1546.9(16, 1.31)&  1474.4(19, 1.25)\\
\hline
 $\vert \chi^n \vert$(GeV$^{-1})$& 26.& 35.& 22.5\\
 $\varphi_{\chi^n}$& 1.63 $^{+0.20}_{-0.18}$& 4.19 ${\pm0.10}$&2.22$^{+0.82}_{-0.98}$ \\
$\chi^s$(GeV$^{-1})$ &26.&26. &22.5\\
  $\vert F_0^{K^0f_0}(m_{D^0}^2)\vert $ & 0.42$^{+0.03}_{-0.04}$ &0.35${\pm0.03}$&2.22$^{+0.26}_{-0.17}$ \\
  $\varphi_{F_0^{K^0f_0}}$ &2.40$^{+0.05}_{-0.04}$ &1.94 ${\pm0.05}$ &2.21${\pm0.10}$\\
 $s'$(GeV$^{-2}$) &  0.&  -3.&1.56$^{+0.02}_{-0.01}$ \\
  $s''$(GeV$^2$) &  1.5&  1.5&\\
  $ \chi^1 $(GeV$^{-1})$  & 8.2 &15&   \\
 $c_1$(GeV$^{2}$)  &-15.38 $^{+0.41}_{-0.45}$&-7.53 ${\pm0.14}$& \\
 $c_2$(GeV$^{-2}$)  & 8.16$^{+0.66}_{-0.65}$&2.88$^{+0.35}_{-0.34}$& \\
 $c_3$(GeV$^{-4}$)&37.20 $^{+1.99}_{-1.94}$&18.24$^{+0.91}_{-0.89}$&  \\
 $\vert F_0^{K^-a^+} (m_{D^0}^2)\vert $ &0.23${\pm0.01}$&0.28${\pm0.01}$& 0.25$^{+0.02}_{-0.03}$\\
 $\varphi_{F_0^{K^-a^+}}$  &5.82${\pm0.05}$ & 5.61$^{+0.05}_{-0.04} $& 5.33$^{+0.12}_{-0.08}$\\
 $ \vert A_0^{K^0\phi}(m_{D^0}^2) \vert $ &  0.99$\pm{0.01}$& 0.99${\pm0.01}$&  0.99${\pm0.01}$ \\
 $\varphi_{ A_0^{K^0\phi} }$ & -0.89${\pm0.02}$&-0.97${\pm0.02}$& 3.67$^{+0.12}_{-0.09}$ \\
 $M_\phi $(MeV) & 1019.55$\pm0.02$& 1019.56$\pm0.02$& 1019.58$\pm0.02$\\
 $\Gamma_\phi $(MeV) & 4.69$\pm$0.04& 4.70$\pm$0.04 & 4.72$\pm$0.04\\
$ \vert A_0^{K^-\rho^+} (m_{D^0}^2)\vert $ & 5.78$^{+0.22}_{-0.25}$& 7.94$^{+0.34}_{-0.37}$& $9.38^{+0.63}_{-0.58}$\\
$\varphi_{A_0^{K^-\rho^+}}$  & 1.18 $\pm{0.03}$& 1.06${\pm0.02}$&5.01$^{+0.06}_{-0.05}$ \\
$\vert P_D\vert $ &15.71$^{+0.78}_{-0.80}$ &14.16$^{+0.69}_{-0.70}$ & $5.52^{+1.25}_{-1.24}$ \\
$\varphi_{P_D}$ &1.13$\pm{0.09}$&0.97$^{+0.10}_{-0.09}$&3.97$^{+0.23}_{-0.25}$ \\
    \hline\hline
  \label{Tabparam}
  \end{tabular} \end{center} 
\end{table}
  
We also find that improved $\chi^2$  are obtained with the $\delta_{12}$ parameter of the isovector-scalar 
$F_{0}^{[\overline{K}^0K^+]^1} $ form factor equal to 150$^\circ$ (see Fig.~\ref{ModFs}). 
With the $N_p$=16 free parameters displayed in Table~\ref{Tabparam}, we obtain a fit, denoted as MO$_{P1}$, with a total $\chi^2$ of 1559.7 which corresponds to a $\chi^2/ndf$=1.32, not as good as that found in the best fit model of Sec.~\ref{results}.
In this fit, the phenomenological function multiplying  the $\Gamma_2^{n(s)}(s_0)$ is chosen to be 
\begin{equation}
P_1(s_0) \equiv P(s_0)
\label{P1},
\end{equation}
with the zero $s'$ of the function $P(s_0)$ [Eq.~(\ref{polygamma})] at 0~GeV$^2$. 
Fixing  $\vert \chi^n \vert$ to 35~(GeV$^{-1})$, $ \chi^1 $ to 15~(GeV$^{-1})$ a slightly better fit,  denoted MO$_{P2}$, with $N_p=16$ and a total $\chi^2$ of~1546.9~($\chi^2/ndf$=1.31) is obtained with
\be
P_2(s_0)=1-s'(s_0-s_{th}).
\label{P2}
\ee

Table~\ref{Tabparam} gives then a comparison of all parameter values with their uncertainties  (when these parameters are fitted) for the MO$_{P1}$, MO$_{P2}$ models together with  the corresponding parameters of the best fit  R$_{{\rm BF}}$ presented in Sec.~\ref{results}.
The variations, in the $s_0$ physical region, of the different functions $P_i(s_0)$ for the MO$_{P1}$ ($i=1, \ s'=0$~GeV$^2$) , MO$_{P2}$  ($i=2$, \  $s'=~{-3}$~GeV$^2$) and R$_{\rm BF}$ ($i=1,\ s'=1.56$~GeV$^{-2}$) fits
are displayed in Fig.~\ref{PlotPandPFs0}(a).
As already indicated in Sec.~\ref{results} [see second sentence below Eq.~(\ref{polyG1})]   Fig.~\ref{PlotPandPFs0}(b) compares the fitted polynomial $W(s_0)$ to the phenomenological polynomials $P_F(s_0)$ [see Eq.~(\ref{PFs0})] multiplying the isovector-scalar form factor $F_{0}^{[\overline{K}^0K^+]^1}(s_0)$  for the  two solutions  MO$_{P1(P2)}$.

\begin{table}[ht]  
\caption{As in Table~\ref{Tabparam} but for the branching fractions in percent of the amplitudes ${\cal{M}}_i$, $i=1$ to $7$.
 For each branching fraction we indicate the dominant resonances (see text) of the quasi-two-body channels of the different amplitudes (see Table~\ref{TabBr}).
For the MO fits, the given uncertainties, $\Delta {\rm Br}_i, i=1,7$ and of their sum are calculated from Eqs.~(\ref{DeltaBri}) to (\ref{defSigmBrbis}) using the average positive and negative uncertainties of the free parameters displayed in Table~\ref{Tabparam}.
See Sec.~\ref{results} for the  calculation of the $\Delta {\rm Br}_i$ of the R$_{\rm {BF}}$ model.
The uncertainties of the sum of the ${\rm Br}_i$ 
 are obtained through the formula given in Eq.~(\ref{defSigmBrbis}).
}
\begin{center} \begin{tabular}{c c c c c }
\hline\hline
Fit &MO$_{P1}$&MO$_{P2}$&R$_{{\rm BF}}$\\
\hline
$\chi^2(N_p, \chi^2/ndf)$ & 1559.7(16, 1.32)& 1546.9(16, 1.31)&  1474.4(19, 1.25)\\
\hline
$ {\rm Br}_1$[$f_0$'$s$, $a_0^0$'$s$]
& 44.9$\pm$8.3 &63.0$\pm$  15.8& $60.9^{+24.4}_{-10.6}$\\
$ {\rm Br}_2$[$\phi$]
&44.9$\pm$0.5\ &44.8$\pm$ 0.5 & 45.5$\pm$ 0.7\\
$ {\rm Br}_3$[$a_0^+$'$s$]
&25.9$\pm$ 0.8&40.1$\pm$1.9 & $20.7^{+9.4}_{-6.0}$\\
$ {\rm Br}_4$[$\rho^+$'$s$]
& 7.7$\pm$ 0.6&13.1 $\pm$1.2 & $21.5^{+3.1}_{-2.8}$\\
$ {\rm Br}_5$[$a_0^-$'$s$]
& 2.6$\pm$ 0.1&2.7$\pm$ 0.1& $0.76^{+0.18}_{-0.15}$\\
$ {\rm Br}_6$[$\rho^-$'$s$]
& 0.03$\pm$0.002 &0.06$\pm$ 0.005&  0.08$\pm$ 0.01\\
$ {\rm Br}_{7}$[$f_2(1270)$]
& 0.37$\pm$ 0.04&0.30$\pm$ 0.03& 0.05$\pm$0.02
\\\hline
$\sum_{i=1,7} {\rm Br}_i$ &126.3$\pm$7.6&164.1$\pm$ 13.7& 149.5$^{+26.9}_{-12.3}$\\
\hline\hline
 \label{TabBr45Req}
\end{tabular} \end{center} 
\end{table}

The different branching fractions  Br$_i, i=1$ to 7, of these two fits are compared to those of the best fit model in Table~\ref{TabBr45Req}.
Following Eq.~(\ref{d2Br}) their given uncertainties $\Delta {\rm Br}_i$ are calculated as 
\begin{equation}
 \Delta {\rm Br}_i=  \frac{1}{32 (2 \pi)^3 m_{D^0}^3 \Gamma_{D^0}}  \left [\sum_{k, l=1,N_p} \frac{\partial f_i}{\partial r_k}\ \frac{\partial f_i}{\partial r_l}\  \rho_{kl} \ \Delta r_k \ \Delta r_l\right ]^{1/2}
\label{DeltaBri}
\end{equation}
where,
\be
\frac{\partial f_i}{\partial r_k}=\int\int ds_+ds_0 \ 2 \left (R_i \frac{\partial R_i}{\partial r_k} + I_i \frac{\partial I_i}{\partial r_k}\right ) \ \rm{with} \ 
f_i=\int\int ds_+ds_0 \ \vert {\cal M}_i(s_0,s_+) \vert^2,
\label{firk}
\ee
and ${\cal M}_i= R_i +i \ I_i$.
 In Eqs.~(\ref{DeltaBri}) and (\ref{firk}) $r_{k(l)}$ are the free parameters entering the amplitude ${\cal M}_{i}$.
In Eq.~(\ref{DeltaBri}), $\Delta r_{k(l)}$  are the $r_{k(l)}$ average uncertainties and $\rho_{kl}$ are the $kl$  correlation coefficients of the MINUIT program.
These quantities are given in the minimization output with  $\rho_{kl}=\rho_{lk}$ and $\rho_{kl} =1$ if $k=l$.
 The uncertainty $\Delta$Br  of the sum of the Br$_i$  is calculated  as\footnote{This, with $i=j$, reduces to $ \Delta{\rm Br}=\left [ \sum_{i=1,7} (\Delta {\rm Br}_i)^2\right]^{1/2}$.}
\be
  \Delta{\rm Br}= \frac{1}{32 (2 \pi)^3 m_{D^0}^3 \Gamma_{D^0}}  \left [\sum_{i,j=1,7} \left (\sum_{k, l=1,N_p} \frac{\partial f_i}{\partial r_k}\ \frac{\partial f_j}{\partial r_l}\  \rho_{kl} \ \Delta r_k \ \Delta r_l \right ) \right ]^{1/2}.
\label{defSigmBrbis}
\ee

\begin{figure}[h]  \begin{center}
\includegraphics[scale=0.36]{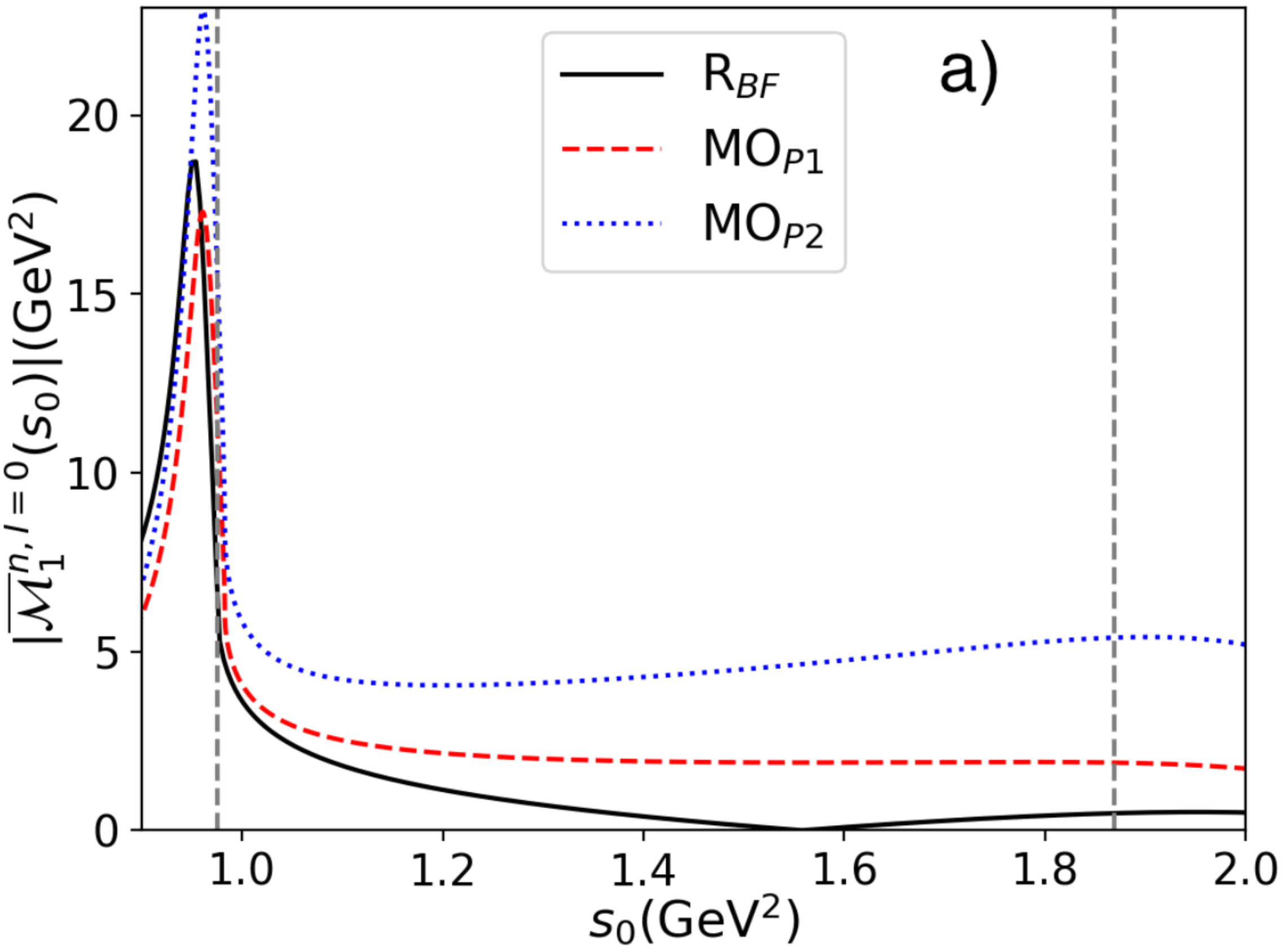}
\includegraphics[scale=0.36]{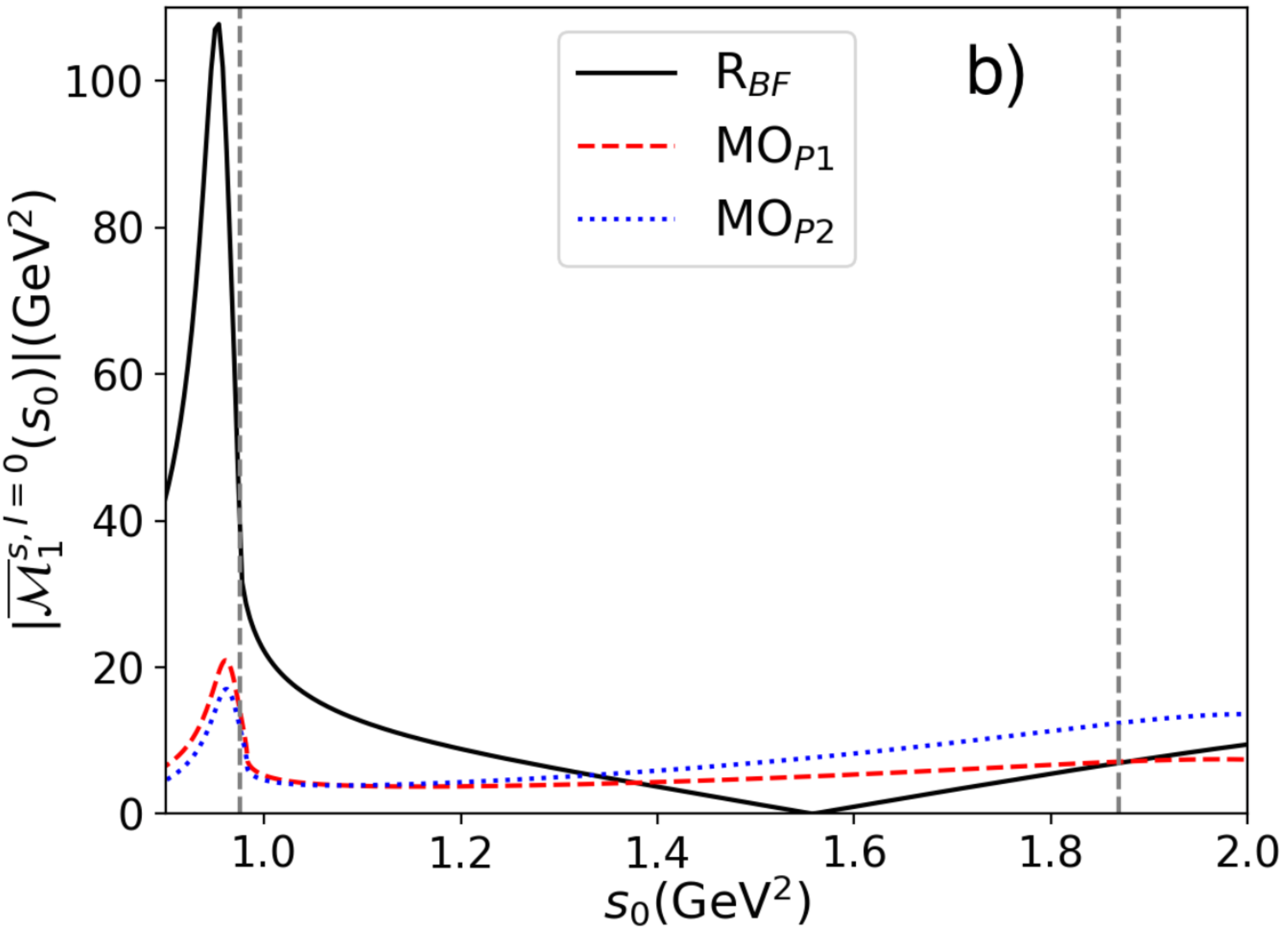}
\includegraphics[scale=0.359]{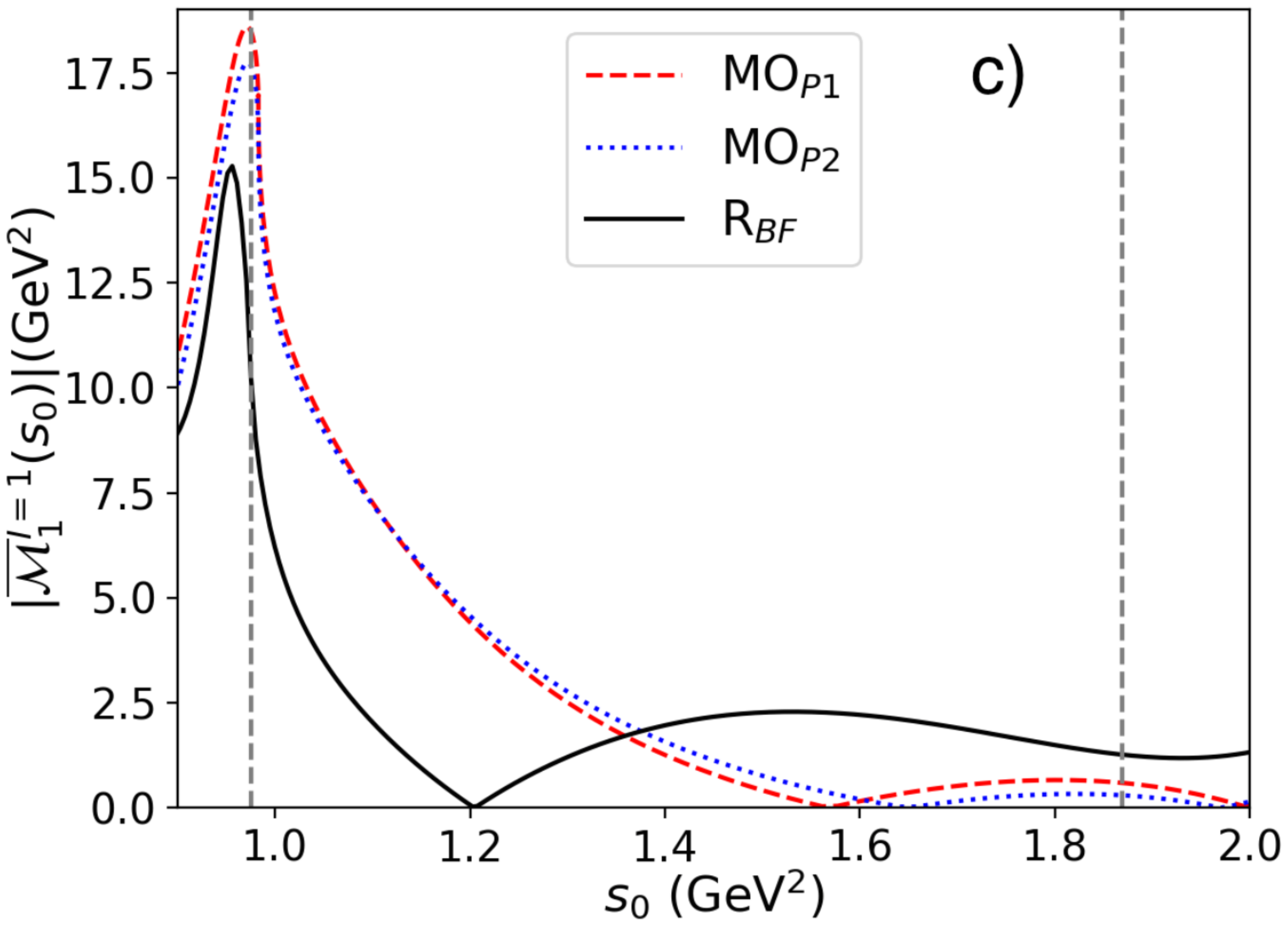}
\includegraphics[scale=0.359]{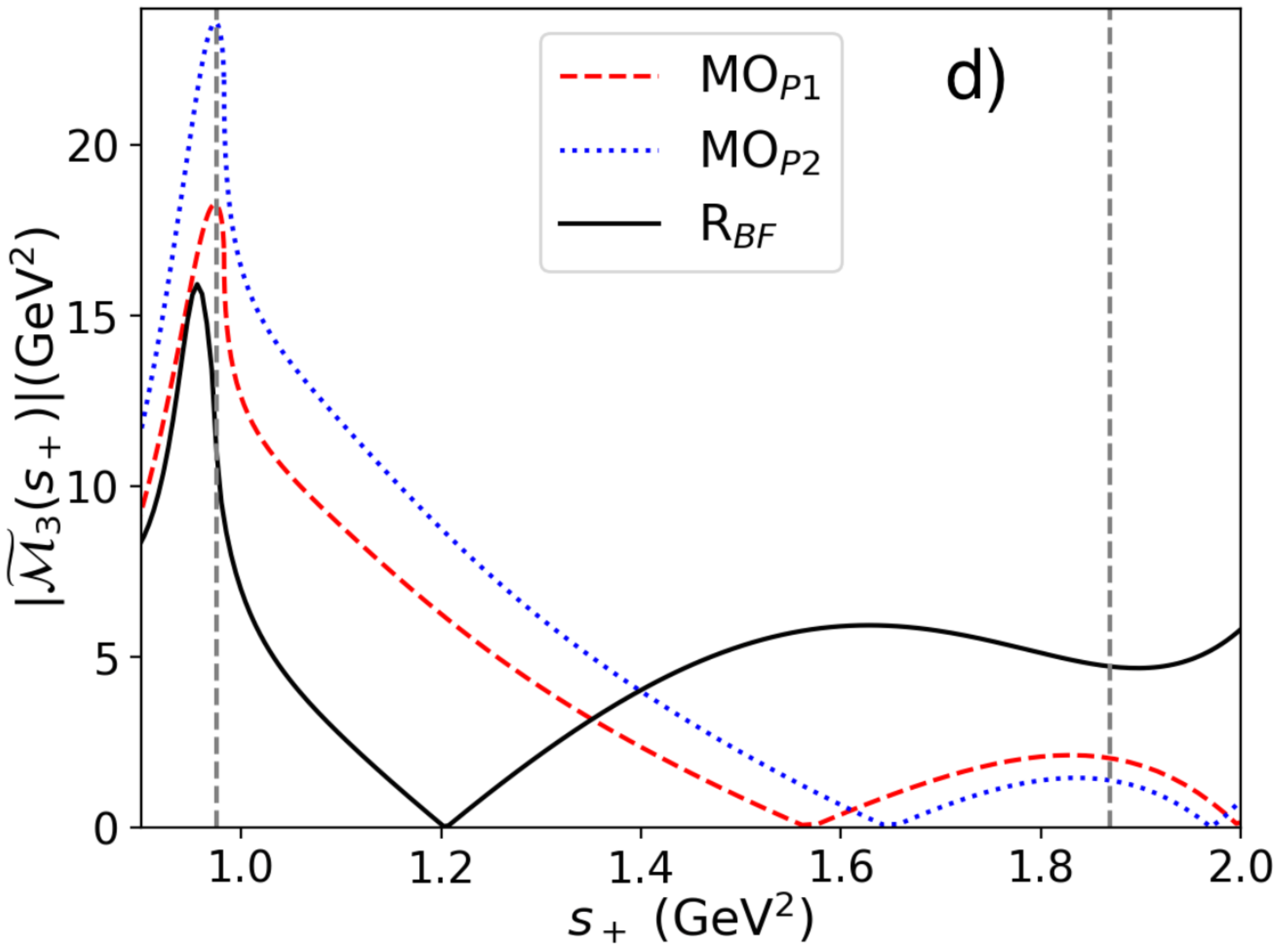}
 \caption{ The black continuous line, denoted by R$_{{\rm BF}}$ represents in a) $|{\overline{\cal{M}}}^{n,I=0}_1(s_0)|$, in b) $|{\overline{\cal{M}}} ^{s,I=0}_1(s_0)|$,  and in c)  $|{\overline{\cal{M}}}^{I=1}_1(s_0)|$ [see Eqs.~(\ref{modoverM1i})]
 and in d) $|{\widetilde{\cal{M}}}_3(s_+)|$ [see Eq.~(\ref{renorM3splus})].
 The red dashed curve, the blue dotted one represent the corresponding moduli for the alternative MO$_{P1}$, MO$_{P2}$  fits, respectively. As in Fig.~\ref{ModFs} for the two vertical dashed lines (the $s_+$ limits are very close to the $s_0$ ones)}. 
\label{M1is} 
\end{center}  \end{figure}

  The branching fractions of the three components of the ${\cal{M}}_1$ amplitude, viz.,  ${\cal{M}}^{n,I=0}_1$, ${\cal{M}}^{s,I=0}_1$, ${\cal{M}}^{I=1}_1$ [see Eqs.~(\ref{M1T})-(\ref{M13})],  given in Table~\ref{TabBr45Reqa} show that the isoscalar $f_0$ and isovector $a_0^0$ resonance contributions can be quite different.  However, taking into account the large uncertainties in the Br$_1$ values of the MO$_{P1}$, MO$_{P2}$ and R$_{{\rm BF}}$  fits (see Table~\ref{TabBr45Req})  the total scalar-resonance contribution in the ${\cal{M}}_1$ amplitude is similar.
The corresponding results can be qualitatively interpreted from,  the expressions of the amplitudes given in Sec.~\ref{scaamp},  the values of the different parameters given in Tables~\ref{parameters}, \ref{Tabparam} and  those of the kaon scalar-form factors in use. 
For the three terms of the  ${\cal M}_1$  amplitude  [see Eqs.~(\ref{M1T})-(\ref{M13})]  one can define the renormalized amplitudes,
\be \label{modoverM1i}
{\overline{\cal{M}}}^{n,I=0}_1(s_0)=\frac{{{\cal{M}}}^{n,I=0}_1(s_0)}{F_W}, \ \ \ {\overline{\cal{M}}}^{s,I=0}_1(s_0)=\frac{{{\cal{M}}}^{s,I=0}_1(s_0)}{F_W} \ \ \ {\rm and} \ \ \ {\overline{\cal{M}}}^{I=1}_1(s_0)=\frac{{{\cal{M}}}^{I=1}_1(s_0)}{F_W},
\ee
where $F_W= -G_F\ (\Lambda_1+\Lambda_2)\  a_2/2$. 
In the case of the amplitude ${\cal{M}}_3(s_+)$ [see Eq.~(\ref{M3Tb})], with $F'_W=G_F\  \Lambda_1 \ a_2/2$, one  define the renormalized amplitude
\begin{equation}
{\widetilde{\cal{M}}}_3(s_+)= \frac{{\cal{M}}_3(s_+)}{F'_W}.
\label{renorM3splus} 
\end{equation}
The coupling $\Lambda_2$ (=$-0.05)$ is small and  $\Lambda_1+\Lambda_2$ (=0.90) is close to  
$\Lambda_1$ (=0.95)  [see Eq.~(\ref{lambda})], consequently $F'_W$ is close to $F_W$.
One can then compare the amplitudes (\ref{modoverM1i}) and (\ref{renorM3splus}) because the contribution of the $a_1$ term in ${\cal{M}}_3(s_+)$ is negligible since the factor $ m_{K}^2- m_{K^0}^2 = -0.0039$~GeV$^2$ is very small. The moduli of these amplitudes are plotted, for the best fit as the black continuous lines denoted by R$_{{\rm BF}}$, the red dashed curve for the MO$_{P1}$ model and the blue dotted one for the MO$_{P2}$ one, in Figs.~\ref{M1is}(a), (b), (c) and~(d).

The comparison, shown in Fig.~\ref{M1is}, of the resulting $s_0$ behavior of the moduli $|{\overline{\cal{M}}}^{n,I=0}_1(s_0)|$,  $|{\overline{\cal{M}}} ^{s,I=0}_1(s_0)|$ and  $|{\overline{\cal{M}}}^{I=1}_1(s_0)|$ allows furthermore to understand qualitatively the different branching fractions displayed in Table~\ref{TabBr45Reqa}. The branching fractions can also be partly compared to the fit fractions\footnote{  Br[$f_0(1370)$]=1.7\%, Br[$a_0(980)^0$]+Br[$a_0(1450)^0$]=71.1\%, Br[$\phi(1020)$]=44.1\%, Br[$a_0(980)^+$]+Br[$a_0(1450)^+$]=45.1\%, Br[$a_0(980)^-$]=0.7\% and Br[$f_2(1270)$]=0.7\%} of the \textit{BABAR} isobar-model experimental analysis~\cite{supmaPRL105}.

 The dominance of the branching fraction associated to the isoscalar-scalar amplitude ${\cal M}_1^{s, I=0}$ for the best fit and to a less extent for the MO$_{P1}$ one,  can be understood as their moduli $|{\overline{\cal{M}}}^{s,I=0}_1(s_0)|$ [Fig.~\ref{M1is}(b)] are larger than the moduli  $|{\overline{\cal{M}}}^{n,I=0}_1(s_0)|$  [Fig.~\ref{M1is}(a)] and  $|{\overline{\cal{M}}}^{I=1}_1(s_0)|$ [Fig.~\ref{M1is}(c)]. Comparison of Fig.~\ref{M1is}(a) and Fig.~\ref{M1is}(c) can also explain qualitatively, for the best fit,  the difference between the ${\cal M}_1^{n, I=0}$ (1.19\%) and ${\cal M}_1^{I=1}$ (4.48\%) branching fractions.

\begin{table}[ht]  
\caption{As in Table~\ref{Tabparam} but for the branching fractions in percent of the amplitudes   ${\cal{M}}^{n,I=0}_1$, ${\cal{M}}^{s,I=0}_1$ and ${\cal{M}}^{I=1}_1$ [see Eqs.~(\ref{M1T}-\ref{M13})]. Lines 8 to 10 give the contribution of the interferences between these amplitudes.
}
\begin{center} \begin{tabular}{c c c c c }
\hline\hline
Fit &MO$_{P1}$&MO$_{P2}$&R$_{{\rm BF}}$\\
\hline
$\chi^2(N_p, \chi^2/ndf)$ & 1559.7(16, 1.32)& 1546.9(16, 1.31)&  1474.4(19, 1.25)\\ 
\hline
$ {\rm Br}_1$[$f_0, a_0^0$]& 44.88&63.01&60.93
\\
\hline
$ {\rm Br}[{\cal{M}}^{n,I=0}_1$: $f_0$ in $\Gamma^n(s)$] & 4.75& 20.66& 1.19 \\
$ {\rm Br}[{\cal{M}}^{s,I=0}_1$: $f_0$ in $F_0^{K^0 f_{0}}(m_{D^0}^2)$ $\Gamma^s(s)$] & 22.69 & 45.91 & 59.82 \\
$ {\rm Br}[{\cal{M}}^{I=1}_1$: $a_0^0$ in $\chi^1 P_F(s)$  $F_{0}^{[\overline{K}^0K^+]^1}(s)\ {\rm or} \ G_1(s)$]& 16.47 & 16.47&  4.48 \\
\hline
${\rm Br}({\cal{M}}^{n,I=0}_1)+{\rm Br}({\cal{M}}^{s,I=0}_1)+{\rm Br}({\cal{M}}^{I=1}_1)$ & 43.90 & 83.04& 65.49\\
\hline
2 ${\rm Br}_{1112}$ & -12.95& -37.98& -14.6\\
2 $ {\rm Br}_{1113}$ & 1.86& 16.44& -0.72 \\
2 ${\rm Br}_{1213}$ & 12.03&0.43
 & 10.8\\
\hline
 $\sum_{i,j=1,3}^{i\neq j}{\rm Br}_{1i1j}$ &0.94&-20.02& -4.57\\
\hline\hline
 \label{TabBr45Reqa}
\end{tabular} \end{center} 
\end{table}

 For the $S$-wave $K \overline K$ contributions there are important differences between the models MO$_{P1}$, MO$_{P2}$  and R$_{{\rm BF}}$.
The kaon isoscalar-scalar form factors are similar (see Figs.~\ref{Gamns} and~\ref{ModGamma2}) but multiplication by the function $P_i(s_0)$  [see Eqs.~(\ref{polygamma}), (\ref{P1} and (\ref{P2})] implies different modifications. 
For the MO fits, the kaon isovector-scalar form factor  $F_{0}^{[\overline{K}^0K^+]^1} (s_0)$ is multiplied by different phenomenolo\-gical polynomials $P_F(s_0)$ [Eq.~(\ref{PFs0})] compared, in the Fig.~\ref{PlotPandPFs0}(b), to the fitted polynomial $W(s_0)$~[Eq.~(\ref{polyG1})] entering the $G_1(s_0)$ function of the best fit.

In the MO$_{P2}$ fit ($ {\rm Br}({\cal{M}}^{n,I=0}_1)$= 20.66~\%) the $f_0$ contribution in $\Gamma_2^n(s)$ is enhanced by a larger $\vert \chi^n \vert$ (35~GeV$^{-1}$) and by the function $P_1(s_0)$  while it is more suppressed in the best fit (1.19~\%) than in the  MO$_{P1}$ one (4.75~\%). 

A striking difference arises for the modulus of the phenomenological transition form factor $F_0^{K^0 f_{0}}(m_{D^0}^2)$: it is quite large, 2.22, for the best fit solution R$_{\rm BF}$ as compared to its magnitude for the  two other fits where it is smaller than 0.43.
 This leads to large value of the branching fraction (59.82~\%) for the ${\cal{M}}^{s,I=0}_1(s_0)$ amplitude  arising from $D^0$ annihilation via $W$ exchange in the  R$_{\rm BF}$ best fit solution (see Table~\ref{TabBr45Reqa}).
 The $\vert \chi^s \vert$  being the same (26~GeV$^{-1}$), the difference between the MO$_{P1}$ (22.69~\%)  and  MO$_{P2}$ (45.91~\%) branching fractions arises from smaller $P_1(s_0)$ enhancement (for $s_0 \gtrsim 1.3$~GeV$^2$) than that of $P_2(s_0)$, as can be seen in Figs.~\ref{PlotPandPFs0}(a) and~\ref{M1is}(b).

The values of the $ {\rm Br}({\cal{M}}^{I=1}_1)$  in the fith line of Table~\ref{TabBr45Reqa} indicate the isospin-1 $a_0^0$ resonances content in $\chi_1 P_F(s) F_{0}^{[\overline{K}^0K^+]^1}$ or $G_1(s)$. 
This branching fraction small (4.48~\%) in the R$_{\rm BF}$ fit (some suppression because of $W(s_0)$) is the same (16.47~\%) in the MO$_{P1}$  and MO$_{P2}$, the larger $P_F(s_0)$ [Fig.~\ref{PlotPandPFs0}(b)] is conpensated by a smaller $
\chi^1 $  (8.2 versus 15).  

The  branching fraction  Br$_3$, which indicates the  isovector $a_0^+$ resonances contribution, has values  of 25.9, 40.1 and 20.7~\% for the MO$_{P1}$, MO$_{P2}$ and R$_{BF}$ fits, respectively (see Table~\ref{TabBr45Req}). 
The modulus of the transition form factor $F_0^{K^- a_{0}^+}(m_{D^0}^2)$, entering in  Eq.~(\ref{M3Tb}) is equal to 0.23, 0.28 and  0.25 for the MO$_{P1}$, MO$_{P2}$ and best fits, respectively.
 The branching fractions depend on the $\chi^1$ values and $P_F(s_+)$ behavior  for the MO fits and on the role of $G_1(s_+)$ in the R$_{\rm BF}$  solution and their values are in qualitative agreement with the corresponding 
$|{\widetilde{\cal{M}}}_3(s_+)|$ curves shown in Fig.~\ref{M1is}(d).

The role of the $[\rho(770)^+ + \rho(1450)^+ + \rho(1700)^+]$
resonances is different in our models: the Br$_4$ of the MO$_{P1}$, MO$_{P2}$ and R$_{\rm BF}$ fits are equal to 7.7, 13.1 and 21.5~\%, respectively (see Table~\ref{TabBr45Req}). 
The large contribution in the R$_{\rm BF}$ solution is partly due to the magnitude, 9.38, of the modulus of the transition form factor  $\vert A_0^{K^-\rho^+}(m_{D^0}^2)\vert$  to be compared to 5.78 and 7.94  for the MO$_{P1}$ and MO$_{P2}$ fits, respectively.
The Br$_4$ ratio between that of the R$_{BF}$ and those of the MO$_{P1}$ and MO$_{P2}$ fits is close to the square of the corresponding   $\vert A_0^{K^-\rho^+}(m_{D^0}^2) \vert$ ratios.
It can be seen that, to improve the $\chi^2$ of the fits, it seems necessary to increase the
$\rho^+$ resonances contributions.

The small  isospin-1 $a_0^-$ and $\rho^-$ resonances contents in $ {\rm Br_{5(6)}}$  come from the fact that the ${\cal M}_{5(6)}$ amplitudes [see Eqs.~(\ref{M2Tb})  and (\ref{M6})] are proportional to the $V_{CKM}$ coupling $\Lambda_2$ with $\vert \Lambda_2/\Lambda_1 \vert~\simeq~5~\times~10^{-2}$, while all other amplitudes are proportional either to $\Lambda_1+\Lambda_2$ [${\cal M}_1$, Eq.~(\ref{M1T}) with Eqs.~(\ref{M11}), (\ref{M12}), (\ref{M13}),
 ${\cal M}_2$, Eq.~(\ref{M2}) and ${\cal M}_7$, Eq.~(\ref{PD})] or to $\Lambda_1$ [${\cal M}_3$ Eq.~(\ref{M3Tb}) and ${\cal M}_4$, Eq.~(\ref{M6Tb})].

The  $f_2(1270)$ resonance contributions in Br$_7$ for our three fits follow the evolution of the square of the $\vert P_D \vert$  parameter in each fit. They are very small and  even smaller than in the \textit{BABAR} analysis~\cite{supmaPRL105}. 

 The negative total interference contributions are equal to -26.3~\%, -64.1~\% and  -49.5~\%,  for the  MO$_{P1}$, MO$_{P2}$  and R$_{BF}$ fits, respectively, compared to that of the isobar \textit{BABAR} model  of -63.4~\%~\cite{supmaPRL105}.

The comparison of the off-diagonal elements ${\rm Br}_{ij}$, $i \neq j$ shows large interferences between the amplitudes giving large or sizable branching fractions (see for instance 
  Table~\ref{tabBrij} of the best fit). This is in particular the case between ${\cal M}_1(s_0) $ and  ${\cal M}_{4}(s_+) $.  These values can be qualitatively expected by inspecting the different branching fractions given in Table~\ref{TabBr45Req}.


\begin{thebibliography}{99} 

\bibitem{Zupanc2009}
A. Zupanc \textit{et al.} (Belle Collaboration), Measurement of $y_{CP}$ in $D^0$ meson decays to the $K^0_S K^+K^-$ final state, Phys. Rev. D \textbf{80}, 052006 (2009).

\bibitem{B10}  
 P. del Amo Sanchez \textit{et al.} (\textit{BABAR} Collaboration), 
    Measurement of $D^0-\bar D^0$ Mixing Parameters Using $D^0 \to K^0_S \pi^+ \pi^-$ and $D^0 \to K^0_S K^+ K^-$ Decays, Phys. Rev. Lett.  {\bf 105}, 081803 (2010).

\bibitem{B5}
 B. Aubert \textit{et al.} (\textit{BABAR} Collaboration), Dalitz plot analysis of $D^0~\to\bar K^0 K^+ K^-$, Phys. Rev. D {\bf 72,} 052008 (2005). 

\bibitem{Aubert2008}
 B. Aubert \textit{et al.} (\textit{BABAR} Collaboration), 
 Improved measurement of the CKM angle $\gamma$ in $B^{\pm} \to D^{0} K^{\pm}$ decays with a Dalitz plot analysis of D decays to $D \to K^0_{\rm S} \pi^+\pi^-$ and $D \to K^0_{\rm S} K^+K^-$ ,
Phys. Rev. D \textbf{78}, 034023 (2008).
  
\bibitem{supmaPRL105} 
 Supplemental Material of P. del Amo Sanchez \textit{et~al.} (\textit{BABAR} Collaboration),  Measurement of $D^0-\bar D^0$ Mixing Parameters Using $D^0 \to K^0_S \pi^+ \pi^-$ and $D^0 \to K^0_S K^+ K^-$ Decays,
 Phys. Rev. Lett. \textbf{105}, 081803 (2010), http://link.aps.org/supplemental/10.1103/PhysRevLett.105.081803)
and F.  Martinez-Vidal (private communication).

\bibitem{Weidenkaff}
P. Weidenkaff, Analysis of the decay $D^0 \to K^0_S K^+K^-$ with the BESIII experiment, Ph.D. thesis, Mainz University, 2016.

\bibitem{Ablikim2006.02800}
M. Abilikim \textit{et al.} (BESIII Collaboration), Analysis of the decay $D^0 \to K^0_S K^+K^-$ ,  arXiv: 2006.02800.

\bibitem{SanchezPRL105121801} P. del Amo Sanchez \textit{et al.} (\textit{BABAR} Collaboration),
 Evidence for Direct CP Violation in the Measurement of the Cabbibo-Kobayashi-Maskawa 
Angle $\gamma$ with $B^{\pm} \to D^{(*)} K^{(*)\pm}$ Decays,
 Phys. Rev. Lett. \textbf{105}, 121801 (2010).

\bibitem{Poluektov2010} 
A. Poluektov \textit{et al.} (Belle Collaboration), 
 Evidence for direct CP-violation in the decay $B^{\pm} \to D^0 K^{\pm}$, $D^0 \to  K^0_{\rm S} \pi^+\pi^-$ and measurement of the CKM phase $\phi_3$, Phys. Rev. D \textbf{81}, 112002 (2010).

\bibitem{J.P.Lees_PRD87_052015_BABAR}
J. P. Lees \textit{et al.} (\textit{BABAR} Collaboration), 
Observation of direct CP violation in the measurement of the Cabibbo-Kobayashi-Maskawa angle $\gamma$ with $B^{\pm} \to D^{(*)} K^{(*)\pm}$ decays, Phys. Rev. D \textbf{87}, 052015 (2013).

\bibitem{LHCbNP2014}
R. Aaij \textit{et al.} (LHCb Collaboration),
 Measurement of CP violation and constraints on the CKM angle $\gamma$ in $B^\pm \to D K^\pm$ with $D \to K^0_{\rm S} \pi^+\pi^-$ decays, Nucl. Phys. B\textbf{888}, 169 (2014).

\bibitem{J.Libby_PRD82_CLEO}
J. Libby \textit{et al.} (CLEO Collaboration),
 Model-independent determination of the strong-phase difference between $D^0$ and $\overline  D^0 \to K^0_{S,L} h^+ h^- (h=\pi,K)$ and its impact on the measurement of the CKM angle $\gamma/\phi_3$, Phys. Rev. D \textbf{82}, 112006 (2010).

\bibitem{H.Aihara_Belle2012} 
H. Aihara \textit{et al.} (Belle Collaboration),
 First measurement of $\phi_3$ with a model-independent Dalitz plot analysis of $B^\pm \to D K^\pm, D\to K_S^0 \pi^+ \pi^-$ decay, Phys. Rev. D \textbf{85}, 112014 (2012).

\bibitem{R.Aaij_JHEP2018}
R. Aaij \textit{et al.} (LHCb Collaboration),
 Measurement of the CKM angle $\gamma$ using $B^{\pm} \to D K^{\pm}$ with $D \to K^0_S \pi^+\pi^-$, $D \to K^0_S K^+K^-$ decays, J. High Energy Phys. \textbf{08} (2018) 176.

\bibitem{BES2007}
M. Abilikim \textit{et al.} (BESIII Collaboration),
Improved model-independent determination of the strong-phase difference between $D^0$ and $\overline D^0 \to K^0_{S,L} K^+ K^-$ decays, arXiv:2007.07959v1.

\bibitem{JPD_PRD89}
J.-P. Dedonder, R. Kami\'nski, L. Le\'sniak and B. Loiseau, 
 Dalitz plot studies of $D^0 \to K_S^0 \pi^+ \pi^-$ decays in a factorization approach,
Phys. Rev. D {\bf 89}, 094018 (2014).

\bibitem{Boito} 
D. Boito, J.-P. Dedonder, B. El-Bennich, R. Escribano, R. Kami\'nski, L. Le\'sniak, and B. Loiseau, 
 Parametrizations of three-body hadronic $B$- and $D$-decay amplitudes in terms of analytic and unitary meson-meson form factors, Phys. Rev. D {\bf 96}, 113003 (2017). 

\bibitem{DedonderPol}J.-P. Dedonder, A. Furman, R Kami\'nski, L. Le\'sniak and B. Loiseau, 
 Final state interactions and CP violation in $B^{\pm}\to \pi^+ \pi^- \pi^{\pm}$ decays,
Acta Phys. Pol. B {\bf 42}, 2013 (2011). 

\bibitem{PLB699_102}
A. Furman, R. Kami\'nski, L. Le\'sniak, and P. \.{Z}enczykowski, 
 Final state interactions in $B^\pm \to K^+K^- K^\pm$ decays, Phys. Lett.  B {\bf 699}, 102 (2011).

\bibitem{KKK}
L. Le\'sniak and P. \.Zenczykowski,  Dalitz-plot dependence of $CP$ asymmetry in $B^{\pm}\to K^{\pm} K^+K^-$ decays, 
Phys. Lett. B {\bf 737}, 201 (2014).

\bibitem{KLL}
R. Kami\'nski, L.~Le\'sniak, and B.~Loiseau,
Three channel model of meson-meson scattering and scalar meson spectroscopy,
Phys. Lett. B {\bf 413}, 130 (1997).

\bibitem{EPJ}
R.~Kami\'nski, L.~Le\'sniak, and B.~Loiseau,  Scalar mesons and multichannel amplitudes,  
Eur. Phys. J.  C {\bf 9}, 141 (1999). 

\bibitem{AFLL}
A. Furman, and L. Le\'sniak,  Coupled channel study of $a_0$ resonances,
Phys. Lett. B {\bf 538}, 266 (2002).

\bibitem{AFLL2}
A. Furman, and L. Le\'sniak,  Properties of the $a_0$ resonances,
Nucl. Phys. B (Proc. Suppl.) {\bf 121}, 127 (2003).

\bibitem{Bruch2005}
C. Bruch, A. Khodjamirian, and J. H. K\"{u}hn, Modeling the kaon form factors in the timelike region,
 Eur. Phys. J. C. {\bf 39}, 41 (2005).

\bibitem{PDG2020}  P. A. Zyla \textit{et al.} (Particle Data Group), Review of particle physics, Prog. Theor. Exp. Phys. 2020, 083C01 (2020).

\bibitem{MO}
N. I. Muskhelishvili, in Singular integral equations (P. Noordhoff Ltd., Grˆningen, The Netherlands 1953), Chaps; 18 and 19;
R. Omn\`es, On the solution of certain singular integral equations of quantum field theory,  Nuovo Cimento {\bf 8}, 316 (1958).

\bibitem{Moussallam_2000}
B.~Moussallam,  $N_f$ dependence of the quark condensate from a chiral sum rule, Eur.\ Phys.\ J.\  C {\bf 14}, 111 (2000) and  private communication.

\bibitem{Moussallam_2019}
B.~Moussallam, (private communication).

\bibitem{Bachir2015} M. Albaladejo and B. Moussallam,
Form factors of the isoscalar-scalar current and the $\eta\pi$ phase shifts,  Eur. Phys. J. C {\bf 75}, 488 (2015).

\bibitem{Buchalla1996}
G. Buchalla, A. J. Buras and M. E. Lautenbacher,  Weak decays beyond leading logarithms,
Rev. Mod. Phys. \textbf{68}, 1125 (1996).

\bibitem{BurasNPB434_606} 
A. J. Buras,  QCD factors $a_1$ and $a_2$ beyond leading logarithms versus factorization in non leptonic heavy meson decays, Nucl. Phys. B{\bf 434}, 606 (1995).

\bibitem{AliPRD58} A. Ali, G. Kramer, and Cai-Dian L\"u, Experimental tests of factorization in charmless nonleptonic two-body B decays, Phys. Rev. D  \textbf{58}, 094009 (1998).

\bibitem{Beneke2003} 
M.~Beneke and M.~Neubert,  QCD factorization for $B\to PP$ and $B\to PV$ decays, 
Nucl. Phys. B\textbf{675}, 333 (2003).

\bibitem{El-Bennich_PRD79}
B. El-Bennich, O. Leitner, J.-P. Dedonder, and B. Loiseau,  Scalar meson $f_0(980)$ in heavy-meson decays,
Phys. Rev. D {\bf 79}, 076004 (2009).

\bibitem{Melichow}
D. Melikhov,   Dispersion approach to quark-binding effects in weak decays of heavy mesons,
Eur. Phys. J. direct  {\bf 4}, 1 (2002).

\bibitem{B7}    
 B. Aubert \textit{et al.}  (\textit{BABAR} Collaboration),
    Amplitude analysis of the decay $D^0 \to K^- K^+ \pi^0$,  Phys. Rev. D  {\bf 76}, 011102(R) (2007).
    
\bibitem{B11}
 P. del Amo Sanchez \textit{et al.} (\textit{BABAR} Collaboration), 
     Dalitz plot analysis of $D_s^+ \to K^+ K^- \pi^+$, Phys. Rev. D  {\bf 83}, 052001 (2011).

\bibitem{Ball2007}
P. Ball and G.W. Jones,  Twist-3 distribution amplitudes of K* and phi Mesons,
J. High Energy Phys. 03 (2007) 069.

\bibitem{chi2}
S. Baker, R.D. Cousins, Clarification of the use of $\chi^2$ and likelihood functions in fits to histograms,
 Nucl. Instrum. Methods Phys. Res. {\bf 221}, 437 (1984).

\bibitem{AF}
 A. Furman, R. Kami\'nski, L. Le\'sniak, and B. Loiseau,    
 Long-distance effects and final state interactions in $B \to \pi\pi K$ and $B \to K {\bar K} K$ decays,
Phys. Lett. B {\bf 622}, 207 (2005).

\bibitem{LL}
L. Le\'sniak,  Meson spectroscopy and separable potentials,
Acta Phys. Pol. B {\bf 27}, 1835 (1996), https://www.actaphys.uj.edu.pl/R/27/8/1835/pdf.

 \bibitem{Be2010}
Y. Nakahama \textit{et al.} (Belle Collaboration),  Measurement of CP violating asymmetries in $B^0 \to K^+ K^- K_S^0$ decays with a time-dependent Dalitz approach, Phys. Rev. D \textbf{82},  073011 (2010).

 \bibitem{BA2012}
J. P. Lees \textit{et al.} (\textit{BABAR} Collaboration),   Study of CP violation in Dalitz-plot analyses of $B^0 \to K^+ K^- K_S^0,  B^+ \to K^+ K^- K^+$ and $B^+ \to  K_S^0 K_S^0 K^+$,  Phys. Rev. D \textbf{85}, 112010 (2012).

\bibitem{NA48} 
J.R. Batley \textit{et al.} (NA48/2 Collaboration), 
 Precise tests of low energy QCD from $K_{e4}$ decay properties, Eur. Phys. J. C {\bf 70}, 635 (2010).

\bibitem{Moussallam_Nov2018} 
B.~Moussallam (private communication).

\bibitem{KLRyb}
R. Kami\'nski, L. Le\'sniak, K. Rybicki, 
 Separation of the $S$-wave pseudoscalar and pseudovector amplitudes in $\pi^+ \pi^- n$ reaction on polarized target, Z.~Phys. {\bf C 74}, 79 (1997).

\bibitem{Cohen}
D. Cohen \textit{et al.},  Amplitude analysis of the $K^- K^+$ system produced in the reaction $\pi^- p \to 
K^- K^+ n$ at 6 GeV/c, Phys. Rev. D {\bf 22}, 2595 (1980).

\bibitem{Etkin}
A. Etkin \textit{ et al.},
 Amplitude analysis of the $K^0_S K^0_S$ system produced in the reaction $\pi^- p \to 
K^0_S K^0_S n$ at 23 GeV/c, Phys. Rev. D {\bf 25}, 1786 (1982).

\end{thebibliography}
\end{document}